\documentclass[12pt]{article}

\usepackage{euscript}
\usepackage{amssymb}
\usepackage{amsfonts}
\usepackage{amsbsy}
\usepackage{amsmath}
\usepackage{epsfig}
\usepackage{amsthm}
\usepackage{amscd}
\usepackage{amstext}

\def\ben{\begin{equation}}
\def\een{\end{equation}}

\let\a=\alpha  \let\g=\gamma \let\d=\delta 
  \let\q=\theta \let\k=\kappa

 \let\r=v
 \let\t=\tau

\let\w=\omega \let\G=\Gamma

\def\nn{\nonumber}

\let\pa=\partial
\def\be{\begin{eqnarray}}
\def\ee{\end{eqnarray}}
\def\ba{\begin{array}}
\def\ea{\end{array}}

\def\dalemb#1#2{{\vbox{\hrule height .#2pt
       \hbox{\vrule width.#2pt height#1pt \kern#1pt
               \vrule width.#2pt}
       \hrule height.#2pt}}}

\newcommand{\bea}{\begin{eqnarray}}
\newcommand{\eea}{\end{eqnarray}}

\def\R{{{\Bbb R}}}

\def\ocal{{\mathcal{O}}}

\def\2apl{\left(\frac{2 \pi
\a'}{L^2}\right)}

\newcommand{\eqn}[1]{(\ref{#1})}
\newcommand{\ap}{\alpha^\prime}
\begin{document}

\begin{flushright}
NSF-KITP-09-208, SU-ITP-09/51, SLAC-PUB-13847, DAMTP-2009-80
\end{flushright}

\vspace{0.2cm}

\begin{center}

{ \LARGE {\bf Towards strange metallic holography}}

\vspace{0.2cm}

Sean A. Hartnoll$^{\sharp, \natural}$, Joseph Polchinski$^{\natural}$, Eva Silverstein$^{\natural,\dagger}$ and David Tong$^{\flat, \natural}$

\vspace{0.2cm}

{\it ${}^\sharp$ Department of Physics, Harvard University,
\\
Cambridge, MA 02138, USA \\}
\vspace{0.2cm}

{\it ${}^\natural$ Kavli Institute for Theoretical Physics and Department of Physics, \\
University of California, Santa Barbara, CA 93106, USA \\}

\vspace{0.2cm}

{\it ${}^\dagger$ on leave from SLAC and Department of Physics, Stanford University, \\
Stanford, CA 94305, USA \\}

\vspace{0.2cm}

{\it ${}^\flat$ Department of Applied Mathematics and Theoretical Physics, \\
University of Cambridge, Cambridge, CB3 OWA, UK \\}

\vspace{0.3cm}

{\tt  hartnoll@physics.harvard.edu, joep@kitp.ucsb.edu,\\ evas@stanford.edu,
d.tong@damtp.cam.ac.uk} \\

\vspace{0.2cm}

\end{center}

\begin{abstract}

We initiate a holographic model building approach to `strange metallic' phenomenology.
Our model couples a neutral Lifshitz-invariant quantum critical theory, dual to a bulk
gravitational background, to a finite density of gapped probe charge carriers,
dually described by D-branes. In the physical regime of temperature
much lower than the charge density and gap, we exhibit anomalous scalings of the
temperature and frequency dependent conductivity. Choosing the dynamical critical exponent $z$ appropriately we can
match the non-Fermi liquid scalings, such as linear resistivity, observed in
strange metal regimes. As part of our investigation we outline
three distinct string theory realizations of Lifshitz
geometries: from F theory, from polarised branes, and from a
gravitating charged Fermi gas.
We also identify general features of renormalisation group flow in Lifshitz theories, such as the appearance
of relevant charge-charge interactions when $z \geq 2$.
We outline a program to extend this model building approach to other anomalous observables of interest such as the Hall conductivity.

\end{abstract}

\pagebreak
\setcounter{page}{1}

\newpage
\tableofcontents
\newpage

\section{Introduction}
\setcounter{equation}{0}

Some of the most interesting challenges in condensed matter physics involve
strongly interacting systems of fermions and other components. The difficulty is to understand `non-Fermi liquid' (NFL) behavior, which is widely believed to require physics going beyond weakly interacting fermions. Of particular interest
are the thermodynamic and transport properties of the `strange metal' phases of heavy fermion compounds \cite{stewart} and high temperature superconductors \cite{hussey1, hussey2}.  A prime example of this is DC resistivity linear in temperature over several decades of temperature $T$, with $T$ much less than the chemical potential $\mu$ of the system, e.g. \cite{martin}.  Other aspects of strange metal phenomenology include possible nontrivial power-law tails in the AC conductivity ($\sigma(\omega)\sim\omega^{-\nu}$ with $\nu\ne 1$ over a range of scales according to \cite{optical})
and anomalous behavior of the Hall conductivity, e.g. \cite{hall}.

Even at the theoretical level, few (if any) calculations reproduce the observed behavior in a controlled quantum field theory.  In this work, we present some basic results in this direction, exhibiting non-Fermi-liquid behaviors such as linear resistivity in a controllable -- though unrealistic -- class of field theories with a holographic dual description.  Another, complementary, class of holographic systems with strange metallic behaviors appears in \cite{MIT}. We will comment on some similarities and differences between the two classes below.\footnote{In particular, we will address the backreaction of the bulk fermi sea in \cite{MIT}\ on the black hole solution used in the analysis, and find a significant effect.}

The holographic correspondence \cite{Aharony:1999ti}\ provides powerful techniques for analyzing a class of strongly coupled quantum field theories.  It is natural to explore these theories at finite charge density.  At the very least this allows us to understand certain strongly correlated many-body systems much better at a theoretical level, and this investigation may ultimately lead to mechanisms for real world phenomena.\footnote{For introductions to the holographic approach to finite density systems see \cite{Hartnoll:2009sz, Herzog:2009xv, McGreevy:2009xe, Hartnoll:2009qx}.}
Therefore, although current holographic technology applies only to extreme limits of special quantum field theories, it is worthwhile to study the physics of strongly interacting fermions and to investigate mechanisms for strange metal behaviors in this context where reliable calculations can be made.  What we learn this way may also back react on our understanding of holography and string theory.

\begin{figure}[h]
\begin{center}
\includegraphics[height=150pt]{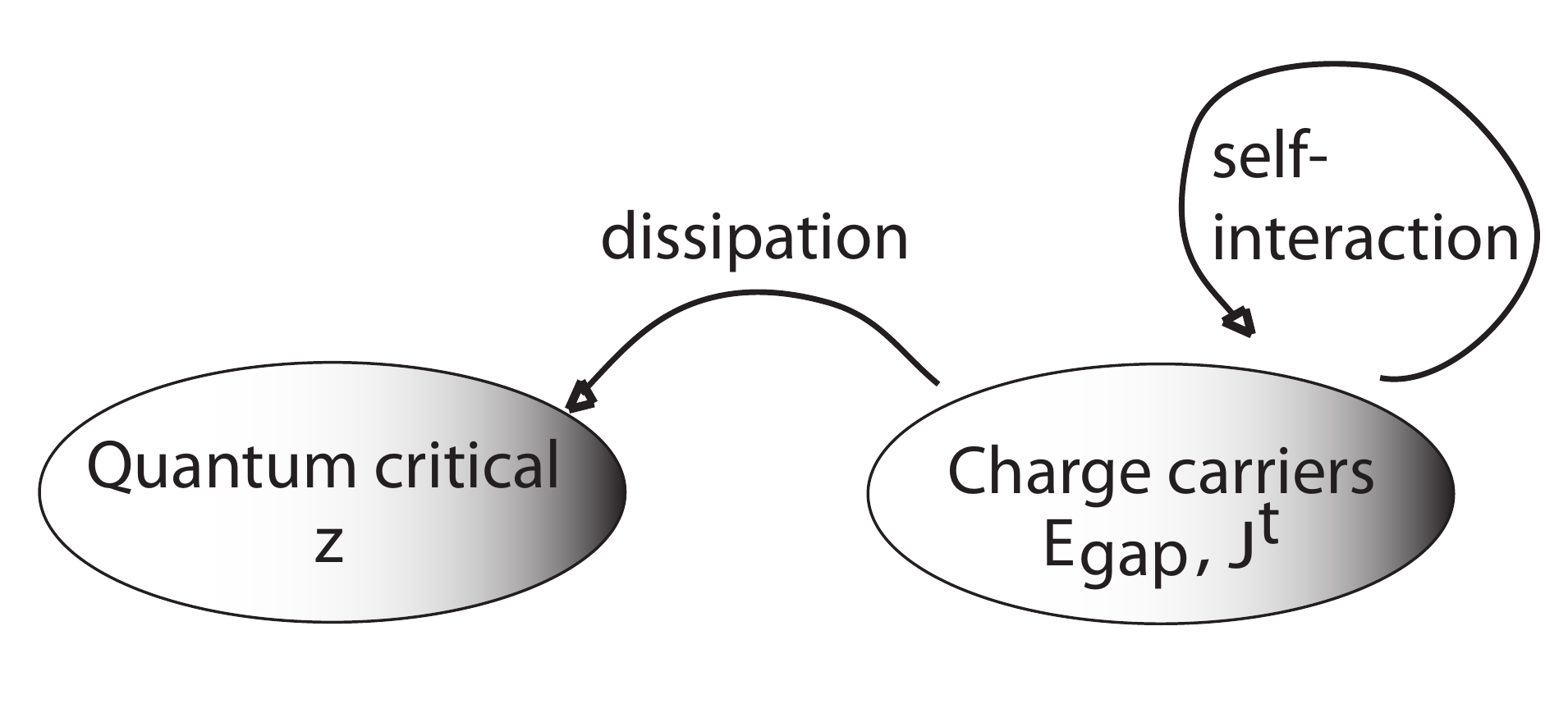}
\end{center}
\caption{Our model will describe probe charge carriers interacting with a quantum critical Lifshitz bath. Parameters include the dynamical scaling exponent $z$, the energy gap $E_\text{gap}$ and density $J^t$ of the carriers. Ultimately the charge-charge interactions are mediated by the Lifshitz sector.}\label{fig:intro}
\end{figure}

The theories we will study involve a sector of (in general massive) charge carriers, in a state of nonzero charge density $J^t$, interacting amongst themselves and with a larger set of neutral quantum critical degrees of freedom. The logical structure is illustrated in figure \ref{fig:intro}.
The quantum critical sector has Lifshitz scale invariance with dynamical critical exponent $z$.  We begin with a brief summary of the scaling properties and renormalization-group (RG) structure of such quantum field theories in \S2.  In a dilute limit, $J^t \ll T^{2/z}$, this structure leads to a simple formula for their resistivity in any dimension, $\rho\propto T^{2/z}/J^t$, which is linear in temperature for $z=2$. This however is not the regime of physical interest -- we will later recover the same formula dynamically in the opposite, physical, limit $J^t \gg T^{2/z}$. The scaling symmetry also implies that for $z$ greater than or equal to the spatial dimensality, a marginal or relevant interaction $\int dtd^d\vec x J^t J^t$ arises among charge carriers.  We then turn to a holographic analysis of such systems.  We use probe `flavor branes' to model the sector of charge carriers, along the lines of the earlier work \cite{kob,KulaxiziParnachev,Karch:2009eb}, but now applied to a bulk theory with Lifshitz scaling \cite{Kachru:2008yh}\ (see also \cite{Koroteev:2007yp}).  (In our final section, for (UV-)completeness we provide three methods for constructing Lifshitz fixed points from the top down, obtaining $z=2$ in the simplest examples.)  After analyzing the holographic manifestation of the renormalization-group structure, we compute the specific heat and the DC, Hall, and AC conductivities of our system and comment on their similarities and differences with respect to the corresponding results for strange metals.

This opens up some new directions, which we outline at various points in the present paper.
For example, we can analyze the basic scales in holographic superconductors in this context, exploring the relationship between $T_c$, the dynamical critical exponent $z$ which determines the strange metallic behaviors, and other parameters.  New model-building possibilities suggest themselves as generalizations of our basic setup.
In particular, having determined the results for the basic transport coefficients in our Lifshitz field theories coupled to charged flavors, we will find it useful to consider generalizations with running couplings arising from radially rolling scalars on the gravity side of the holographic duality.  This suggests mechanisms for mixing and matching non-Fermi-liquid behaviors such as
\be\label{eq:exponents}
\rho\sim T^{\nu_1} ~~~~ {\rm and} ~~~~ \sigma(\omega)\sim \omega^{-\nu_2} \,,
\ee
for different nontrivial exponents $\nu_1$ and $\nu_2$ (though in our simplest setup, $\nu_1=\nu_2$).  Moreover, there are many possibilities for multiple flavor sectors subject to gauge and global symmetries which organize them into composites that might mock up various scenarios for fractionalization of the electron.  We leave for future work the detailed construction of theories based on these mechanisms.

A key limitation of current holographic theories vis \`a vis the real world is that our theoretical control arises in the unrealistic limit of a large-rank gauge symmetry, for example $U(N_c)$ Yang-Mills theory at large $N_c$.  In the present case, we use an expansion in $N_f/N_c$, where $N_f$ is the number of charged flavors, in order to control the calculations.   One would ultimately hope for control of more realistic theories with mutually interacting sectors without such large disparities.

\section{Dimensional analysis, $z$ and renormalization}
\label{sec:dimension}
\setcounter{equation}{0}

We wish to study the thermodynamic and transport properties of
charge carriers interacting with a strongly coupled and scale invariant quantum field theory.
The quantum critical theory will be neutral under the charge.
We will work in a limit in which the neutral quantum critical theory has many more degrees of freedom than the charged `flavor' sector. This can be measured for instance using the free energy. So long as we stay within a range of density and scales where this `probe flavor' description is valid, then the charge-carrying flavors have a negligible effect on the state of the neutral sector. We will discuss regimes of validity below, as well as give a critical assessment of the phenomenological relevance of this limit.

Spatially isotropic scale invariance is characterized by the dynamical critical exponent $z$ \cite{hertz}.
The theory is invariant under space and time rescaling of the form
\be\label{eq:scaling}
t \to \lambda^z t \,, \qquad \vec x \to \lambda \vec x \,.
\ee
This scale invariance is often called a Lifshitz invariance.
Invariance under this scaling forces physically meaningful observables to appear in specific ratios in order to be dimensionless. This is usefully implemented by assigning time and space the following dimensions of momentum
\be
[t] = -z \,, \qquad [\vec x] = -1\,.
\ee
We can now work out the scaling dimension of various quantities of interest, which we collect here for future reference. Throughout we work with $\hbar = k_B = e = 1$. The charge and current densities have
\be
[J^t] = d \,, \qquad [\vec J] = d + z - 1 \,,
\ee
where $d$ is the number of space dimensions. The former follows from the definition of $J^t$ as a density while the latter follows from charge conservation $\dot J^t + \nabla \cdot \vec J = 0$. The dimensions of external scalar ($\Phi$) and vector ($\vec A$) potentials are fixed by the fact that these appear gauging derivatives. The dimensions of electric and magnetic fields then follow as
\be
[\Phi] = z \,, \qquad [\vec A] = 1 \,, \qquad [\vec E] = z+1 \,, \qquad [\vec B] = 2 \,.
\ee
The temperature and free energy both have dimensions of energy. This leads to the following dimensions for the specific heat and the magnetic susceptibility
\be
[T] = z \,, \qquad [F] = z \,, \qquad [c_V]= d \,, \qquad [\chi] = z - 4 \,.
\ee
Finally, the dimensions of conductivity follow from Ohm's law to be
\be
[\sigma] = d-2 \,.
\ee
In particular, the conductivity is dimensionless in $d=2$ spatial dimensions.

This simple dimensional analysis leads to the following statement.  Consider a system with an energy gap $E_{\rm gap}$ to exciting charge carriers which is large compared to the temperature.  {\it If} the conductivity in a Lifshitz system is linear in the density $J^t$ of charge carriers, then by the scaling given above we can conclude that the resistivity $\rho = 1/\sigma$ scales like
\be\label{eq:resistivity} \rho\propto \frac{T^{2/z}}{J^t} \,. \ee
This result is independent of the spatial dimension $d$.  Here we are using the fact that increasing the energy gap should not lead to larger conductivity in order to exclude significant $E_{\rm gap}$-dependent contributions to (\ref{eq:resistivity}). (Contributions to the conductivity which decrease with $E_{\rm gap}$ are negligible in the limit of large $E_{\rm gap}/T$.)

%We will obtain this result in the regime of interest, $T\ll\mu$, for attempting to connect to strange %metal phenomenology (with $\mu$ the chemical potential), as well as in an essentially trivial dilute %regime with $\mu\ll T$.

When the chemical potential $\mu \ll T$, linearity of the conductivity in the density is immediate: this regime corresponds to very low density, where the conductivity is linear in $J^t$ because it is simply the sum of the individual contributions of non-interacting charge carriers.
%Here $\mu$ is the chemical potential, equivalent to a constant background potential $\Phi$.

However, we will also find, using the approach of \cite{kob}, that in an extreme limit of
holographic systems with probe flavor branes the result (\ref{eq:resistivity}) persists
for  $\mu\gg T$, which is the regime of interest for strange metal phenomenology. Here
the self interactions of the charge carriers are non-negligible. The linearity of the conductivity as a function of charge density in these more general cases may arise because in the probe limit $\mu J^t \ll F_\text{QCT}$, where $F_\text{QCT}$ is the free energy of the quantum critical theory (QCT) into which the momentum of the charge carriers is dissipated.  Roughly speaking, the interactions among the charge carriers may be a subdominant effect on the (DC) resistivity, even though these interactions are important enough to preclude a quasiparticle interpretation of the charge carriers. We will make this statement a little more precisely below, suggesting that it is related to the fact that without the neutral QCT `medium'  to carry away momentum, the DC conductivity would be infinite.  The mobility $\sigma/J^t$ as a function of doping has been studied experimentally in e.g. \cite{Ando}, exhibiting weak dependence that may be consistent with (\ref{eq:resistivity}).\footnote{We thank S. Kivelson for pointing this paper out to us.}  Note that in contrast to single-scale models such as that discussed in \cite{Phillips}, where $J^t$ is taken to scale with temperature as $T^{d/z}$, in our system $J^t$ is an independent scale.

%{\tt does this really make sense? Again, sometimes things scale the same at weak and strong coupling for no clear %reason.  --ES}

%\label{subsec:RGQFT}
In fact, independently of the holographic correspondence, we can see from the RG structure of our theory that interactions among charge carriers will necessarily be important in the case $z\ge d$.
The dimension of $J^t$ being $d$, the operator $J^tJ^t$ becomes marginal at $z=d$, and relevant for $z>d$.  For $d=2$ -- the dimensionality of interest for many unconventional real materials such as high-$T_c$ superconductors -- this transition happens at $z=2$, the value of $z$ for which the resistivity is linear.  In general, for $z\ge d$, this operator is important at low energies in our theory, leading to additional interactions among charge carriers.  As a relevant operator for $z>d$, its coefficient is naturally at the UV cutoff scale of the system.\footnote{One could formally introduce counterterms to cancel this divergence, but this would constitute a fine tuning in our system. We discuss this fact in some detail in section {\ref{sec:notes}} below.}

\section{Probe D-branes in IR scaling geometries}
\setcounter{equation}{0}
\label{sec:scaling}

We are primarily interested in the low temperature and low energy behaviour of the theory. Low temperatures and energies will be defined with respect to some energy scale: $T,E \ll \Lambda_\text{UV}$. In particular, we will restrict attention to theories for which the neutral sector we defined above becomes quantum critical at these low energies. Here $\Lambda_\text{UV}$ should presumably be
of order the lattice scale (i.e. electron volts), although this may be larger than the melting temperature of the solid, allowing scaling laws to persist up to the melting point, as in e.g. \cite{martin}. Quantum criticality means that there are no intrinsic scales in the low energy (IR) theory, the only scales will be external: temperature $T$, electric and magnetic fields $E, B$ and the density of charge carriers $J^t$. Later we will add an energy gap scale $E_\text{gap}$ for the charge carriers.  In the systems we study, we will see that for $J^t \ne 0$, our window of control in which the charge carriers do not back react on the geometry does not extend all the way into the infrared (as long as all parameters are finite), but still covers a wide range of scales in our probe approximation.

For concreteness, and with a view to ultimately connecting to interesting experimental systems, we focus on 2+1 dimensional field theories, with 3+1 dimensional bulk duals. The dual IR geometry therefore takes the following form at zero temperature \cite{Kachru:2008yh}
\be\label{eq:IRmetric}
ds^2_\text{IR} = L^2 \left( - \frac{dt^2}{\r^{2z}} + \frac{d\r^2}{\r^2} + \frac{dx^2 + dy^2}{\r^2} \right) \,.
\ee
This metric realises the scaling symmetry (\ref{eq:scaling}) as an isometry, together with $\r \to \lambda \r$. The radial coordinate therefore has dimensions of length and extends from the (singular) `horizon' $\r=\infty$ to the `boundary' $\r = 0$. We will require the above metric to give the correct physics for a window of radial positions $\r$ satisfying
\be\label{eq:window}
\r_\text{br}\gg \r \gg \epsilon \equiv \frac{1}{\Lambda_\text{UV}^{1/z}} \,.
\ee
where $\r_\text{br}$ is an infrared `backreaction' radial scale at which our probe approximation breaks down; we will quantify this shortly. We will implement the UV cutoff approximately by taking the metric (\ref{eq:IRmetric}) to be valid up to $\r = \epsilon$ and imposing boundary conditions there.

\
The full background will generally have nonzero matter fields supporting the metric (\ref{eq:IRmetric}), such as those described in \cite{Kachru:2008yh}.
%compatible with the scaling symmetry.
%We will give some examples in the following section.
Furthermore,
when embedded into a consistent quantum gravity theory, such as string theory, there may be additional spatial dimensions to those shown.  We will outline three classes of examples of string-theoretic constructions of infrared Lifshitz geometries later in the paper.

When placed at a finite temperature the metric can be written as
\be\label{eq:IRmetricT}
ds^2_\text{IR} = L^2 \left( - \frac{f(\r) dt^2}{\r^{2z}} + \frac{d\r^2}{f(\r) \r^2} + \frac{dx^2 + dy^2}{\r^2} \right) \,.
\ee
The precise form of $f(\r)$ will depend on the theory and various solutions of this form have been constructed \cite{lifbh,ohmann,bertoldi,bm}. All we will require is the presence of a horizon, $f(\r_+)=0$, which defines the temperature
\be\label{eq:temp}
T = \frac{|f'(\r_+)|}{4 \pi \r_+^{z-1}} \propto \frac{1}{\r_+^z} \,.
\ee
In the second relation there is an order one number which we do not know explicitly unless $f(\r)$ is given. Our normalisation is such that at the boundary $f(0) = 1$.
In order to maintain control over our calculations, we may consider cases in which the infrared back reaction scale $\r_\text{br}$ is cloaked by a black hole horizon: $\r_\text{br} > \r_+$.

We now turn to the dynamics of a probe D-brane in the background (\ref{eq:IRmetricT}). The background describes a quantum critical theory; we are interested in the physics of a small number of charge carriers interacting with this theory. By a `small number' here we mean that the carriers do not backreact on the quantum critical system. As we will emphasize, this does not imply that the charge carriers are weakly interacting amongst themselves; in general they will have significant interactions mediated by the quantum critical sector.  A probe D$q$ brane is described by the Dirac-Born-Infeld (DBI) action
\be\label{eq:dq}
S_q = - T_q \int d\tau d^q\sigma\ e^{- \phi} \sqrt{\left |{}^\star g + 2 \pi \alpha' F \right|} \,.
\ee
The nonlinearity of this action in the field strength $F$ encodes the interactions between carriers.
In general the D$q$ brane can also have Chern-Simons like couplings to bulk field strengths. We will ignore these for the moment. In (\ref{eq:dq}) ${}^\star g$ is the pullback of the metric  (\ref{eq:IRmetricT}), $F=dA$ is the field strength of a worldvolume $U(1)$ gauge field and $e^{-\phi}$ is the dilaton. In order for the background solution to respect the scale invariance, $\phi$ and $T_q$ must be constant in the IR region.

We look for an embedding given by
\be\label{eq:embedding}
\t = t \,, \quad \sigma^1 = x \,,  \quad \sigma^2 = y \,, \quad \sigma^3 = \r \,, \quad \{\sigma^4 \ldots \sigma^q \} = \Sigma \,,
\ee
together with the gauge potential
\be
A = \Phi(\r) dt + B x dy \,.
\label{A1}\ee
In (\ref{eq:embedding}), $\Sigma$ refers to a submanifold of an internal space. If the background spacetime is a direct product of (\ref{eq:IRmetricT}) with an internal space $M$, the simplest way to solve the equations of motion is for $\Sigma$ to be a stationary submanifold of $M$, independent of $\{t,x,y,\r\}$.
Many backgrounds of interest are not direct products and many probe brane embeddings of interest are not constant in the internal directions. Nonetheless, for the moment we will take a `phenomenological' approach and consider that the only effect of internal dimensions is to multiply the overall D$q$ brane action by the volume of $\Sigma$. The effective brane in $3+1$ dimensions thus has tension
\be\label{eq:teff}
\t_\text{eff.} = T_q \text{Vol}(\Sigma) e^{-\phi} \,.
\ee

The assumption in (\ref{eq:embedding}) that the D-brane does not bend into the transverse dimensions will shortly translate into the assumption that the charge carriers are gapless. While this may be relevant for materials with a Dirac-cone dispersion relation for electronic excitations (along the lines of graphene), or other situations in which there are emergent gapless charge carrying excitations, in general we will wish to consider massive charge carriers. We will consider the massless case first for simplicity, and generalise to the massive case in section \ref{sec:massive}.

It is straightforward to solve the equations of motion for $\Phi$ to obtain
\be\label{eq:sol}
F_{\r t} = \Phi' = \frac{1}{\r^{1+z}} \frac{C}{\sqrt{\r^{-4} + \left(\frac{2 \pi \a'}{L^2}\right)^2 (B^2 + C^2)}} \,,
\ee
where $C$ is a constant of integration. Near the boundary $\r \to 0$ the potential is expanded as
\be
\Phi = \mu - \frac{1}{\r^{z-2}} \frac{C}{z-2} + \cdots \,,
\label{seanat}\ee
for $z \neq 2$ and
\be
\Phi = \mu + C \log \frac{\r}{\Lambda} + \cdots \,,
\label{seanot}\ee
when $z=2$. In this last case $\mu$ has a scheme dependence on a scale $\Lambda$.
We think of $\mu$ as the chemical potential, although for $z \geq 2$ it is not the largest mode near the boundary. We will discuss this phenomenon in detail in the following section.
The coefficient $C$ is proportional to the charge density,
\begin{equation}\label{eq:jt}
J^t = \tau_{\rm eff}(2\pi\ap)^2 \, C\ ,
\end{equation}
as one reads off from the boundary term that arises upon varying the action with respect to $\delta A_t^{(0)} = \delta \mu$.

Evaluating the action on this solution gives,
\be\label{eq:free}
\frac{TS_q}{V_2} = - \t_\text{eff.} L^4 \int_{\epsilon}^{\r_+} d \r \frac{1}{\r^{1+z}}\frac{\r^{-4} + B^2}{\sqrt{\r^{-4} + B^2 + C ^2}} \,.
\ee
Here $V_2$ is the spatial volume. In this and the following few expressions, we will drop the factors of $\frac{2 \pi \a'}{L^2}$ that appear multiplying $B$ and $C$.
Expanding the integrand for small $\r$, the contribution from the UV endpoint is
\be\label{eq:wdiv}
 \frac{TS_q}{V_2}=  \t_\text{eff.} L^4 \left( - \frac{1}{z+2} \frac{1}{\epsilon^{z+2}} + \frac{C ^2- B^2}{2(z-2)} \frac{1}{\epsilon^{z-2}}
+ \frac{B^4 - 3 C ^4 - 2 B^2 C ^2}{8 (z-6)} \frac{1}{\epsilon^{z-6}} + \cdots \right)
\ee
for $z \neq 2$. For $z=2$ we have
\be
\frac{TS_q}{V_2}=  \t_\text{eff.} L^4 \left( - \frac{1}{4} \frac{1}{\epsilon^{4}} + \frac{B^2- C ^2}{2} \log \frac{\epsilon}{\Lambda}
+ \cdots \right) \,.
\ee
For all positive $z$ the leading term is divergent as $\epsilon \to 0$.  This term is independent of the temperature and all other parameters, and reflects the fact that the energy density is dominated by UV physics.  For the relativistic case, $z=1$, this is the only divergence, but for $z \geq 2$ the second term diverges as well. The coefficient of this divergence depends on the magnetic field and charge density.  Again, naturalness requires that we include this as representing a UV sensitivity of the physics.  In the next section we will analyze why a divergence appears at $z = 2$, and why additional such effects appear as $z$ is increased further.

When we vary the action to obtain the specific heat and other observable quantities, depending on the application we may wish to hold fixed either the charge density $J^t$ or the chemical potential $\mu$.
%it is the charge density $J^t$ and not the chemical potential $\mu$ that should be held fixed in order to stay as close as possible to the real systems we wish to mock up.  This is because in the real systems, the Coulomb interaction enforces overall charge neutrality: the density of carriers is therefore equal to the density of dopants. Although the $U(1)$ symmetry is not gauged in our field theory, in real materials it is coupled to a physical Maxwell field.
In our setup we can implement a fixed $J^t$ by adding the familiar `Neumannizing' term \cite{hr}.
In the following section we will discuss boundary conditions in some detail and note that for $z>2$ fixed charge is in fact the `natural' boundary condition in a renormalisation group sense. The free energy is then
\be
f \equiv \frac{F}{V_2} = \frac{TS_q}{V_2} +\mu J^t\ .
\ee
By integrating (\ref{eq:sol}) from the horizon, where $\Phi=0$, to near the boundary and comparing to (\ref{seanat}), we obtain
\be \mu J^t = \frac{\t_{\rm eff.}L^4 C^2}{(z-2) \r_+^{z-2} }\, {}_2F_1\left(\frac{1}{2},\frac{2-z}{4},\frac{6-z}{4}, - (B^2 + C^2) \r_+^4  \right) \,.
\ee
%
%Inverting this relation in the low temperature ($r_+ \to \infty$) limit with $B=0$ gives
%
%\be
%C = \frac{(16 \pi \mu^2)^{1/z}}{\left(- \G(\frac{2-z}{4}) \G(\frac{z}{4}) \right)^{2/z}} \left(1 - \frac{2}{z^2 %r_+^z \mu} + \cdots \right) \,.
%\ee
%
For the case $z=2$ one has instead
\be
\mu J^t = \frac{\tau_{\rm eff.}L^4C^2}{2} \log \left(\frac{1 + \sqrt{1+ (B^2+C^2) \r_+^4}}{C\r_+^2} \right) \,.
\ee
where we have partially fixed the scheme dependence by requiring that this quantity remains finite as $\r_+\rightarrow \infty$.

With fixed charge, the divergences appearing in the free energy (\ref{eq:wdiv}) are temperature independent. The following difference of free energies is then finite
\bea
\Delta f & \equiv & f(T) - f(0) \label{eq:line1} \\
&  = & - \t_\text{eff.} L^4 \int_{\infty}^{\r_+} d \r \frac{1}{\r^{1+z}}\frac{\r^{-4} + B^2}{\sqrt{\r^{-4} + B^2 + C ^2}} + (\mu(T) - \mu(0)) J^t \label{eq:line2} \\
& = & \t_\text{eff.} L^4 \left( \frac{1}{z} \sqrt{B^2 + C ^2} \frac{1}{\r_+^z} + \frac{1}{2 (z+4) \sqrt{B^2 + C ^2}} \frac{1}{\r_+^{4+z}} + \cdots \right) \quad  \text{as $\r_+ \to \infty$} \nn \\
& \propto & \t_\text{eff.} L^4 \left( \sqrt{B^2 + C ^2} T + \frac{1}{\sqrt{B^2 + C ^2}} T^{1+4/z} + \cdots \right) \label{eq:lastline} \,.
\eea
In the last line we have not kept track of numerical coefficients, as we do not know the precise relation between $\r_+$ and the temperature $T$. The full integral in (\ref{eq:line2}) may be performed in terms of hypergeometric functions. However it is clear, as emphasised in \cite{Karch:2009eb}, that the low temperature free energy only depends on the zero temperature metric at $\r=\r_+$.

It is now simple from (\ref{eq:lastline}) to compute the specific heat.
The specific heat divided by temperature is
\be
\frac{c_V}{T} = - \frac{\pa^2 f}{\pa T^2} \,.
\ee
The linear term in (\ref{eq:lastline}) will drop out upon taking two derivatives, leaving the second term. Setting $B=0$, this gives the specific heat
\be
\frac{c_V}{T} \propto - \t_\text{eff}^2\, L^6 \, \a' \, \frac{T^{4/z-1}}{J^t}  \,.
\ee
Here we restored the $\alpha'$ factors. This is the leading small-$T$ behavior at fixed $J^t$. Within the probe approximation, to be made precise shortly, this scaling will always be subdominant to the thermodynamics of the Lifshitz sector.

The magnetic susceptibility
\be\label{eq:chi}
\frac{\chi}{V_2} = - \frac{\pa^2 f}{\pa B^2} \,,
\ee
has a UV sensitivity through $f(0)$ if $z \geq 2$. This gives a temperature independent term. At low temperatures from (\ref{eq:line1}) and (\ref{eq:lastline})
\be\label{eq:chieval}
\frac{\chi}{V_2} \propto -  \t_\text{eff.} \, L^2 \, \a'^2 \left( \frac{1}{L^2} \Lambda_{UV}^{1-2/z} + \t_\text{eff.} \, \a' \frac{T}{J^t} \right) \,.
\ee
where we have set $B$ to zero after differentiating. The dependence on the UV scale is logarithmic when $z=2$. In the absence of the UV divergence, the temperature independent term is proportional to  $(J^t)^{z/2-1}$. We have not been careful with the relative normalisation of the two terms in (\ref{eq:chieval}).

Let us consider the regime of control of our system (\ref{eq:dq}).
There are two issues to address. The first is the neglect of backreaction of the brane onto the metric.
In our background, the effective action $S_q$ takes the form (with $2 \pi \a' = 1$)
\be\label{eq:dqexplicit} S_q=- \t_\text{eff.} \int dt\, d^2x\, d\r \sqrt{-g} \sqrt{1+g^{tt}g^{\r\r}F_{\r t}^2+g^{tt}g^{xx}F_{tx}^2
+g^{\r\r}g^{xx}F_{\r x}^2} \,. \ee
To avoid back reaction of the probe on the metric, its stress-energy must be smaller than the stress energy generating the original background (\ref{eq:IRmetric}).  The original energy density is of order $M_4^2|\Lambda|\sim M_4^2/L^2$, where $M_4$ is the four-dimensional Planck mass and $\Lambda$ the four-dimensional cosmological constant.  Varying (\ref{eq:dqexplicit}) with respect to $g_{tt}$, we find this condition to be
\be\label{eq:brcondition} \gamma\equiv \frac{1+g^{\r\r}g^{xx}F_{\r x}^2}{\sqrt{1+g^{tt}g^{\r\r}F_{\r t}^2+g^{tt}g^{xx}F_{tx}^2
+g^{\r\r}g^{xx}F_{\r x}^2}}\ll \frac{M_4^2|\Lambda|}{\t_\text{eff.}} \,. \ee
In our solutions $\gamma$ approaches one at the boundary and grows toward larger $\r$, the region corresponding to the infrared regime of the field theory.  As long as the brane tension is sufficiently small, the right hand side allows for a window of scales in which $\gamma$ can grow larger than one (leading to nontrivial DBI dynamics, corresponding to interactions between the charge carriers) while satisfying the condition (\ref{eq:brcondition}). This is the regime $\r_\text{br} \gg \r$ of equation (\ref{eq:window}).

In the simplest examples of brane probes, such as those discussed in \cite{Karch:2009eb}, the probe limit requires a power law tune $N_c/N_f \gg 1$ where $N_c$ is the rank of the Yang-Mills gauge group of the field theory, and $N_f$ the number of charged matter fields (the number of probe branes).  More generally in the landscape, however, low-tension branes can arise naturally -- via an exponential hierarchy -- in compactifications with strong warping (gravitational redshift) in the extra dimensions. This effectively makes the internal volume $\text{Vol}(\Sigma)$ small in the tension (\ref{eq:teff}).

There is another way that the description can break down at large $\gamma$.  As $\gamma$ gets large the electric field on the brane is approaching its critical value, where the force on string endpoints exactly balances the string tension~\cite{Burgess:1986dw}.  Beyond this point the system is unstable to creation of open strings.  As the critical field is approached, the effective value of the open string tension falls as $1/\gamma^2$~\cite{Gopakumar:2000na,Seiberg:2000ms} and so the effective string length scale grows as $\gamma$.  When this exceeds the typical length scale of the geometry, supergravity will break down in an interesting way as string modes become important.  The condition for supergravity to be valid is then
\begin{equation}
\alpha'\gamma^2 \ll 1/|\Lambda|\ .
\label{eq:br2}
\end{equation}
The conditions~(\ref{eq:brcondition}) and~(\ref{eq:br2}) are compatible with one another but independent.  We can go to a regime where the first is satisfied but the second is not; this appears to correspond to a situation where the backreaction of the charges on the critical sector is small, but their interaction with each other due to finite density has become stronger than their 't Hooft coupling interactions (setting for instance $|\Lambda| \a' = \lambda^{-1/2}$).  In general, as we approach this regime, an expansion in small perturbations about the DBI action (\ref{eq:dqexplicit}) brings down inverse powers of the square root, i.e. powers of $\gamma$, leading to strong non-Gaussian effects \cite{DBIsky}.  It would be interesting to understand the role of these effects in holographic condensed matter systems.

\section{Renormalization and Lifshitz holography}
\label{sec:notes}
\setcounter{equation}{0}

We would like to understand in a more general way the divergences that we encountered with increasing $z$ in section 3.\footnote{For a recent work on Lifshitz holography see \cite{Horava:2009vy}.}  For simplicity we focus on the quadratic Maxwell action, which governs the dominant behavior of the full action (\ref{eq:dqexplicit}) near the boundary,
\begin{equation}
S \propto -  \int dt\, d^dx\, d\r\, \r^{-d-z-1} \left( \frac{1}{2} \r^{2+2z} F_{ti}^2 -\frac{1}{2} \r^4 F_{i\r}^2 + \frac{1}{2} \r^{2+2z} F_{t\r}^2 -\frac{1}{4} \r^4 F_{ij}^2 \right)\ .
\end{equation}
For the dominant boundary behavior we can ignore $i$ and $t$ derivatives in the field equations. With the pure Liftshitz background (\ref{eq:IRmetric}) this gives the equations of motion
\begin{equation}
\partial_\r (\r^{1 + z - d} \partial_\r A_t) = \partial_\r (\r^{3 - z - d} \partial_\r A_i) = 0\ ,
\end{equation}
with solutions
\begin{eqnarray}
A_t = \alpha + \beta  \r^{d-z}\ , \quad A_i = \alpha' + \beta'  \r^{d + z - 2}\ . \label{Aasymp}
\end{eqnarray}
In this limit $A_\r$ is pure gauge.  The field strengths scale as
\begin{eqnarray}
F_{ti} &\sim& \alpha  + \beta \r^{d-z} + \alpha' + \beta' \r^{d+z-2}\ ,
\nonumber\\
F_{ij} &\sim& \alpha' + \beta' \r^{d+z-2}\ ,
\nonumber\\
F_{t\r} &\sim& \beta \r^{d- z-1}\ ,
\nonumber\\
F_{i\r} &\sim& \beta' \r^{d + z - 3}\ . \label{Fasymp}
\end{eqnarray}

We now focus on $d=2$. In the relativistic case $z = 1$, the $\alpha$ and $\alpha'$ solutions are larger at the boundary $\r \to 0$.  We thus take the usual quantization, in which these are fixed while $\beta$ and $\beta'$ are dynamical.  That is, we fix the potentials and field strengths tangent to the boundary.  At $z=1$, all terms in the action are convergent at $\r \to 0$.  As we increase $z$, at $z\geq 2$ the $F_{ij}^2$ term has a divergence proportional to $\alpha'^2$, and the $F_{tr}^2$ term has a divergence proportional to $\beta^2$, and new boundary counterterms are needed to obtain a finite on shell action.

To understand these divergences from the point of view of the field theory, recall the momentum (inverse length) dimensions
from section \ref{sec:dimension}
\begin{equation}
[J^t] = d \ , \quad [ J^i] = d + z - 1  \ , \quad [A_t] = z \ , \quad [A_i] = 1 \ .
\end{equation}
The field theory contains an explicit $A_\mu J^\mu$ interaction, but will generate additional divergences for any gauge-invariant relevant interaction constructed from $A$ and $J$. Here $A$ is treated as a nonfluctuating spurion field, while $J$ is a single trace operator, and the new counterterms will in general involve multiple traces.  The field theory volume element has $[dt\, d^d x] = -d - z$, so an interaction will be relevant if its momentum dimension is less than or equal to $d+z$.
The dimension of $F_{ij}^2$ is 4, so this becomes relevant at $d + z = 4$.  The dimension of $(J^t)^2$ is $2d$, so this becomes relevant when $d = z$. For $d=2$, the critical $z$ is 2 for both operators.

The divergence from $F_{ij}^2$ involves the fixed $\alpha'$, so this is just an additive classical term.  It reflects the fact that the dominant momentum-, temperature-, and frequency-independent magnetic susceptibility will come from the UV when $z >2$. We saw this in equation (\ref{eq:chieval}) above.

The divergence from $F_{tr}^2$ at $z > 2$ is more subtle.  At the same time that $(J^t)^2$ becomes relevant, the $\alpha$ and $\beta$ solutions cross, and the latter dominates at the boundary $\r \to 0$.  Thus we are in the situation discussed for relativistic scalars in Ref.~\cite{Klebanov:1999tb}.  For a generic UV theory we will flow to the more stable boundary condition in which $\alpha$ is dynamical and $\beta$ is fixed.\footnote{Ref.~\cite{Witten:2003ya} studied the 2+1 dimensional relativistic ($z=1$) gauge theory and showed that it had two IR stable realizations, the second corresponding to the gauging of the $U(1)$ symmetry.}

It is tempting to `renormalize' the low energy effective theory, adding boundary counterterms to cancel the divergences. In the range $2 < z < 4$ the following counterterms would do the job
\be\label{eq:bdy}
S_{\rm{bdy.}} = \frac{1}{g^2_\text{eff.}} \int_{\epsilon}dt\,d^2x \sqrt{| {}^\star \gamma|} \left[\frac{1}{z+2} +
\frac{\zeta}{2} \left( F_{ij} F^{ij} - F_{t\r}F^{t\r} \right) \right] \,,
\ee
In this expression $g^2_\text{eff.} = 1 / {\tau_{\rm eff}(2\pi\ap)^2}$,  $\zeta = 1/(z - 2)$, and
${}^\star \gamma$ is the induced metric on the $r = \epsilon$ surface.
For the field-independent and $F_{ij}^2$ terms this just subtracts off the UV contribution and isolates that from the IR.  For the $F_{t\r}F^{t\r}$ term, however, the boundary terms actually change the theory, from the IR stable $\beta = 0$ theory to the tuned $\alpha = 0$ theory.  To see this, perform the variation of the bulk and boundary action with respect to $A_t$, and insert the asymptotics~(\ref{Aasymp}) to obtain
\begin{eqnarray}
\delta S &=& {\rm e.o.m.} - \frac{1}{g^2_\text{eff.}} \int_{\epsilon}dt\,d^2x \sqrt{| {}^\star \gamma| g_{\r\r}}
(\delta A_t + \zeta \partial_\r \delta A_t)  F^{t\r}
\nonumber\\
&\propto& \{ \delta\alpha +(1 + (2-z)\zeta) \epsilon^{2-z}\delta\beta \} \beta\ .
\end{eqnarray}
We see that precisely the value $\zeta = 1/(z - 2)$, that cancels the divergence, also gives the tuned $\alpha$-fixed theory. Throughout this paper we work with the untuned, $\beta$-fixed, theory. This has no boundary counterterms. We argued in the previous section that this fixed charge ensemble is in fact the physically correct one for condensed matter applications of holography, for all values of $z$.

The difference between the $\alpha$- and $\beta$-fixed theories is subtle: in the planar limit, only the correlators of $J^t$ are affected by the double trace deformation \cite{Witten:2001ua}, so most observables are the same.  Intuitively, a large $(J^t)^2$ interaction would inhibit local fluctuations of $J^t$, explaining why $\beta$ must be fixed.

The expansions~(\ref{Aasymp}, \ref{Fasymp}) have have higher order terms, e.g.\ at relative order $k^2 \r^2$, which will lead to further divergences as we increase $z$.  In the field theory, this is reflected by the operators $F_{ij,k}^2$ and $(J^t_{,k})^2$ becoming relevant at $z=4$.  As $z$ is further increased, higher spatial derivatives become relevant, so that in the $z\to \infty$ limit of $AdS_2 \times \R^2$ there is an infinite number of relevant operators.  Note however that terms with additional time derivatives never become relevant.  Higher powers of the fields can become relevant, e.g. $(F_{ij} F_{ij})^2$, $F_{ij} F_{ij} (J^t)^2$ and $(J^t)^4$ at $z=6$.

\section{Massless charge carriers}
\setcounter{equation}{0}

A key observable capturing strange metallic behavior is the conductivity.  We will start by analyzing the DC conductivity of our system, following closely the work of
Karch and O'Bannon \cite{kob}, keeping the full nonlinear dependence of the current resulting from a given constant electric field. Using similar methods \cite{obhall}\ we also compute the Hall conductivity.  Then we will obtain the optical conductivity by computing the linearized response of the system to small oscillating electric field perturbations. We focus first in this section on the massless case, to illustrate the techniques. Then in the next section we will include a mass for the charge carriers, motivated by the large energy gap of carriers relative to temperature in real-world strange metals.

\subsection{DC conductivity}

In order to compute the DC ($\w=0$) conductivity,
the strategy is to turn  on an electric field $E \equiv F_{tx}$ on the D-brane probe and compute the resultant current $\langle J^x\rangle$ in the boundary theory. This then allows us to directly read off the field and temperature dependent conductivity $\sigma(E,T)$ from Ohm's law
\be\label{eq:sigma}
J^x = \sigma(E,T) E \,.
\ee
To this end, we revise the ansatz \eqn{A1} for the worldvolume gauge field, now looking for solutions
of the form
\be A = \Phi(\r)dt + (-Et+h(\r)) dx \,. \ee
With this ansatz, the action \eqn{eq:dq} becomes
\be S = -\tau_{\rm eff}\int dt\,d^2x\,d\r\, \sqrt{g_{xx}}
\sqrt{-g_{tt}g_{xx}g_{\r\r} - (2\pi\ap)^2\left(g_{\r\r}E^2 + g_{xx}\Phi^{\prime\,2}
+g_{tt}h^{\prime\,2}\right)}\ .
\label{s2}
\ee
The action depends only on the derivatives $\Phi^\prime$ and $h^\prime$, resulting in
two quantities which are independent of the radial direction $\r$,
\be C &=& \frac{-g_{xx}^{3/2}\,\Phi^\prime}{\sqrt{-g_{tt}g_{xx}g_{\r\r} - (2\pi\ap)^2\left(g_{\r\r}E^2 + g_{xx}\Phi^{\prime\,2}
+g_{tt}h^{\prime\,2}\right)}} \nn \,, \ee
and
\be H = \frac{-g_{tt}\sqrt{g_{xx}}\,h^\prime}{\sqrt{g_{tt}g_{xx}g_{\r\r} - (2\pi\ap)^2\left(g_{\r\r}E^2 + g_{xx}\Phi^{\prime\,2}
+g_{tt}h^{\prime\,2}\right)}} \,. \ee
These are the same expressions found in \cite{kob} and, as pointed out by those authors, obey the relation $g_{tt}h^\prime C = g_{xx}\Phi^\prime H$. The near-boundary profile of $\Phi$ is once again given by
\eqn{seanat} for $z\neq 2$ and \eqn{seanot} for $z=2$. Meanwhile,
the asymptotic behavior of $h(\r)$ is
\be \label{eq:hlarge} h(\r) \rightarrow h_0 + \frac{H}{z}\,\r^z+\ldots \,. \ee
We set $h_0=0$ and identify the coefficient of the decaying term with the current
\be
J^x = \tau_{\rm eff}(2\pi\ap)^2 \, H \ .
\ee
This equation is completely analogous to that defining the charge density in (\ref{eq:jt}).
The normalization factor of $\frac{1}{z}$ in (\ref{eq:hlarge}) follows from computing $J^x$ as the derivative of the on-shell action with respect to $h_0$.
The on-shell bulk action \eqn{s2} can be written as,
\be S = \tau_{\rm eff} \int dtd^2xdv\ g_{xx}^{3/2}\sqrt{-g_{tt}g_{\r\r}}
\,\left[\frac{g_{tt}g_{xx}+(2\pi\ap)^2E^2}{(2\pi\ap)^2(g_{tt}C^2+g_{xx}H^2)+g_{xx}^2g_{tt}}\right]^{1/2} \,.
\ee
The divergences (UV sensitivities) of this expression are as discussed in the previous sections, and do not involve $E$.

As pointed out in \cite{kob}, both the numerator and the denominator of $[\ldots]^{1/2}$ change sign between the boundary $\r=0$ and the horizon $\r=\r_+$ (recall that $g_{tt}<0$). The reality of the action means that this sign change must take place at the same point in the radial direction, $\r_+>\r_\star>0$, such that both
\be \left.\begin{array}{c}\ -\end{array}g_{tt}g_{xx}\right|_{\r=\r_\star} = (2\pi\ap)^2 E^2 \,,
\label{whereise}\ee
and
\be \left. (2\pi \a')^2 \left(g_{tt} C^2 +g_{xx}H^2 \right) =-g_{xx}^2g_{tt}\right|_{\r=\r_\star}\ ,
\ee
should be satisfied.\footnote{It might appear that a boundary condition is being imposed at the point $\r_\star$, rather than the horizon, but in fact this is equivalent to the usual condition of ingoing b.c.\ on the horizon.  One can study this by approximating the near-horizon geometry as Rindler, for which a finite ingoing wave satisfies the DBI field equation, and then taking the zero frequency limit.  Outgoing b.c.\ give the opposite sign for $H$.}
Using the finite temperature metric \eqn{eq:IRmetricT}, the first of these equations fixes $\r_\star$ in terms of the electric field,
\be f(\r_\star) = {\textstyle{ {\2apl^2}}} E^2\,\r_\star^{2z+2}\ .
\label{rstar}\ee
Meanwhile, the second equation can be rewritten to give the sought-after equation for the conductivity,
\be \sigma(E,T) = \sqrt{(2\pi\ap)^4 \tau_{\rm eff}^2 +  {\textstyle{\2apl^2}}\r_\star^4(J^t)^2}
\ .\label{conduct}\ee
The right-hand-side of \eqn{conduct} is the root-mean-square of two terms: the first is a constant piece and arises from
thermally produced pairs of charge carriers. It is expected to be Boltzmann suppressed when the
charge carriers have large mass, as we will see shortly.
The surviving term exhibits the simple power-law (\ref{eq:resistivity}) for the DC resistivity, namely
\be\label{eq:resistivityII} \rho\sim \frac{T^{2/z}}{J^t} \,. \ee
As discussed in \S2, one situation in which this behavior is generic is a regime of dilute charge carriers which are coupled to a Lifshitz bath in such a way as to inherit its scaling symmetry (\ref{eq:scaling}).  The diluteness of the charge carriers implies that the conductivity is approximately linear in $J^t$ and the rest follows from dimensional analysis. However, here we see that the linearity in $J^t$ arises in the massless case only in the concentrated (i.e. non-dilute) regime $J^t \gg T^{2/z}$, while in the very massive case it arises for all $J^t/T^{2/z}$ in the DC conductivity.  We will discuss this further below.

The first term in (\ref{conduct}) is independent of both temperature $T$ and electric field $E$. This is due to the fact we are in a $3+1$ dimensional bulk, rather than any Lifshitz scaling. This same constant term was seen in Section 5 of \cite{kob}. The second term on the right hand side contains the dependence on the temperature and on the electric field. Both of these arise through $\r_\star$ defined in \eqn{rstar}. To compute the nonlinear, $E$ dependent corrections to the conductivity, we need to know the  specific function $f(\r)$, which will depend on the the matter content sourcing the Lifshitz background. On dimensional grounds, such corrections depend on the ratio $\frac{(2\pi\ap)^2}{L^4}\,\frac{E^2}{T^{2+2/z}}$. In the relativistic case ($z=1$), the nonlinearities of the conductivity in the electrical field could be elegantly understood by considering the drag force on a single string and using Lorentz invariance \cite{kob}. This does not appear to be the case at general $z$.

A translationally invariant medium with a net charge density should have an infinite DC electrical conductivity. Specifically, the real part should have a delta function and the imaginary part should have a pole: $\sigma(\w) \sim \delta(\w) + i\, \w^{-1}$ at small $\w$. This can be seen either directly from hydrodynamics or via the holographic correspondence (e.g. \cite{Hartnoll:2007ih, Hartnoll:2007ip}). The underlying reason is that the conserved momentum cannot relax and, combined with a net charge density, this gives a current that does not relax. Yet we have just obtained a finite DC conductivity. The reason for this \cite{kob} is that in the probe limit the momentum can be transferred to the quantum critical `bath' without any backreaction on the charged probe system. Technically, the coefficient of the divergence goes to zero in the probe limit, so that the probe and $\w \to 0$ limits do not commute. A physical circumstance in which this probe approximation could be legitimate is if
the quantum critical excitations are more efficient at dissipating heat into the environment (via interaction with impurities etc.) than the charge carriers.

The fact that the DC conductivity would diverge in the absence of the Lifshitz bath suggests a heuristic understanding of why this conductivity is linear in the charge density, if the first, constant, term in (\ref{conduct}) is taken to be Boltzmann suppressed. This is not an immediate result as interactions between the charges are important as evidenced, for instance, by the nonlinear dependence of the free energy (\ref{eq:free}) on the charge density. The fact that the Lifshitz medium is responsible for making the potentially infinite conductivity finite suggests that medium-carrier interactions are playing a dominant role in the DC limit. The diluteness of the carriers with respect to the medium then suggests that this interaction will be proportional to the density of carriers, leading to the linear dependence of the finite DC conductivity on $J^t$.

\subsection{DC Hall conductivity}

The techniques of \cite{kob} can be extended to compute the conductivity tensor,
\be J^i = \sigma^{ij}E_j \,.
\ee
The calculations are identical to those of \cite{obhall} and we present only the results. The
conductivity is once again expressed in terms of a function
$\r_\star(T,E,B)$, defined by the requirement that
\be
-g_{tt}g_{xx}^2= (2\pi\ap)^2(g_{tt}B^2+g_{xx}E^2) \Big|_{\r = \r_\star} \,,
\ee
which generalizes \eqn{whereise}. For $E$ and $B$ small, this gives $\r_\star \sim \r_+ \sim 1/T^{1/z}$.
Corrections to this expression
are functions of the dimensionless ratios $E/T^{1+1/z}$ and $B/T^{2/z}$. The Hall conductivity
has a simple expression in terms of $\r_\star$,
\be \sigma^{xy} = -
\frac{J^t B\r_\star^4}{({\textstyle \frac{L^2}{2\pi\ap}})^2+ B^2\r_\star^{4}} \,. \ee
Notice that the Hall conductivity is automatically linear in charge density. When both $B$ and $E$ are small,
this becomes $\sigma^{xy} \sim T^{-4/z}$. The expression for $\sigma^{xx}$ is
\be \sigma^{xx} = \frac{1}{1+(\frac{2 \pi \a'}{L^2})^2 B^2\r_\star^4}\,\sqrt{(2 \pi \a')^4
\t_\text{eff}^2 [1+({\textstyle \frac{2 \pi \a'}{L^2}})^2
B^2\r_\star^4]+({\textstyle\frac{2\pi\ap}{L^2}})^2(J^t)^2\r_\star^4} \,.
\ee
It's simple to see that this reduces to our previous expression when $B=0$. In particular, when
the $J^t$ term dominates in the square root,  and $B$ is small, then we reproduce the result $\sigma^{xx} \sim \r_\star^2 \sim T^{-2/z}$.

Among the interesting, anomalous, results exhibited by strange metals is the ratio $\sigma^{xx}/\sigma^{xy}$.
The anomalous behavior $(\sigma^{xx})^{-1}\sim T$ is accompanied in the cuprates by the scaling
$\sigma^{xx}/\sigma^{xy}\sim T^2$ (e.g. \cite{hall}). This is to be contrasted with Drude theory\footnote{Recall that in Drude theory: $\sigma^{xx} = \frac{e^2 n}{m \tau} \frac{1}{\tau^{-2} + e^2 B^2 m^{-2}}$ while $\sigma^{xy} = \frac{e^3 n B}{m^2} \frac{1}{\tau^{-2} + e^2 B^2 m^{-2}}$. Thus $\frac{\sigma^{xx}}{\sigma^{xy}} = \frac{m}{eB\t} = \frac{B e n}{\sigma^{xx}(B=0)}$.} which implies $\sigma^{xx}/\sigma^{xy}\sim(\sigma^{xx})^{-1}$.  Within our probe calculation, this ratio is given by
\be\label{hallratio}
\frac{\sigma^{xx}}{\sigma^{xy}} = -\left(\frac{L^2}{2 \pi \a'}\right)^2
\frac{1}{J^t B\r_\star^4}\, \sqrt{(2 \pi \a')^4 \t_\text{eff}^2 [1+({\textstyle \frac{2 \pi
\a'}{L^2}})^2 B^2\r_\star^4]+({\textstyle\frac{2\pi\ap}{L^2}})^2(J^t)^2\r_\star^4} \,.
\ee
The relevant experimental limit for the ratio is when the first term is subdominant in the square root, leading to
%$B$ dominates in the square root, leading to
$\sigma^{xx}/\sigma^{xy} \propto T^{2/z}\sim(\sigma^{xx})^{-1}$.
%\sim B T^{2/z}/J^t$.
We see that this aspect of the probe computation does not reveal strange behavior, but rather mimics the Drude result.
In a later model-building section we will consider generalizations of the setup which might evade this conclusion.

\subsection{AC conductivity}
\label{sec:massless}

%\section{Gauge field fluctuations}

Let us next calculate the frequency dependent conductivity.  In this case, we will focus on the linear response rather than working out the full nonlinear dependence on the electric field as we did in the DC case.  To do this, we will expand in small fluctuations about the background (\ref{eq:sol}), working at zero magnetic field ($B=0$) and zero momentum for simplicity.  As before we extract the conductivity from the ratio of non-normalizable and normalizable modes of $A_x$ near the boundary $\r \to 0$, having introduced a background electric field $E_x(t)\equiv \text{Re}\, E_x(\omega)e^{-i\omega t}$:
\be\label{eq:condexpansion} A_x(\omega) = \frac{E_x(\omega)}{i\omega}+\frac{J_x(\omega)}{z \t_\text{eff.} (2 \pi \a')^2}v^z  + \cdots \,.
\ee
The coefficients in this expansion will be determined by solving the bulk equations of motion, with the ratio between the response $J_x$ and the source $E_x$ obtained from a boundary condition at the horizon ensuring that the former is determined causally (via the retarded Green's function) from the latter. The implementation of ingoing boundary conditions at the horizon \cite{Son:2002sd} is by now standard, see e.g. \cite{Hartnoll:2009sz} for a discussion.

The fluctuations of the probe gauge fields take the form
\be\label{eq:deltaA}
\delta A = (A_t(\r) dt + A_x(\r) dx + A_y(\r) dy) e^{- i (\w t -
k x)} \,.
\ee
The quadratic action for fluctuations about the background solution (\ref{eq:sol}) is found to be
\be
S^{(2)} = - \frac{\t_\text{eff.} (2 \pi \a')^2}{2} \int d\r
d^3x \,
\r^{1-z} \g \left( f F_{\r i}^2
- \frac{\r^{2z-2}}{f} F_{ti}^2 - \r^{2z-2} \g F_{t\r}^2 + \frac{1}{\g} F_{xy}^2
\right) \,,
\ee
where $\gamma$ is defined as
\be
\g = \sqrt{1 +  {\textstyle{\2apl^2}} C^2 \r^4} \,,
\label{1gamma}\ee
as in (\ref{eq:brcondition}) and $i$ runs over $x$ and $y$.

For simplicity, we specialize to the case $k=0$, applicable when the applied field has wavelength much longer than the mean free path of charge carriers.  In this case, the equations of motion for the transverse and longitudinal fields are the same; let us focus on the longitudinal ($x$-) component.
\be
\r^{1-z}  f  \left(\r^{1-z} \g f A_x' \right)'  = - \g \w^2 A_x \,.
\ee
%For the longitudinal mode, one needs to eliminate $A_t$ when $k \neq 0$ (at $k=0$ it's the same as the other mode). %This satisfies
%\be
%\w \, r^{2z-2} \g A_t' + k f A_\parallel' = 0 \,.
%\ee
%The usual way to do this is to obtain a second order equation for $A_\parallel'$.
%Setting
%\be
%F_\parallel = r^{1-z} f A_\parallel' \,,
%\ee
%then the equation is found to be
%\be
%r^{1-z} \sqrt{ \g} f  \left(r^{1-z} \sqrt{\g} f F_\parallel' \right)'
%+ 2 f (\g-1) r^{-2z} \left(r f' - z f + \frac{4f}{\g}\right) F_\parallel
%= \left(r^{2-2z} f k^2 - \g \w^2\right) F_\parallel
%\,.
%\ee
It is useful to map this equation into a Schr\"odinger form
\be
- \frac{d^2\Psi}{ds^2} + U \Psi = \w^2 \Psi \label{erwin}\,.
\ee
This is achieved with the change of variables
\be\label{eq:varchange} A_x = \frac{1}{\g^{1/2}} \Psi \,, \ee
with
\be \frac{d}{d\r} = \frac{\r^{z-1}}{f} \frac{d}{ds} \,, \ee
%
%\be
%\left.
%\begin{array}{c}
%A_\parallel
%\end{array}
%\right\}
% = \frac{1}{\g^{1/4}} \Psi \,, \qquad \frac{d}{dr} = \frac{r^{z-1}}{f} \frac{d}{ds} \,,
%\ee
leading to the potential
\be\label{eq:V}
U = \frac{(\g-1) f}{\g^2 \r^{2z}} \left((\g+1) \left(1- z + \frac{3}{\g^2} \right) f + \r f' \right) \,.
\ee
%V = \frac{f}{\g^2 \, v^{2z}} \left(\left( 2+z + (1-z) \g^2 - \frac{3}{\g^2} \right) f  + (\g-1)v f' \right) \,,
%and
%\be
%V_\parallel = \frac{f}{\g \, r^{2z}} \left(\left( - 6 - z + (1+z) \g + \frac{5}{\g} \right) f  + (1-\g)r f' + r^2 k^2 %\right) \,.
%\ee
%
The AC conductivity is obtained by imposing ingoing boundary conditions at the horizon $s\to\infty$, ensuring a causal relationship between $E_x$ and $J_x$.  We will shortly present numerical results for the conductivity, after commenting on some of the physics evident from the above formulae.

The potential $U$ (\ref{eq:V}) exhibits different behavior for different ranges of charge density $C$ and $z$.  At zero charge density, $C=0$, we have $\gamma=1$ and the potential vanishes.  It is then straightforward to solve analytically for $\sigma(\omega)$, which is nonzero and constant at all temperatures
\be\label{eq:sigma0}
\sigma(\w) = \t_\text{eff.} (2 \pi \a')^2 \equiv \sigma_0  \, .
\ee
In the absence of any ambient charge density, the current can only arise from thermal fluctuations or particle production. The constancy of the result, technically following from the absence of scattering when $U=0$, reflects more than just the scaling symmetry of our system. General quantum critical theories in 2+1 dimensions have a conductivity $\sigma(\w/T)$ that tends to different constant values as $\w \to 0$ and $\w \to \infty$ \cite{sachdev}. The frequency independence here is related to additional symmetries of the Maxwell and DBI actions \cite{Herzog:2007ij}.

The above $C=0$ result is for massless charge carriers. However, in the real world systems we are ultimately interested in modeling, the effective energy gap $E_\text{gap}$ of the charge carriers is often greater than the temperature, and the current resulting from their thermal production should be Boltzmann suppressed.  Their mass is also often greater than the frequency scale $\omega\sim 10^3-10^4 \text{cm}^{-1}\sim 10^{-1}-1 \text{eV}$ of the electric field applied in the relevant measurements (see Fig 14 of \cite{ACreview} for an example in the cuprates).  So their dynamical production by the oscillating electric field should also be suppressed.  As a result, we do not expect the order one constant value we just obtained in our $C=0$ massless calculation to survive in a more realistic treatment.  We will study the massive case in the next section.

Having made this cautionary remark, let us continue to analyze the physics of the solution.
The potential $U$ of (\ref{eq:V}) approaches zero at the horizon.
As we approach the boundary $\r\to 0$, $U$ becomes of order $C^2\r^{4-2z}$.
For $z<2$, the potential is everywhere bounded, and approaches zero at the boundary.  For $z=2$, $U$ approaches a constant value of order $C^2$ at the boundary, and for $z>2$ the potential blows up at the boundary.
Again $z=2$ is a marginal case separating two behaviors.

The regime $\frac{(2 \pi \a')^2}{L^4} C \ll 1$ is the dilute limit, in the sense that the charge density is small compared to the temperature scale. Specifically
\be\label{eq:lll}
\frac{1}{\t_\text{eff.} 2\pi \a' L^2} \frac{J^t}{T^{2/z}} \ll 1 \,.
\ee
In this limit it is immediate that the potential (\ref{eq:V}) becomes proportional to $C^2$. The conductivity $\sigma(\w)$ is directly related to the reflection amplitude for scattering off the potential \cite{Horowitz:2009ij}. The correction to the constant result (\ref{eq:sigma0}) will therefore be proportional to $C^2$. This simplifies the general scaling form of the conductivity when the charge
density is small,
\be\label{eq:dilute}
\sigma\left(\frac{\w}{(J^t)^{z/2}}, \frac{J^t}{T^{2/z}}\right) \sim \t_\text{eff.} (2 \pi \a')^2 + \frac{1}{\t_\text{eff.}^2 (2\pi \a')^2 L^4} \frac{(J^t)^2}{T^{4/z}} {\mathcal F}\left(\frac{\w}{T}\right) + \cdots \,.
\ee
In this limit, DBI nonlinearities due to the charge density have been made small, the interactions between charge carriers are negligible and one might have expected the conductivity to be proportional to the density. However, in this case of massless charge carriers, the leading correction to the constant conductivity is found to be quadratic in the density. The observable we are computing here would be perhaps best characterized as a mobility rather than a conductivity. The result (\ref{eq:dilute}) is consistent with the previous DC result (\ref{conduct}). In (\ref{conduct}) the linear dependence on the charge density arises in the opposite limit to (\ref{eq:lll}), in which interactions between the charges are important. It is worth emphasizing again therefore that the calculation of the conductivity which we are using does capture nonlinearities in the charge density, and the linearity emerging in the DC limit at large density is nontrivial.

We now move away from the dilute limit and explore numerically the dependence of the dissipative conductivity $\text{Re}\, \sigma(\omega)$ for different values of $z$. To proceed,  we need to make a choice for $f(\r)$. This will depend on the particular matter fields sourcing the Lifshitz background. For generic examples, one can expect the asymptotic behavior
$f \sim 1 - \r^{2+z}$ near the boundary $\r \to 0$. This is because the normalisable mode of $g_{tt}$ is dual to the energy density $T^{tt}$, which has scaling dimension $[T^{tt}] = z + 2$. Indeed such asymptotic behavior was found in the simple models of \cite{Bertoldi:2009vn, Ross:2009ar}, which used a massive vector field coupled to gravity. Thus, for illustration we will take
\be\label{eq:fform}
f = 1 - \left(\frac{\r}{\r_+}\right)^{2+z} \,.
\ee
The temperature of these backgrounds is, from (\ref{eq:temp}), $T = (2+z)/4\pi\r_+^z$.
We will comment below on the sensitivity of the results to this choice of $f$. The resulting conductivities are shown in figure \ref{fig:sigma} below.

\begin{figure}[h]
\begin{center}
\includegraphics[height=130pt]{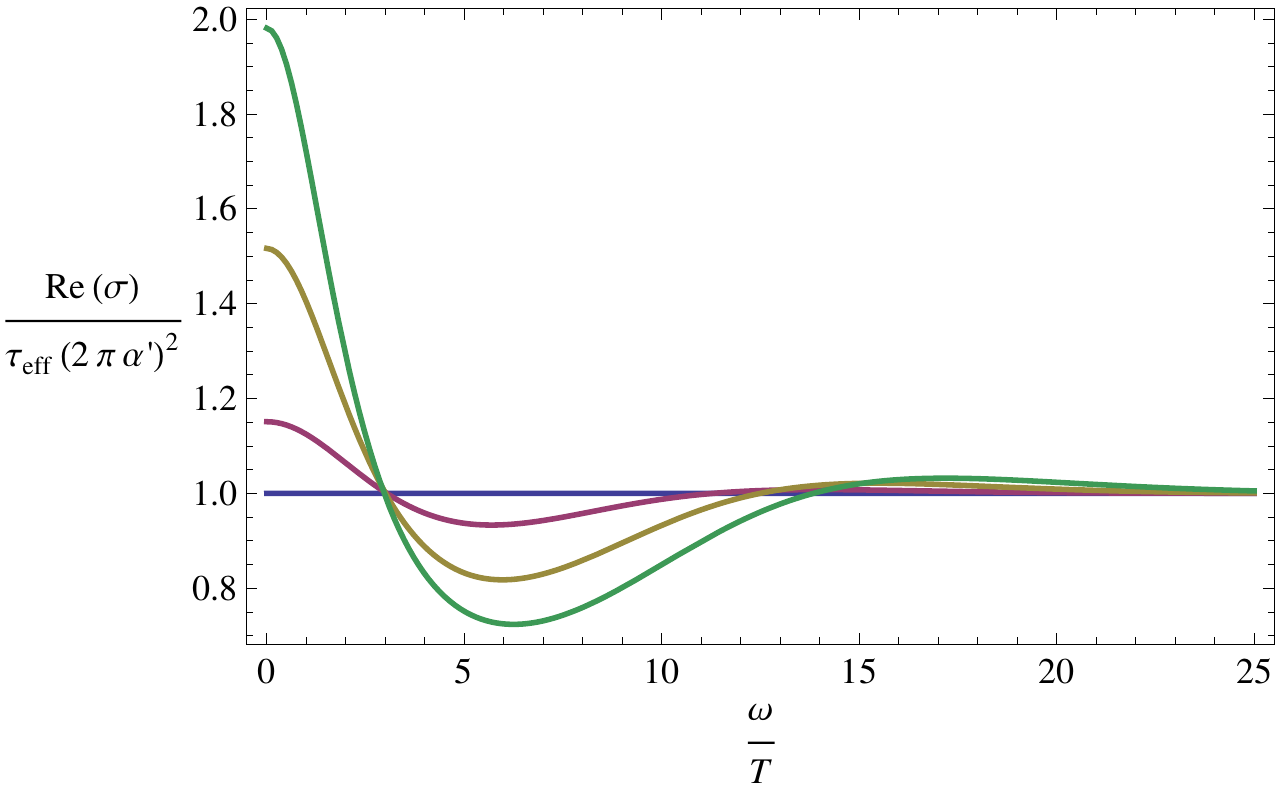}\includegraphics[height=130pt]{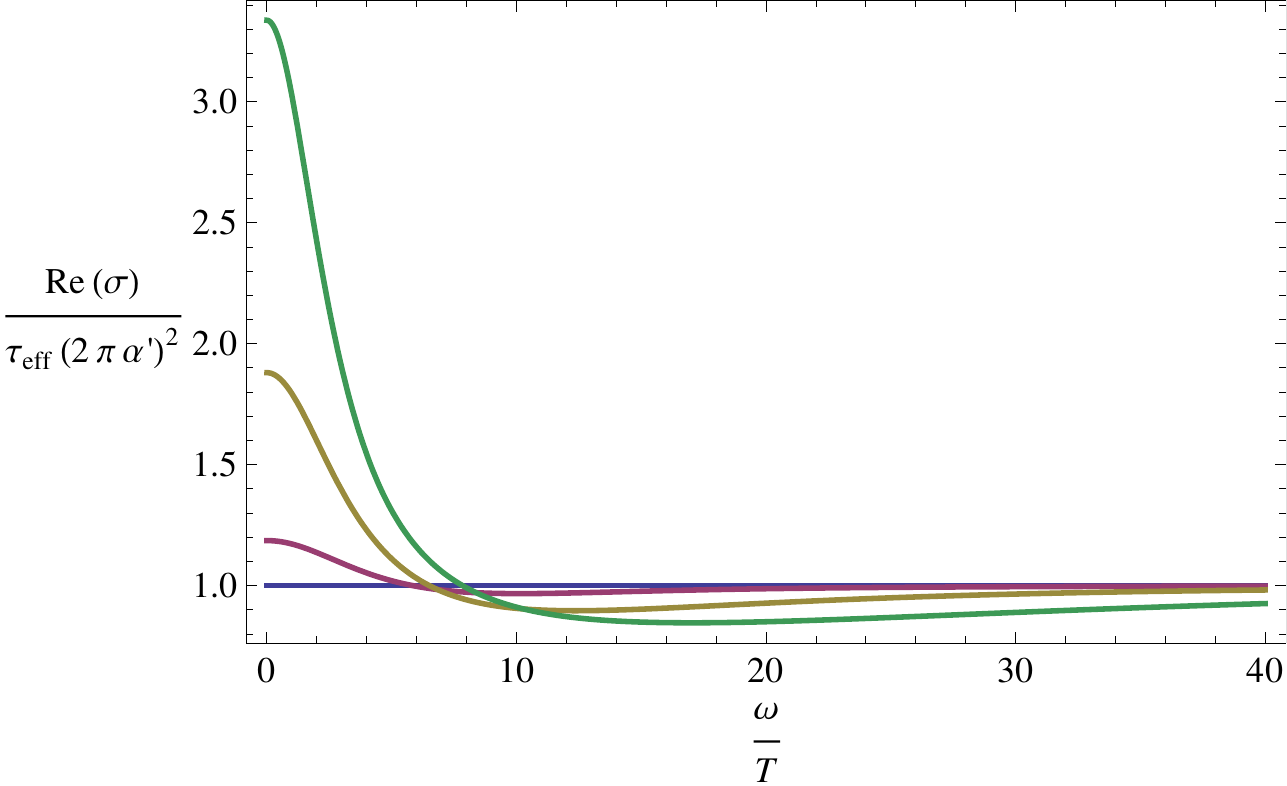}
\end{center}
\caption{The real part of the conductivity as a function of frequency for $z=1$ (left) and $z=2$ (right).
The four curves in each graph correspond to $\frac{1}{\t_\text{eff.} 2\pi \a' L^2} \frac{J^t}{T^{2/z}}$ equal to $\{0,10,20,30\}$ (left) and $\{0,2,5,10\}$ (right), with $J^t=0$ giving the expected constant lines.}\label{fig:sigma}
\end{figure}

For all values of $z$, the conductivity exhibits a peak reminiscent of Drude theory at $\omega=0$, approaches a nonzero constant value $\sigma_0$ at $\omega\to \infty$, and exhibits a dip in the middle.
Using the form for $f$ in (\ref{eq:fform}), it is easy to evaluate (\ref{conduct}) in the large charge density limit to obtain
\be
\frac{\sigma(\w=0)}{\t_\text{eff.} (2\pi \a')^2} = \frac{1}{\t_\text{eff.} 2 \pi \a' L^2} \frac{J^t}{T^{2/z}} \left(\frac{2+z}{4 \pi} \right)^{2/z} \,.
\ee
The final $z$ dependent term leads to the peak being bigger for a given $J^t/T^{2/z}$ at larger z, as seen in the figure. All the conductivities exhibit a dip at intermediate frequencies. This can be understood from a sum rule: one can straightforwardly show using the Kramers-Kr\"onig relations that $\int_0^\infty d\omega \text{Re}\, \sigma(\omega)$ is independent of the dimensionless ratio $J^t/T^{2/z}$. In using the Kramers-Kr\"onig relations it is important that the conductivity tends to its asymptotic value $\sigma_0$ sufficiently quickly in $\w$. For $1<z<2$, $\text{Re}\, \sigma(\omega)$ approaches $\sigma_0$ from above; that is, it has a second peak (albeit much smaller than the Drude-like peak).  For $z>2$, $\text{Re}\, \sigma(\omega)$ approaches $\sigma_0$ from below.

Plotting the Schr\"odinger potential one can see that there is dip close to the horizon where the potential becomes negative. The dip is not sufficiently big to allow negative energy bound states (which would lead to an instability of the spacetime), but it does allow for a low energy resonance. This is the `Drude peak'. This observation gives some indication of how sensitive our numerical results are to the form of $f$ chosen in (\ref{eq:fform}). Experimentation shows that the overall shape of the conductivity plots is fairly robust if we do not modify $f$ drastically. That is, if we do not modify the minimal form of the potential in which there is a dip at the horizon that is then connected smoothly onto the asymptotic boundary behavior discussed above. However, if we introduce oscillations into $f$ in such a way that additional dips are introduced into the potential, then we can get additional peaks in the conductivity. Typically one finds a second or more peaks at low frequencies that are smaller than the peak at $\w=0$. Curiously, such additional peaks also arise if one takes (\ref{eq:fform}) with $z<1$.

In real-world strange metals, $\text{Re}\, \sigma(\omega)$ has a Drude-like peak at $\omega=0$, and  approaches zero at large $\omega$ more slowly than in Drude theory (see e.g. Figure 14 of \cite{ACreview}).  In our case, as discussed above, the massiveness of charge carriers as compared to the temperature and $\omega$ should lead to suppression of the $C=0$ conductivity $\sigma_0$.
Before turning to the massive case, we make one final observation.

Although we have set the momentum $k$ to zero in our computations, in it straightforward in principle to work with a finite momentum. One interesting observation is that the combination of momentum and energy appearing in the Schr\"odinger equation is
\be
v^{2-2z} f k^2 - \g^2 \w^2 \,.
\ee
The feature of interest in this combination is that the coefficient of $k^2$ goes to zero at the horizon, while that of $\w^2$ does not. This leads to the existence of low energy modes with arbitrary momentum, manifested for instance in a nonzero spectral density at zero temperature. This has something of the flavor of a Fermi surface; with a weakly coupled Fermi surface there are zero energy modes with finite momentum connecting different points on the Fermi surface. These exist for $k < 2 k_F$ and lead to a sharp feature at $k = 2 k_F$. While interesting properties of the finite $k$ perturbation spectrum were found in \cite{KulaxiziParnachev,Karch:2009eb,Karch:2008fa}, no such sharp feature was observed.
This suggests that D-brane probe theories in the relativistic regime  $\gamma\gg 1$ do not describe weakly coupled fermions.  It is worth scrutinizing these systems more closely, taking into account effects that appear as $\gamma$ grows, cf (\ref{eq:window}) and (\ref{eq:brcondition}).  In addition to the back reaction on the metric, there are important effects in the open string sector that arise as the electric field $F_{tr}$ approaches the critical value, $\gamma\to\infty$ \cite{Seiberg:2000ms,Gopakumar:2000na}.  Fermi surfaces have been found directly in other holographic systems in \cite{MIT}\ (see \cite{Rey}\ for an early approach to this problem).

\section{Massive charge carriers}\label{sec:massive}
\setcounter{equation}{0}

We will now study the effect of including a nonzero mass $m$ for the charge carriers described by the flavor brane, as in \cite{kob}.  As discussed above, this is the case of interest in modeling some features of real-world strange metals, whose energy gap is large compared to the temperature:
\be
E_{\rm gap}\gg T \,.
\ee
For instance $E_\text{gap}$ might be at the lattice eV scale which is larger than the melting temperature of the relevant materials.

\begin{figure}[h]
\begin{center}
\includegraphics[height=160pt]{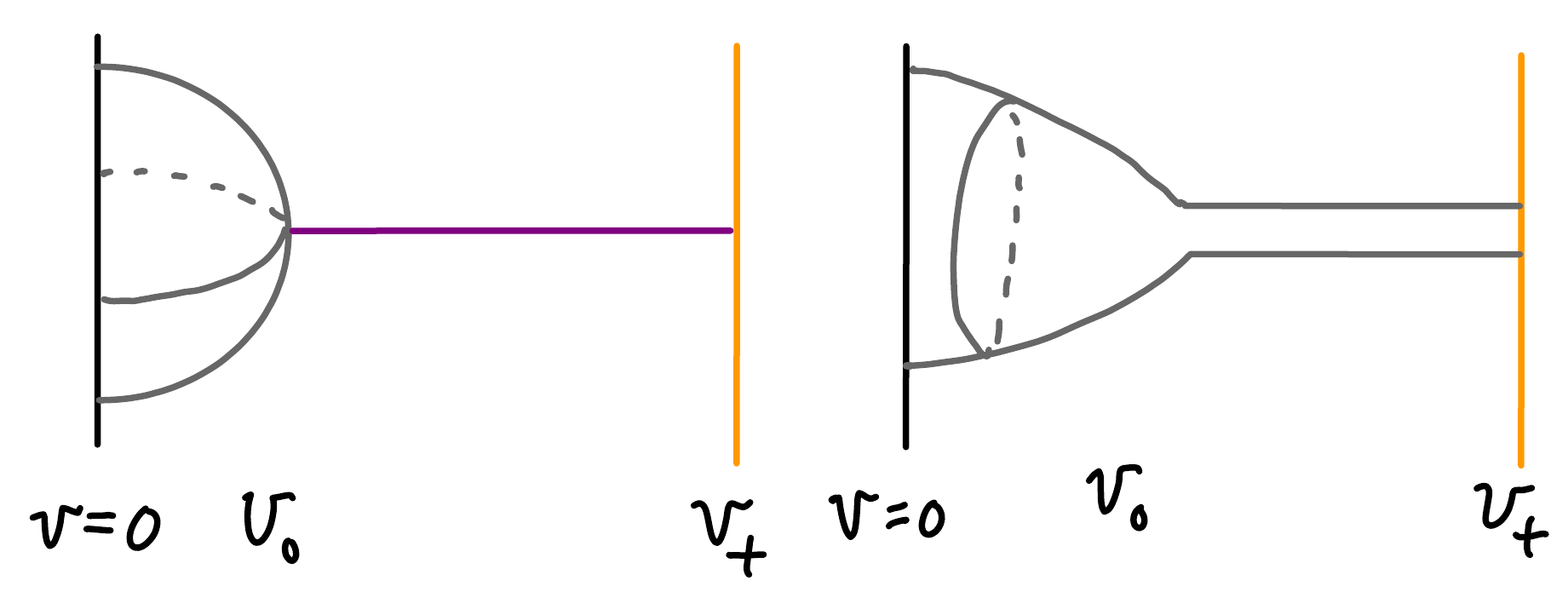}
\end{center}
\caption{Schematic depiction of the massive flavor brane in the string regime (left) and spike regime (right).}
\label{fig:spike}
\end{figure}

As discussed in previous works on flavor branes \cite{KarchKatz}\ and on finite-density holography \cite{myers}, a finite mass scale involves a configuration in which the volume of the internal cycle the flavor brane wraps varies with radial position, shrinking toward the horizon.  In the case of massive flavors at $T=\mu=0$  \cite{KarchKatz}, the volume shrinks smoothly to zero at a finite radial position $v=v_0$; the brane forms a cigar-like shape with its tip at $v_0$.  Charge carriers correspond to strings stretching from the tip of the cigar down to the Poincar\'e horizon $v=\infty$.  At finite temperature, a black hole horizon arises at a finite radial position $v_+$.  For large enough mass (i.e. small enough $v_0$), the flavor brane still shrinks to a point outside the black hole horizon.  Charge carriers in this $0 < T \ll E_\text{gap}, \mu=0$ theory correspond to strings stretching from the flavor brane at $v_0$ to the horizon at $v_+$.  To obtain finite density and temperature, in a dilute limit one can simply introduce a small density of such strings and ignore back reaction on the brane configuration.  We will refer to this as the string regime.  For larger densities, the back reaction of the charge density on the brane is important; the upshot of this will be that the brane forms a `spike' or `tube' stretching to the horizon from $v_0$ in place of the bundle of strings that pertained in the dilute limit \cite{myers}. While the spike is string-like in some senses, it has a finite (order one in general) extent into the transverse internal space. An artist's rendering of these two possibilities is shown in figure \ref{fig:spike}.

In the following subsections, we will generalize these considerations to our Lifshitz backgrounds and use the resulting brane and string configurations to compute the AC conductivity for massive charge carriers.  First we will consider the dilute limit in which the strings do not back react on the shape of the flavor brane,
%($\mu \ll T$)
and extend standard calculations of drag forces \cite{Herzog:2006gh, Gubser:2006bz} to the Lifshitz case.  This will lead to analytic results exhibiting a tail in the AC conductivity going like $\sigma(\omega)\sim\omega^{-2/z}$ over a range of frequencies for $z\ge 2$, giving nontrivial exponents for $z\ne 2$. Next we will analyze the problem for the %more realistic
case of larger densities,
%($T \ll \mu$),
exhibiting
again the scaling $\sigma(\omega)\sim\omega^{-2/z}$ as well as analytic and
numerical solutions to the equations for the brane embedding and Maxwell field fluctuations.  This latter regime includes nontrivial interactions among charge carriers.
Finally, we will comment on the potential for rolling scalar backgrounds to shift this exponent and to generalize our results for the Hall conductivity. These last comments will point the way towards a model-building approach to holographic condensed matter physics.

We are working at fixed charge density, where the difference between the string and spike regimes is a factor of $L^2/\alpha'$ in the charge density.
In terms of the chemical potential $\mu$, in order for the density not to be exponentially suppressed by the Boltzmann factor
$e^{-(E-\mu)/T}$, one needs $\mu > E_\text{gap}$. The effect of Boltzmann suppression has been discussed holographically in, for instance, \cite{Mateos:2007vc, kob2}.

\subsection{Drag calculation: conductivity in the dilute regime}

In the extreme dilute limit, one can think of the carriers as individual strings stretching between the flavor branes and the thermal horizon. There are no interactions between the carriers in this limit. The flavor brane itself is given by the zero density solution, a cigar-like shape with $v = v_0$ at the tip.  This framework has been extensively employed for DC calculations in the relativistic case. e.g. \cite{kob, Karch:2009eb}.  Here we apply it to DC and AC conductivities, in our Lifshitz background. We will obtain our first instance of scaling $\sigma(\w) \sim \w^{-2/z}$.

Expanded to quadratic order in transverse fluctuations and in the gauge $t = \tau, v = \sigma$, the Nambu-Goto string action becomes
\be
S_\text{N-G.} = {L^2\over{2\alpha'}}\int dt \int_{v_0}^{v_+} dv\, (f v^{-1-z} x^{i\prime} x^{i\prime}  - f^{-1} v^{z-3} \dot x^i \dot x^i) + \left. \int dt\, \dot x^iA_i(t,x)\right|_{\r = v_0}\ .
\ee
The surface term in the field equation is then
\be
 -{L^2\over\alpha'}f v^{-1-z}\partial_v x^i + F_{i0} + F_{ij} \dot x^j = 0\ \ {\rm at}\ v_0\ .
\ee
At zero frequency the field equation is easily integrated
%, for $f = 1 - (v/v_+)^{2+z}$:
%\be
%x^i =  V^i \biggl(t + { v_+^{z}\over{2+z}} \ln f \biggr)\ ,
%\ee
\be
x^i =  V^i \biggl(t + \frac{1}{v_+^2} \int^\r \frac{u^{1+z}}{f(u)} du \biggr)\ ,
\ee
where $V^i$ is an integration constant and ingoing boundary conditions at the horizon have been used to fix the relative normalization of the two terms.  At the boundary, assuming $v_0 \ll v_+$, we get
\be
v_+^{-2}{L^2\over\alpha'} V^i =  F_{i0} + F_{ij} V^j   \label{xbound} \,,
\ee
which is Drude's law with $m/\tau \propto (L^2/\alpha')T^{2/z}$.  The mass $v_0^{-1}$ drops out as in Ref.~\cite{kob}.  We are evaluating the electric and magnetic fields at the tip of the cigar.  In the zero frequency case these are equal to their asymptotic values. The DC conductivity in the dilute limit of these massive carriers is therefore
\be
\sigma = \frac{\tau J^t}{m} \propto \frac{J^t}{T^{2/z}} \,.
\ee
As we anticipated, the constant term in the massless result (\ref{conduct}) is no longer present.

Now consider the AC case.  First we establish the range of scales of interest.  The mass of the charge-carrying fundamental string hanging down from $v_0$ to the horizon is
\be
E_{\rm gap}=\int_{v_0}^{v_+}dv\sqrt{-g_{vv}g_{tt}}/\alpha' = {L^2\over{z\alpha' v_0^z}}\ .
\ee
where we used the relation $E_{\rm gap}=Mc^2$.
Another natural scale is $\omega_0 = v_0^{-z}$.  This corresponds to the energy scale of bulk excitations at the radius $v_0$.  Also, $v_0^z \gg T$ corresponds to the flavor brane being outside the horizon.  The existence of the two scales $E_{\rm gap}$ and $\omega_0$, differing by a power of the 't Hooft coupling, is similar to the existence of distinct hadronic scales for supergravity and string excitations in \hbox{AdS/QCD}.  In the present case, we are presumably interested in the lowest scales, below $\omega_0$, but it is interesting to look in all ranges for interesting behaviors.

A related point, made in \cite{kob2}\ in the AdS ($z=1$) case, is that our drag calculation
is valid for densities small enough that the Nambu-Goto action is subdominant to the brane action.
The former scales like $L^2/\alpha'$ times the string density, while the latter scales like, for instance $\t_\text{eff.} L^4 \sim N_fN_cL^4/\alpha'^2$.  So in terms of scalings with $L^2/\alpha'$, as long as the density of strings is much less than of order $L^2/\alpha'$, their back reaction on the brane will be small.

The bulk equation of motion for $x(v,t)=\text{Re}(X_\omega(v)e^{-i\omega t})$ is
\be\label{bulkeom}
\partial_v(f v^{-1-z}\partial_v X_\omega)=-\omega^2 f^{-1}v^{z-3} X_\omega(v)\,.
\ee
At zero magnetic field, with $F_{01} = E$, the boundary condition~(\ref{xbound}) implies that if we evaluate the conductivity at $\r = \r_0$ then
\be
 \sigma = { J^t V_\omega(v_0)\over E} = {i\omega J^t X_\omega(v_0)\over E} = {{i\omega J^tX_\omega(v_0) v_0^{1+z }\over{(L^2/\alpha')} f \partial_vX_\omega(v_0)}}\,. \label{eq:sigdrag}
\ee

Consider first the very high frequency limit.
For $\omega\gg \omega_0$ we can use a WKB approximation in the whole range $v_0\le v\le v_+$; the derivatives acting on the exponents dominate.  As noted above, this can be consistent with $\omega < E_{\rm gap}$ when $L^2/\a'$ is large.  The leading WKB solution to the bulk equation of motion is
\be
X_\omega(v)\approx C_1 e^{-i\int_{v_0}^v\omega v^{z-1}/f}+C_2 e^{i\int_{v_0}^v\omega v^{z-1}/f} \,.
\ee
The conductivity~(\ref{eq:sigdrag}) in this regime is then
\be
\sigma_{\rm WKB}={J^t\alpha'\over{ L^2 T^{2/z}}}\left({v_0\over v_+}\right)^2+{\cal O}({1/{\omega v_0^z}})
={J^t\alpha'\over{ L^2 }} v_0^2+\dots \,.
\ee
This goes to zero as the gap is taken to infinity.  Note that in this regime there will be additional propagation effects in connecting the fields at $v_0$ to the fields at the boundary, which may give additional structure. In particular, at frequencies sufficiently higher than the scale $\r_0$ one should expect to recover the massless dilute result $\sigma_0$ of (\ref{eq:sigma0}).

We can also treat the regime $T \ll \omega \ll \omega_0$.  For $v \ll v_+$  we can approximate $f = 1$ and solve the equation of motion directly in terms of a Bessel function.  For $v \gg \omega^{-1/z}$ we can again use WKB.  These ranges of $v$ overlap when $T \ll \omega$.
The Bessel solution is
\be
X_\omega = v^{z\zeta}  H_\zeta^{(1)}(\omega v^z/z), \quad \zeta = {1\over2} + {1\over z}\ , \quad v \ll v_+ \ . \label{eq:hank}
\ee
We do not need the explicit WKB result, but just the fact that WKB gives zero reflection, so that the ingoing Bessel function~(\ref{eq:hank}) is still ingoing at the horizon $v_+$.  To evaluate the conductivity in the range $\omega \ll E_{\rm gap}$, expand the solution for small argument:
\be\label{eq:joeexpand}
X_\omega \propto 1 -  {1\over\Gamma(-\zeta)} \left({\omega v^z}\over{2z}\right)^2
+ e^{i\pi\zeta} \left({\omega v^z}\over{2z}\right)^{2\zeta} \ .
\ee
So this is interesting: if $z < 2$ then $\zeta > 1$ and the $\omega^2$ term dominates $\partial_v X_\omega$.  The conductivity (\ref{eq:sigdrag}) is then
\be\label{eq:zless2}
\sigma = 2 z \Gamma(-\zeta) {{J^t  \alpha' v_0^{2-z}} \over {L^2}} \frac{1}{i \w} \,.
\ee
This is a nice Drude result, matching in magnitude the WKB result.

If $z>2$ then $\zeta < 1$ and the $\omega^{2\zeta}$ term in $\partial_v X_\omega$ dominates.
The conductivity acquires a nontrivial scaling with frequency and is proportional to $\omega^{1 - 2\zeta}
= \omega^{ - 2/z}$,
\be\label{eq:zbig2}
\sigma =  4(2z)^{2/z} e^{-i\pi/z} {{J^t  \alpha' } \over { L^2}} \frac{1}{\omega^{2/z}}  \,.
\ee
The crossover of scaling behavior at $z=2$ originates as follows: the inertial mass $\int dt\, \dot x^2$ has dimension $z - 2$, and so becomes irrelevant for $z > 2$.  In this regime the inertia of the probe is dominated by the bulk degrees of freedom it drags around.  Note that the conductivity here is independent of the scale $v_0$.

The value $z=3$ looks interesting; it gives a falloff in frequency like that seen in strange metal data (e.g. \cite{optical}).  However, this is not for the same $z$ as gives the observed linear DC resistivity in our simplest setup, since we found that $\rho\sim T^{2/z}$, which would need $z=2$. Below, after generalizing our analysis to larger densities, we will outline a way to generalize the model to obtain different exponents for these quantities.

In the above calculation we have not included the propagation of the source electric field from the boundary through the brane to the cigar tip $v_0$.  This may be justified in the very large mass limit corresponding to very small $v_0$. In fact, the energy gap could be at or beyond the lattice UV scale, in which case the string solution would be correct throughout the regime in which the background Lifshitz geometry is applicable.  In the next section, we will analyze the full problem for large densities, for which the brane extends all the way to the horizon.

\subsection{Finite densities}

Now let us analyze the conductivity for massive charge carriers in a regime where their back reaction on the flavor brane cannot be neglected.
We want to introduce an extra scalar field into our DBI action that is dual to a relevant `mass'
operator in the boundary field theory. This scalar field determines the volume of the internal cycle wrapped by the brane, and
will roughly dictate how far the brane sits from the horizon in the `Minkowski' embedding (when $J^t=0$) \cite{KarchKatz}.

The action for this scalar field is not universal, but depends on the internal space of
the geometry. The brane wraps an $n$-dimensional submanifold $\Sigma_n$ of this internal space and, for
simplicity, we will focus attention on the motion of the brane in a single direction,
parameterized by $\theta$, orthogonal to $\Sigma_n$. We take the volume of the submanifold wrapped by the brane to be $V(\theta)^n$. The effective DBI action reduced to $3+1$ dimensions is then given by,
%
%We will assume that the metric
%for the full internal space takes the form,
%%
%\be ds^2 = d\theta^2 + V(\theta)^2\,d{\Sigma}_n + \ldots
%\ee
%%
%In particular, the volume of the submanifold wrapped by the brane is given by $V(\theta)^n\,{\rm %Vol}({\Sigma}_n)$.
%If we need to use a specific form of $V(\theta)$ in what follows, a good choice is $V(\theta)= \cos %\theta$.
%In particular, this is the choice that arises if we consider, say, the D3-D7 system where the probe %brane wraps a sphere.
%
%The presence of this extra mode changes the DBI action in two ways: firstly, the pull
%back metric on the brane,
%${}^\star g$,  now includes kinetic terms of $\theta$. Secondly, there is an overall
%term of ${\rm Vol}(\Sigma)$
%multiplying the DBI action, reflecting the changing of tension of the brane as it moves
%in the $\theta$ direction. We have
%
\be S = -\tau_{\rm eff} \int d\tau d^3\sigma \,V(\theta)^n \sqrt{|{}^\star g + 2\pi\ap\,F|} \,,
\ee
where the pull-back metric on the brane, ${}^\star g$, now includes kinetic terms for $\theta$. In what
follows, we make the choice of normalization $g_{\theta\theta}=L^2$. For much of the calculation below, we leave $V(\theta)$ unspecified, but when required for definiteness we choose $V(\theta)=\cos\theta$, as befits a brane wrapped on a sphere ${\bf S}^n$ inside ${\bf S}^{n+1}$ \cite{KarchKatz}. In general the range of $\theta$ will also remain unspecified; for example, we could take more generally $V(\theta)=\cos c\theta$ for any $c$.

%The path to determine the AC conductivity requires two steps. We first solve for the
%solution in the background
%of a charge density $J^t$, but no electric field $E$. We then consider small
%fluctuation around this background.
%To take the first step we have to find solutions to the equations of motion
%within the ansatz $A_t=\Phi(\r)$
%and $\theta = \theta(\r)$.

As in the massless analysis above,  the equation of motion for $\Phi$ can
easily be integrated once to give
\be \Phi^\prime = \frac{ \sqrt{-g_{tt}(g_{\r\r}+L^2\theta^{\prime\,2})}\,C}{\sqrt{g_{xx}^2V(\theta)^{2n}+(2\pi\ap)^2C^2}} \,.
\label{phi}\ee
Here, as previously, $C$ is a constant of integration.
There is no such luxury for the background profile $\theta(\r)$. To write the second order
equation of motion, it is useful to first define the `boost factor'
\be\label{eq:gammaC} \gamma = \sqrt{1 + \2apl^2\frac{C^2\r^4}{V(\theta)^{2n}}}\label{gamma} \,.
\ee
This agrees with our previous definition \eqn{1gamma} for the massless case when $V(\theta)=1$.
To avoid clutter, we will drop factors of $\2apl$ throughout the following calculation, restoring them only in the final answer for the conductivity. The background profile
of the brane must satisfy,
%
%\be \tilde{g}_{\r\r}=g_{\r\r}+\theta^{\prime\,2}\ee
%
%
%\be
%\frac{d}{d\r}\left\{\frac{V^n(\theta)\,g_{xx}g_{tt}\theta^{\prime\,2}}
%{\left[-g_{tt}\tilde{g}_{\r\r}
%-\Phi^{\prime\,2}\right]^{1/2}}\right\} + \frac{\partial V^n}{\partial \theta}\,g_{xx}\,
%\left[-g_{tt}\tilde{g}_{\r\r}-\Phi^{\prime\,2}\right]^{1/2}=0\nn\ee
%
%
\be \frac{\partial}{\partial \r}\left\{\frac{g_{xx}g_{tt}\,\gamma\, \theta^\prime}
{\sqrt{-g_{tt}(g_{\r\r}+L^2\theta^{\prime\,2})}}\right\} +\frac{1}{V^n(\theta)}
\frac{\partial V^n}{\partial \theta}\frac{g_{xx}}{\gamma}
\sqrt{\frac{-g_{tt}}{(g_{\r\r}+L^2\theta^{\prime\,2})}}
\left[(\gamma^2-1)\theta^{\prime\,2}-\frac{g_{\r\r}}{L^2}\right]
=0\ \ \  \ \ \ \ \label{teom}\ee
This equation looks somewhat formidable, but we can get  insight from numerical solutions.
In order to perform numerics we need to make a choice for $V(\theta)$ and for $f(\r)$. As in our numerical studies of the massless case in section \S\ref{sec:massless}, we will choose for simplicity $f = 1 - (\r/\r_+)^{2+z}$. We will furthermore take $V(\theta) = \cos(\theta)$ and $n=2$, which corresponds for instance to the brane wrapping a two dimensional sphere inside a three dimensional sphere of unit radius.

\begin{figure}[h]
\begin{center}
\includegraphics[height=150pt]{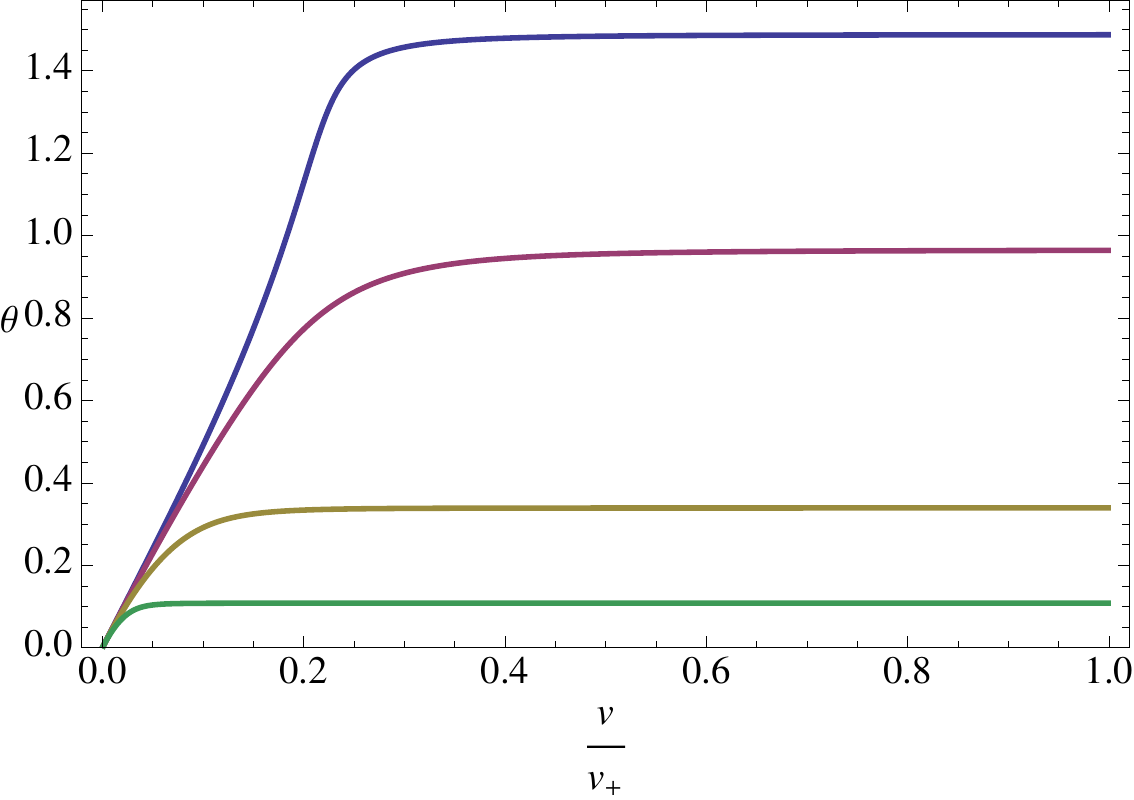}\hspace{0.4cm}\includegraphics[height=150pt]{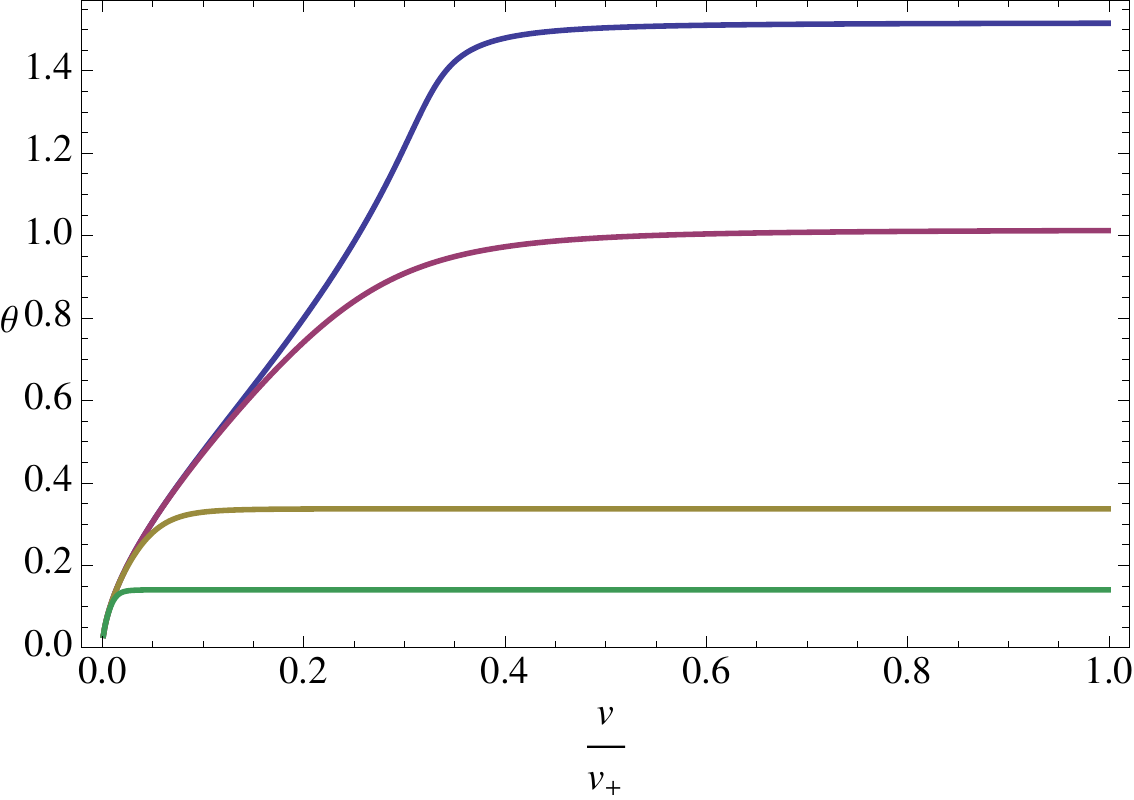}
\end{center}
\caption{The brane profile for $z=1$ (left) and $z=2$ (right) with $\frac{1}{\t_\text{eff.} 2\pi \a' L^2} \frac{J^t}{T^{2/z}}$ equal to $\{10,100,1000,10000\}$ and $\{0.5,10,500,10000\}$ respectively. In all plots $\frac{m}{T} = 20$ and $n=2$.}\label{fig:t1t2plot}
\end{figure}
Figure~\ref{fig:t1t2plot} shows the profile for the exponents $z=1$ and $z=2$, with $m/T$ fixed at the moderately large value of 20 and scanning over a range of charge densities. Here the mass scale $m$ of the charge carriers is determined by the coefficient of the non-normalisable mode of $\theta$ near the boundary. We will define $m$ more precisely very shortly.
In all cases shown in the figure, $\theta$ is approximately constant over a wide range $\r_0 < \r < \r_+$,
\be\label{eq:constant}
\theta(\r) \approx \theta_+
%= \theta_0 + \frac{\pi}{2 z}
\,, \qquad (\r > \r_0) \,.
\ee
The constant regimes arise because $\gamma$ is large in this range, allowing a solution with $\theta' = 0$. This behavior will allow us to obtain analytic results for the conductivity in the low frequency range $\omega v_0^z\ll 1$.
The constant solution and its large charge density regime of applicability has been discussed previously in the $z=1$ case (see e.g. equation (2.37) of  \cite{myers}).
%In any case, the result below for the conductivity will not be sensitive to the precise form of $\theta(\r)$ in this %range.

At low densities, the top-most curves, the profile follows the zero density solution until close to $\theta = \pi/2$, where it rapidly transitions to a narrow spike.  For the special case of $n = z+1$, which happens to apply to the left-hand graph, there is an analytic solution at zero density, $\sin \theta = v/v_0$.  The value of $v_0$ is fixed by the asymptotics as discussed below.  At large densities, the bottom-most curve, one observes that $\theta$ never becomes large.  A linearized solution is then available.

Of course, sufficiently close to the boundary we can always linearise \eqn{teom} in $\theta$; the scalar field must go to zero near the (UV) boundary because it is dual to a relevant operator. The potential must satisfy $V'(0)=0$ in order for there to be a solution. Since we are at $\r\ll\r_+$, we can set $f=1$.
The two solutions near the boundary are then
\be
\theta(\r) =  \sum_{\pm} c_\pm \r^{(z+2 \pm \a)/2} + \cdots \,,
\ee
where $\a = \sqrt{(2+z)^2 + 4 n \frac{V''(0)}{V(0)}}$.
The coefficient $c_-$ of the non-normalisable mode is a parameter in the field theory action, which sets the scale of the carrier mass. Specifically
\be\label{eq:EofC}
m \equiv c_-^{2z/(z+2-\a)} \,,
\ee
which follows from relating the coupling of the operator $\ocal$ to $m$ by dimensional analysis.
Meanwhile the normalisable mode determines the expectation value: $\langle \ocal \rangle \sim  c_+$. In order for the operator $\ocal$ to be relevant, we need $z + 2 > \a$, which requires $V''(0) < 0$.

As long as $\theta$ remains small we can extend the linearized solution to larger $v$. The solution to equation (\ref{teom}) linearised is\footnote{When $z=2$ the
form simplifies to $\theta(\r) =  \sum_{\pm} d_\pm \r^{2 \pm \frac{\a}{2}} \left(1 + \sqrt{1+C^2 \r^4 V(0)^{-2n}} \right)^{\mp \frac{\a}{4}} \,.$}
\be\label{eq:nearb}
\theta(\r) = \sum_{\pm} c_\pm v^{(2+z \pm \a)/2} {}_2F_1\left(\frac{2-z\pm \a}{8}, \frac{2+z \pm \a}{8}, 1 \pm \frac{\a}{4}; - Y \right) \,,\ \ \ \ \ \ \ \
\ee
where $Y =  C^2 \r^4/V(0)^{2n}$.  Using the hypergeometric transformation formulae, we can express this in terms of functions of $1/Y$.  The condition that the growing mode at the horizon, i.e. for $C \r^2 \gg 1$, vanish then gives
\be
c_+ = - c_- C^{\alpha/2} V(0)^{-n/2} I_{-\alpha, z} / I_{\alpha, z} \,,
\ee
where
\be
I_{\alpha,z} = {{\Gamma(1+\alpha/4)\Gamma(z/4)}\over{\Gamma([2+z+\alpha]/8)\Gamma([6+z+\alpha]/8)}} \,.
\ee
One then obtains the horizon value (for small temperatures)
$$
\theta_+ = (I_{-\alpha, -z} - I_{\alpha, -z} I_{-\alpha,z}/I_{\alpha, z}) (m^{1/2z}  C^ {-1/4} V(0)^{n/4})^{z+2-\alpha }.
$$
The linear approximation is self-consistent when this is small, $C^z \gg m^2$ up to factors of order 1.
In this limit we can see that $v_0 \sim C^{-1/2}$, rather than being tied to the mass scale $m$.

Before moving on to solve for the conductivity this background, we can briefly discuss the thermodynamics of these solutions. The on-shell action is easily evaluated to give
\be\label{eq:massiveaction}
\frac{T S}{V_2} = -\tau_\text{eff}L^4\int_\epsilon^{\r_+} dv { V(\theta)^n \over v^{5+z}}
\sqrt{\frac{1+v^2 f\theta'^2}{v^{-4} + C^2 V(\theta)^{-2n}}} \,.
\ee
This formula generalises (\ref{eq:free}) to include a mass. (Just as in \eqn{eq:free}, we have
dropped powers of $\2apl$; we have also set the magnetic field $B$ to zero). The `spike' or `tube' region (\ref{eq:constant}) then gives the contribution
\be\label{eq:freespike}
\frac{T S_\text{spike}}{V_2}  = -\tau_\text{eff} L^4 V(\theta_+)^n \int_{\r_0}^{\r_+} dv { 1 \over v^{5+z}}
\frac{1}{\sqrt{v^{-4} + C^2 V(\theta_+)^{-2n}}} \,.
\ee
This expression is the same as (\ref{eq:free}), but with the effective tension and the constant $C$ multiplied by powers of
$V(\theta_+)$. This contribution in fact contains almost all of the temperature dependence of the free energy in the limit in which we are working, through the endpoint $\r_+$. This then leads to the same thermodynamic temperature scalings we found in section \S3. There is additional temperature dependence through the $\mu J^t$ term. In the large charge density limit, where the linearised solution (\ref{eq:nearb}) is applicable, $\mu(T)$ is given to leading order by the $\theta=0$ result, again recovering the previous massless temperature scalings.

%Our solutions for $\r < \r_0$ have $f=1$ and therefore no explicit temperature dependence in
%(\ref{eq:massiveaction}). The background solution (\ref{eq:bigC}) for $\theta$ in this range depends on two constants, %$\theta_0$ and $\r_0$. We noted above that in our limit $\theta_0 \approx 0$ while $\r_0$ is fixed in terms of the %mass $m$ and $C$. Therefore at fixed mass and charge density, $\theta$ does not have any temperature dependence in %this regime. It follows that the various low temperature scaling laws for thermodynamic quantities that we found in %the massless case in section \ref{sec:scaling} above will also hold for this massive solution.

\subsection{DC conductivity}

The DC calculation presented in section 5.1 generalizes in a straightforward manner to the massive
case; equation \eqn{conduct} becomes,
\be \sigma(E,T) = \sqrt{(2\pi\ap)^4 \tau_{\rm eff}^2 V(\theta_\star)^{2n} +
{\textstyle{\2apl^2}} \r_\star^4(J^t)^2} \,.
\ee
Here $V(\theta_\star)$ is evaluated at $v_\star$ defined in \eqn{whereise}. When the mass of the
charge carriers is large in comparison to the charge density, $V(\theta(v_\star)) \rightarrow 0$
ensuring that the first term in the conductivity, which is independent of $J^t$, is suppressed
as expected \cite{kob}. We can see an instance of this in figure \ref{fig:t1t2plot}; in the upmost curve,
the mass is larger than the charge density and $V(\theta) \sim 0$ over a large range.

When the charge density is large compared to the mass, then $V(\theta(v_\star))$ is finite and order one,
as we see in the lower curves of figure \ref{fig:t1t2plot}. The second term in the above equation for $\sigma$
still dominates because $J^t$ is large.

\subsection{AC conductivity}

The AC conductivity can be understood as a competition between the four dimensionful quantities $\{T, m, J^t, \w\}$.
We are primarily interested in the low temperature regime, in particular $T \ll m, (J^t)^{z/2}$, and are therefore well away
from phase transitions of the sort described in e.g. \cite{myers}. Furthermore, we are working at fixed charge rather than
fixed chemical potential. At fixed chemical potential, there can be a phase transition to a phase with no charge density when $\mu \lesssim m$, as described in e.g. \cite{Mateos:2007vc}. In short we are mainly interested in the effect of $m$ versus $J^t$ at low $T$.

\subsubsection{Numerical results}

\begin{figure}[h]
\begin{center}
\includegraphics[height=150pt]{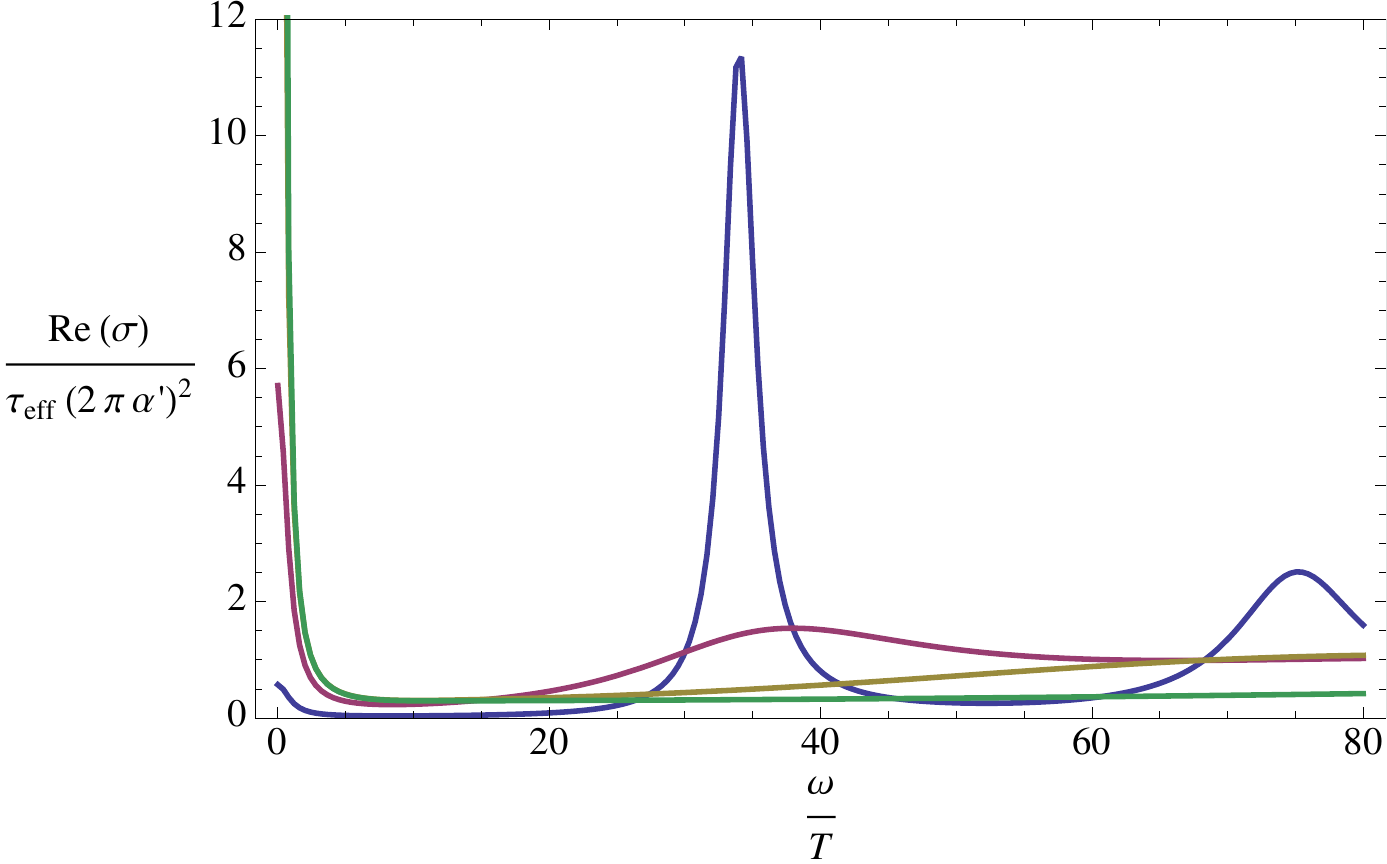}\includegraphics[height=150pt]{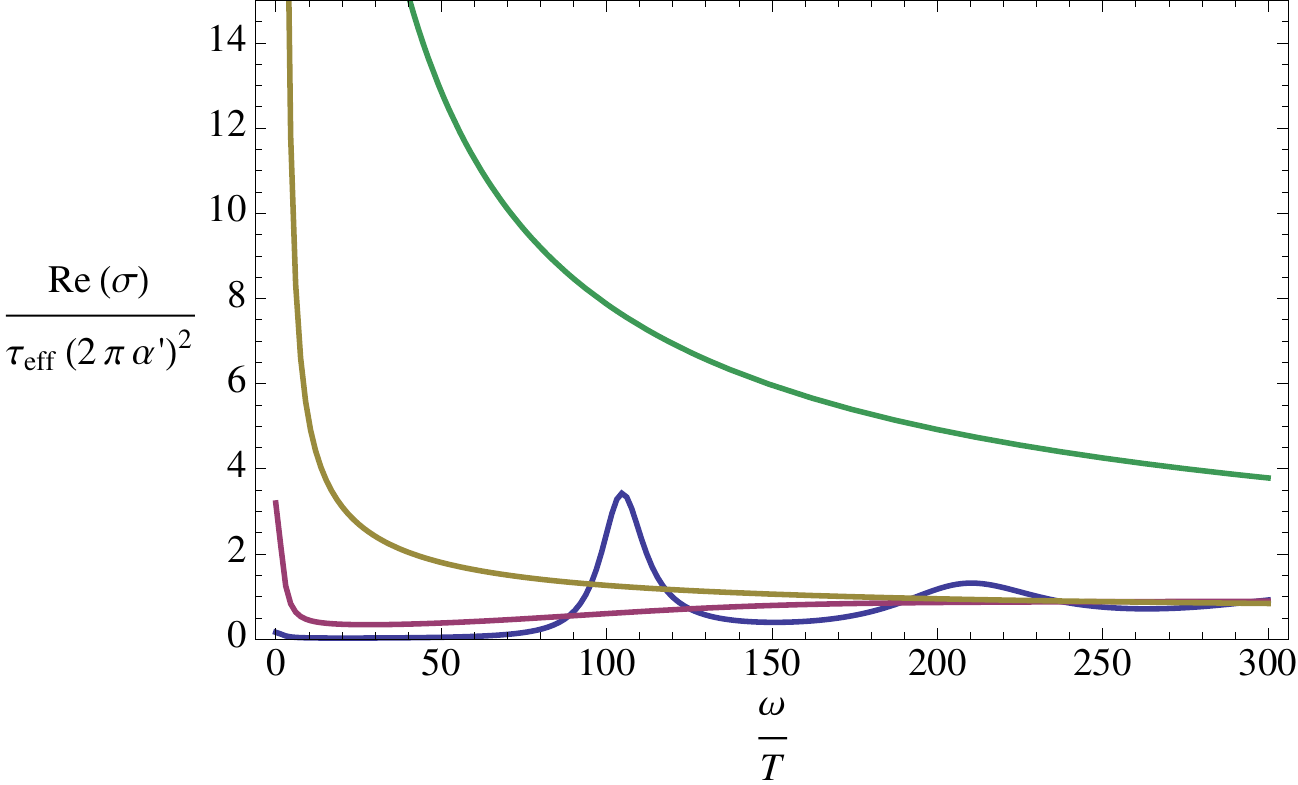}
\end{center}
\caption{The dissipative conductivity for $z=1$ (left) and $z=2$ (right), with the same parameter values as in figure~\ref{fig:t1t2plot}}\label{fig:z1z2plot}
\end{figure}

Figure \ref{fig:z1z2plot} shows the conductivities for the same parameter values as for the profiles in figure \ref{fig:t1t2plot}.
One observes resonances at low charge density; these become discrete excitations of the brane in the zero density limit.  There is also an approach to a constant value at high frequency; at frequencies much larger than the mass scale the massless result is recovered. There are dips in between the peaks, as well as between the Drude peak and the asymptotically constant behavior, as required by the Kramers-Kr\"onig relations (to see that the sums rules hold for the large density cases with $z=2$, one needs to integrate out to much large frequencies than those shown in the above plot).  We are most interested in the behavior at frequencies below these natural scales of the system, where the drag calculation gave a power law.  Here too we can perceive such a power law in the numerics, the decay in the left hand region of the plots, as we now proceed to derive analytically.

\subsubsection{Maxwell fluctuations}

Our goal now is to study the Maxwell fluctuations in the background
solution $\theta(\r)$. The low frequency behavior arises from the regime $v_0 < v < v_+$ where $\theta$ is constant. We expand the gauge field in fluctuations \eqn{eq:deltaA} and, for
simplicity, consider zero momentum, $k=0$.
%
%Here we restrict ourselves to transverse fluctuations of the form
%%
%\be \delta A_y = A_\perp(\r) e^{-i(wt-kx)}\ee
%%
In principle, we should also expand the scalar profile around the background
$\theta(\r)$ but rotational invariance ensures that there are no mixing terms.
%
%To quadratic order in the fluctuations, the Lagrangian for the longitudinal fluctuations reads
%
%\be {\cal L}= -\tau_{\rm eff}\, V(\theta)^n g_{xx} \left[ %\left[-g_{tt}\tilde{g}_{\r\r}-\Phi^{\prime\,2}\right]\left(1 +
%\frac{F_{xy}^2}{g_{xx}^2}\right)+\frac{g_{tt}F_{y\r}^2}{g_{xx}} + %\frac{\tilde{g}_{\r\r}F_{ty}^2}{g_{xx}}+\ldots\right]^{1/2}\label{lag}\ee
%
%With this definition,
%the terms in the Lagrangian involving only the field strengths can be written as
%
%\be {\cal L} = \frac{1}{2}V(\theta)^n \gamma\sqrt{-g_{tt}\tilde{g}_{\r\r}}
%\left[\frac{1}{\tilde{g}_{\r\r}}F_{y\r}^2+\frac{1}{\gamma g_{xx}}F_{xy}^2+\frac{1}{g_{tt}}
%F_{ty}^2\right]\ee
%
%From this we get the equation of motion for $A_\perp$,
%
%
The linearized equation for the longitudinal fluctuation takes the form,
\be \partial_\r\left(V(\theta)^n\gamma\sqrt{\frac{-g_{tt}}{({g}_{\r\r}+L^2\theta^{\prime\,2}) }}A_x^\prime\right) = - V(\theta)^n\gamma \sqrt{\frac{-(g_{\r\r}+L^2\theta^{\prime\,2})}{g_{tt}}}\omega^2 A_x\label{fl} \,.
\ee
We can again put this fluctuation equation into Schr\"odinger form \eqn{erwin}, now with the
change of variables $\Psi = (V(\theta)^n\gamma)^{1/2}\,A_\perp$ and
\be \frac{d}{ds} = \frac{\r^{1-z}f}{\sqrt{1+\theta^{\prime\,2}\r^2f}}\,\frac{d}{d\r} \,.
\ee
The potential in the Schr\"odinger  equation \eqn{erwin} is given by
\be U = \frac{1}{2}\frac{1}{\sqrt{V(\theta)^n\gamma}}\frac{d}{ds}\left(\frac{1}{\sqrt{V(\theta)^n\gamma}}
\frac{d\left(V(\theta)^n\gamma\right)}{ds}\right) \,.
\ \ \ \
\ee

We now solve the Maxwell equations in various regimes that will overlap at large $\w$. Firstly, away from the asymptotic boundary ($\r = 0$), the Schr\"odinger potential $U$ is bounded, and in this regime we can solve the Schr\"odinger equation for $A_x$ with $\w^2 \gg U$. The answer is
\be\label{eq:WKB}
A_{x}(\r) \propto  \frac{e^{i \w s(\r)}}{V(\q)^{n/2} \g^{1/2}} \,,
\ee
where
\be
s(\r) = \int_0^\r \frac{u^{z-1}}{f} \sqrt{1 + \q'^2 u^2 f} du \,.
\ee
There is no assumption about large $C$ here. We have imposed ingoing boundary conditions at the horizon, which fixes the sign of the exponent. This solution is to be matched onto a solution that extends closer to the boundary.

Secondly, consider the large $C$ limit and use the constant solution (\ref{eq:constant}) for $\r > \r_0$. The Schr\"odinger equation for $A_x$ can be solved for $\r_0 < \r \ll \r_+$, where $f=1$. The general solution is
\be
A_x(\r) = a_1 \r^{\frac{z}{2}-1} J_{\frac{1}{2} - \frac{1}{z}}\left(\frac{\w \r^z}{z} \right) + a_2 \r^{\frac{z}{2}-1} J_{-\frac{1}{2} + \frac{1}{z}}\left(\frac{\w \r^z}{z} \right) \,.
\ee
When $z=2$ these solutions are degenerate and the second $J_\nu$ function is replaced by a $K_\nu$ function. At $\w \r^z \gg 1$ this solution can be matched onto the `WKB' solution (\ref{eq:WKB}) which then picks out the mode that is ingoing into the horizon (the existence of an overlap region requires $\w \gg T$). Thus we must have
\be\label{eq:h}
A_x(\r) \propto \r^{\frac{z}{2}-1} H^{(1)}_{\frac{1}{2} - \frac{1}{z}}\left(\frac{\w \r^z}{z} \right) \,.
\ee
If we furthermore assume that $\w \r_0^z \ll 1$
%(we will also consider a different limit below)
then close to $\r = \r_0$ we can take the limit $\w \r^z \ll 1$ of the Hankel function to obtain
\be\label{eq:hexpand}
A_x(\r) = A_x(\r) = 1
%- \frac{1}{\frac{1}{z}+\frac{1}{2}} \left(\frac{\w \r^z}{2 z} \right)^2
-  \frac{\pi \left(\tan \frac{\pi}{z} - i \right)}{\G\left(\frac{1}{2} - \frac{1}{z}\right) \G\left(\frac{3}{2} - \frac{1}{z}\right)}\left(\frac{\w \r^z}{2 z} \right)^{1-\frac{2}{z}} + \cdots \,.
\ee
In the case $z=2$ logarithms appear. The expansions thus require $T \ll \w \ll v_0^{-z}$, which is consistent with our previous assumption $v_+ \gg \r_0$.
%Note that due to (\ref{eq:e0}), $\w \r_0^z \ll 1$ does not imply $\w \ll m$.
Equation (\ref{eq:hexpand}) is similar, but not identical, to an expansion appearing in the string computation (\ref{eq:joeexpand}).
The dictionary between the two can be determined by comparing the sources for the bulk field $B_{vx}$, yielding $\dot x = v^{3-z}F_{vx}$ which is found to map the Hankel functions onto each other.  Further, the implication for a `local conductivity' evaluated at $\r = \r_0$ will now be seen to be the same in the two pictures. The (a priori nontrivial) agreement between the two different regimes is due to the spike behavior emerging in the large density limit. As we saw previously for thermodynamics in (\ref{eq:freespike}), the emergent tube behaves effectively like a string with a different tension.
%The dominant behaviour is given by the first and last terms in equation (\ref{eq:hexpand}).
The local conductivity will be
\be\label{eq:ss}
\sigma(\w,\r_0) = \frac{- i J_x(\r_0)}{\w A_x(\r_0)} \propto \frac{\t_\text{eff.} C \r^{3-z} \pa_{\r} A_x(\r_0)}{\w A_x(\r_0)} \,,
\ee
where we used the usual definition
\be\label{eq:jj}
J_x(\r_0) = \Pi_{A_x}(\r_0)= \frac{\pa {\mathcal{L}}}{\pa A'_x(\r_0)} \,,
\ee
evaluated on the constant $\theta$ background solution.
The expansion (\ref{eq:hexpand}) then gives
\be
\label{eq:wpowers}
{\sigma(\w,\r_0)} \, \propto \, \left\{
\begin{array}{cc}
 \t_{\rm eff.}C \r_0^{2-z} \, \w^{-1} & \quad \text{if $z < 2$} \\
\t_{\rm eff.} C \, (\w \log \w \r_0^2)^{-1} & \quad \text{if $z = 2$} \\
\t_{\rm eff.} C \, \w^{-\frac{2}{z}} & \quad \text{if $z > 2$}
\end{array} \right. \,.
\ee
Together with the extensions considered immediately below, these scaling laws are our main result for the conductivity at finite density.

Thirdly, we need to move into the region $\r < \r_0$.
%described by the background solution (\ref{eq:bigC}).
We could avoid this step if we set $\r_0$ to be below our UV cutoff radius $\epsilon$. This would not necessarily be unreasonable, as we might expect the mass scale to be of the order the lattice scale. In this case (\ref{eq:wpowers}) would be the final result. However, we will proceed without this assumption and consider the conductivity as $\r \to 0$.
%As the computation is a little complicated, let us first anticipate the form of the result.
Before presenting a computation, we will first give an argument that anticipates the result.
In essence, we will see that since we are working in a regime $\omega v_0^z\ll 1$, the frequency
dependence just found in (\ref{eq:wpowers}) will persist after continuing our solution to the boundary.

Consider the local conductivity as a function of $v$:
$\sigma(\omega,C,v_0;v)$.  We are in the regime where $f\sim
1$, because $\r < \r_0 \ll \r_+$, so
$v_+$ drops out and we have included all the remaining parameters on which the conductivity could
depend. Writing the local conductivity in terms of its value at the boundary, we have without
loss of generality
\be
\label{gencondlocal} \sigma(\omega,C,v_0;v) = \sigma(\omega,C, v_0; 0)
 \tilde f(\omega v^z, Cv^2,v/v_0) \,,
\ee
where the function $\tilde f$ approaches 1 at the boundary, $v\to 0$.

Let us expand $\tilde f$ in powers of $\omega$, which means powers of
$\omega v^z,\omega v_0^z$,
and $\omega C^{-z/2}$.  These are all small:
everywhere in the range of $0\le v \le v_0$ we have that $\omega v^z
\ll 1$, which, combined with the additional assumption that $C v_0^2 \gtrsim 1$ means that also
$\omega C^{-z/2} \ll 1$.
There cannot be any inverse powers of these quantities since that would ruin the
boundary behavior $\tilde f\to 1$.  We do have the constant $\omega^0$ term
because we know $\tilde f\to 1$ at the boundary.

So in the limits of parameter space in which we are working, i.e. the physically relevant regime
$T \ll \w \ll v_0^{-z} \lesssim C^{z/2}$,
the $\omega$-dependence of $\tilde f$ will
be trivial.  This means that the $\omega$-dependences found above in (\ref{eq:wpowers}) will persist upon continuing the calculation from $v=v_0$ out to the boundary $v=0$. The dependence on $C v^2$ and $v/v_0$ need not
be trivial, and we expect a reshuffling of the dimensionful prefactors in (\ref{eq:wpowers}).

We will now
confirm this argument with an explicit calculation.  For the linearised (large charge density) solution, where $\theta$ remains small, we may set $\theta=0$ in the Maxwell equation.
Furthermore, at the low frequencies of interest $\w^2 \r^{2z} \lesssim \w^2 \r_0^{2z} \ll 1$, the
solution to \eqn{fl} is
\be
A_x(\r) = p_1 + p_2 \, (\sqrt{C} \r)^z \, {}_2F_1
\left( \frac{1}{2}, \frac{z}{4}, 1 + \frac{z}{4}, - C^2 \r^4 \right) \label{basil}\,.
\ee
The conductivity, evaluated on the boundary, is given by
$\sigma(\w) \propto \t_\text{eff.}C^{z/2} \w^{-1} p_2/p_1$. Expanding the solution \eqn{basil}
for $C \r^2 \gg 1$, i.e. into the constant regime, one obtains
\be
A_x(\r) = p_1 + p_2 \frac{\G\left(\frac{1}{2} - \frac{z}{4}\right) \G\left(1 + \frac{z}{4}\right)}{\sqrt{\pi}}
 + p_2 \frac{z}{z-2} \left(\sqrt{C} \r \right)^{z-2} + \cdots \label{sybil}\,.
\ee
which can be matched to \eqn{eq:hexpand}. Restoring factors of $2\pi\ap$ and $L^2$, the conductivity is,
\be\label{eq:smiley}
\frac{\sigma(\w)}{\t_\text{eff.}(2\pi\ap)^2} \, \propto \, \left\{
\begin{array}{cc}
\2apl^{z/2} C^{z/2} \, \w^{-1} & \quad \text{if $z < 2$} \\
\2apl C \, (\w \log \w \r_0^2)^{-1} & \quad \text{if $z = 2$} \\
\2apl C \, \w^{-\frac{2}{z}} & \quad \text{if $z > 2$}
\end{array} \right. \,.
\ee
This is now an exact result for the conductivity evaluated at the boundary, including factors of $C$ and $\r_0$.
This result has required $T \ll \w \ll
\r_0^{-z} \lesssim \textstyle{\2apl^{z/2}}C^{z/2}$.
Roughly this corresponds to a low temperature regime where the frequency is lower than the energy gap and charge density scales. These are both of order eV in the cuprates, and so this regime may be roughly compatible with the
experimentally measured anomalous scalings (e.g. \cite{optical}), depending on the precise numerical relation of $C$ and $\r_0$ to the charge density and energy gap scales.

\subsection{Model building}

So far we have found two strange metal-like behaviors involving nontrivial exponents,
\be\label{SMredux} \rho\sim T^{\nu_1} ~~~ {\rm and} ~~~ \sigma(\omega)\sim \omega^{-\nu_2}  \,, \ee
with
\be\label{expLif} \nu_1=2/z, ~~~~ \nu_2=2/z ~~~(z\ge 2) \,, \ee
in pure Lifshitz backgrounds.  In real-world strange metals, such as the cuprates, $\nu_1\approx 1$ and (according to some analyses, e.g. \cite{optical}) $\nu_2\approx .65$. These values would correspond in our formulae to $z=2$ and $z=3$, respectively. It would be interesting to find a generalization of our computations which produces different exponents for these two quantities. In this subsection we will outline a mechanism for accomplishing this, though it is fair to say that our simplest examples give the scaling (\ref{expLif}).

Our strategy is to consider pseudo-Lifshitz solutions, in which the metric is Lifshitz but there are scalar fields that run in the solution.  One example of this setup was described in \cite{Azeyanagi:2009pr}, and others with interesting properties are under investigation \cite{KachruTrivedi}.  The tension $\tau_\text{eff}$ of the flavor brane generally depends on scalar fields such as the string coupling and internal volumes in the four-dimensional Lifshitz background; in essence it is itself a scalar field.   In one of our regimes of interest, this brane forms a string-like spike realizing a finite density of charge carriers.

Let us consider the case where the tension of this spike varies with radial position $v$ such that
\be
\label{tauvary} \tau\sim v^\kappa \,.
\ee
There are various potential examples of this.
One could consider, for example, a situation where the string coupling runs in the solution.  Then, if the flavor strings involved were D-strings, this would affect the dilute drag calculation of \S5.4.1\ by making the tension $1/g_s\alpha'$ run with scale, as we now show.

We can compute the AC conductivity in the situation just outlined. We will use the dilute limit drag approach of \S5.4.1 for simplicity, bearing in mind our previous observation that the same scalings go through in the large density limit due to the formation of a spike.
The bulk equation of motion (\ref{bulkeom}) on the bundle of strings becomes
\be\label{genbulk}
\partial_v(f v^{-1-z+\kappa}\partial_v X_\omega)=-\omega^2 f^{-1}v^{z-3+\kappa} X_\omega(v) \,.
\ee
For $f\approx 1$,
this has Bessel function solutions like those in (\ref{eq:hank}), but now with index
\be\label{genindex}
\zeta_\kappa = {1\over 2}+{1\over z}-{\kappa\over{2z}} \,.
\ee
This leads to a nontrivial falloff, for $z \geq 2 - \kappa$,
\be\label{genac}
\sigma \sim \omega^{(\kappa-2)/z} \,,
\ee
i.e. the exponent is now $\nu_2=(2-\kappa)/z$.

The next question is what happens to the DC calculation.  The DC calculation in \S5.4.1\ depends on $f(v)$, which in turn depends on the black hole solutions with scalar hair.
As in \S5.4.1\ we impose
ingoing boundary conditions at the horizon, which means that
\be\label{genbc}
\partial_v X|_{v\to v_+} \sim {V\over {(v-v_+)T}} \,,
\ee
in terms of the velocity $V$, with the temperature $T\sim f'(v_+)/v_+^{z-1}$.

The DC ($\omega=0$) equation of motion implies
\be
f v^{-1-z+\kappa}\partial_v X_\omega = \tilde V \,,
\ee
for a constant $\tilde V$.
The conductivity is therefore
\be
\sigma_{DC}\sim \frac{J^t V}{E} \sim J^t{\alpha'\over L^2} \frac{V}{\tilde V} \sim J^t v_+^{2-\kappa} \,,
\ee
where we imposed
the ingoing boundary condition (\ref{genbc}) to obtain
$ {V\over\tilde V} = v_+^{2-\kappa} $.
The question of whether the exponent in the DC resistivity is modified therefore boils down to the question of how $T\sim f'(v_+)/v_+^{z-1}$ scales with $v_+$ in this system with running scalars.  It would be interesting to pursue this in detail in future work.  If one still has $f=1-v^n/v_+^n$, then we get the same answer as before, $T\sim 1/v_+^z$, and then we have $\rho\sim T^{(2-\kappa)/z}$, i.e. $\nu_2=(2-\kappa)/z$.  This would not decouple the two behaviors (DC and AC conductivity), but does generalize our mechanism
from pure Lifshitz, allowing for linear resistivity for different $z\ne 2$ if the scalar determining the tensions runs with nontrivial $\kappa$ as in (\ref{tauvary}) above.

However, now we may use the fact that the DC calculation at $\omega=0$ and the AC one in the regime $\omega\gg T$
involve different ranges of scales.  In a general solution, $\kappa$ may itself depend on radial position. For example, $\kappa$ may jump across a domain wall in the bulk.  So it may be possible to shift the exponent $\nu_2$ relative to $\nu_1$ in order to mimic the strange metallic behaviors.  Strange metal phenomenology requires the nontrivial exponent $\nu_1\ne 2$ at $\omega=0$, for $T$ ranging up to the melting temperature.  At a given temperature $T$, the phenomenology requires $\nu_2\ne 1$ over a range of frequency $\omega$ greater than $T$.
In order to accomplish this using the strategy just outlined, the domain wall would need to remain outside the black hole horizon.

Similarly, let us briefly consider model-building possibilities for addressing the Hall conductivity and generalizing it from the Drude-like result we obtained above in our simplest model.
If we could insert a different power of $v$ in front of
the $F^2$ terms in the action it would change the scaling of the hall ratio (\ref{hallratio}), for the following reason. The Hall ratio (\ref{hallratio}) for $J^t\gg B$, the case which is Drude-like, depends
on $B$ while the DC conductivity $T^{(2-\kappa)/z}/J^t$ does not depend on $E$ or $B$. However, such an extra power would also change the scaling of the $J^t$ terms in various quantities, and a full computation is needed to see if there is a net effect.

One way such a modification might arise is if we had some source of stress energy which
generates an appropriate conformal factor in front of the Lifshitz metric, which would be
a distinct effect from the $v$-dependent tension leading to the shift by $\kappa$.
Another possibility is to consider a bulk theta angle, $\theta\int F\wedge F$, which would shift $J^t$ by $B\theta$,
also affecting the Hall result.

Although these generalizations of our basic structure are somewhat complicated, it seems very interesting to investigate what set of bulk ingredients produce the suite of anomalous strange-metallic behaviors.  We leave detailed model building to future work, and next turn to microscopic constructions of holographic Lifshitz backgrounds.

\section{Lifshitz from string theory}
\setcounter{equation}{0}

In this section, we address an open problem in the literature on holographic duals with Lifshitz symmetry, outlining three string theoretic constructions of such solutions.

%{\tt Discuss sea and landsicknesses somewhere below...}
%The effective theory \cite{Kachru:2008yh}\
%
\be\label{eq:LifshitzEFT} {\cal S}=\int d^4 x\sqrt{-g}M_P^2\left({\cal R}-2\Lambda\right)-{1\over 2}\int\left(F_2\wedge *F_2+H_3\wedge * H_3\right)-c\int B_2\wedge F_2 \ee
generates, for a special value of $\Lambda$, a Lifshitz solution with metric (\ref{eq:IRmetric}) (let us change the radial coordinate to $ r=1/v$ to match the form of the metric given in \cite{Kachru:2008yh}). The form fields in the solution are
\be\label{eq:Lifshitzforms} F_2=\frac{\sqrt{2z(z-1)}}{L}\theta^t\wedge\theta^{r},
~~~~ H_3 = {2\over L}\sqrt{z-1}\ \theta^x\wedge\theta^y\wedge\theta^{r} \,, \ee
in terms of the orthonormal basis of forms
\be\label{eq:orthoforms} \theta^t=Lr^zdt, ~~~~ \theta^x = Lrdx, ~~~~ \theta^y=Lrdy, ~~~~
\theta^{r}=L\frac{dr}{r}. \ee
The condition on $\Lambda$ that is required for this solution is that $c$ and $\Lambda$ be related according to
\be\label{eq:cLambda} c^2={2z\over L^2} \,,~~~~ \Lambda=-\frac{z^2+z+4}{2L^2} \,. \ee
Turning this around, this means that in the Landscape of string vacua, the possible values of the dynamical critical exponent $z$ will be determined by the discretuum of possible values of the cosmological constant and `mass' $c$.

\subsection{Lifshitz solutions from Landscape dual pairs}
\label{subsec:LifshitzLandscape}

As reviewed above and in \cite{Li:2009pf}, in order to obtain the Lifshitz geometries as derived in \cite{Kachru:2008yh}\ from string theory, one requires not just the matter content (\ref{eq:LifshitzEFT}) of \cite{Kachru:2008yh}\ but also the relationship (\ref{eq:cLambda}) between the Chern-Simons coefficient $c$ and the cosmological constant $\Lambda$ in the four-dimensional effective theory.
Another requirement is that the fluxes producing the Lifshitz geometry not destabilize the moduli. It is interesting to ask whether this is possible to obtain in the Landscape, where $\Lambda$ is often tunable, and if so what is the content of the dual field theory which generates this Lifshitz scale invariance. We will now outline such solutions in a corner of the landscape developed recently where the dual field theory is also known implicitly via the low energy limit of a specific brane construction.\footnote{See \cite{Azeyanagi:2009pr}\ for recent constructions of Lifshitz-like theories with radially rolling scalars, and \cite{Aharonyetal}\ for other constructions of Lifshitz geometries that may be related to Chern-Simons theories.}

In string theory, the cosmological term $\Lambda$ in (\ref{eq:LifshitzEFT}) is a potential for dynamical scalar field moduli of the compactification of extra dimensions.  The fluxes $F_2$ and $H_3$ in the above construction may come from a variety of fluxes in the underlying higher-dimensional theory.  Let us study the compactifications \cite{Polchinski:2009ch}\ of F theory on an elliptic fibration over a six-manifold of the form $Y^5\times S^1$, where the 5-manifold $Y^5$ is a Hopf fibration over a four-dimensional base ${\cal B}$.  This model has a moduli potential of the form (with radii given in string units and $g_s\sim 1$\footnote{The string coupling is order one on the mutually nonlocal sets of 7-branes in the simplest constructions \cite{Polchinski:2009ch}. The dilaton per se is typically heavy in F theory compactifications, and the moduli of the 7-branes are flat to very good approximation, leading at worst to allowed tachyons.})
\be\label{Umod} {\cal U}\sim M_P^4 (R_f R_6 R^4)^{-1}\left(\frac{R_f^2}{R^4}-\frac{\epsilon}{ R^2}+\frac{N_c^2}{R^8R_f^2}+\frac{Q_1^2}{R_6^2}
+\frac{Q_3^2}{R^4R_6^2}
+\frac{\tilde Q_3^2}{R^4 R_f^2}
\right)  \ee
where $R\sqrt{\alpha'}$ is the size of ${\cal B}$, $R_f\sqrt{\alpha'}$ the size of the Hopf fiber, and $R_6\sqrt{\alpha'}$ the size of the $S^1$ factor in the geometry.  The term proportional to $\epsilon$ arises from the curvature of the compactification, including the effects of the 7-branes which partially cancel the positive curvature of the $Y^5$ component.
Here $N_c$ is the number of units of 5-form flux along $Y^5$ and $Q_1$ is the 1-form flux quantum number along the  $S^1$ factor.  We have also included RR 3-form flux quantum numbers $Q_3$ and $\tilde Q_3$ threaded through 3-cycles in $Y^5\times S^1$, applicable in the generic case that $Y^5$ contains 2-cycles and dual 3-cycles.
%The simplest cases with this property have topology $S^2\times S^3$;
%let us consider this case for simplicity.
Extremizing the potential
with respect to $R$, $R_f$, and $R_6$, we get a solution with
\be\label{eq:scalingsfour} R_f^2\sim \epsilon R^2, ~~~~~ R^4\sim \frac{N_c}{\epsilon}, ~~~~~
R_{AdS}^2\sim \frac{R^2}{\epsilon}\sim \frac{R_6^2}{Q_1^2} \ . \ee
In deriving this we assume that $Q_3, \tilde Q_3$ are small enough to make these terms subdominant, but if we turn off the 1-form flux, similar scalings result but with $Q_1$ replaced here by $Q_3/R^2, \tilde Q_3/R^2$.

To begin let us identify $p$-form potential fields that descend to the 1-form and 2-form potentials $A_1$ and $B_2$ prescribed in \cite{Kachru:2008yh}.
We cannot immediately take the flux $B_2$ in (\ref{eq:LifshitzEFT}) to be simply the Neveu-Schwarz 2-form field of type IIB string theory in general in F theory, since the type IIB 2-form fields generically undergo monodromy about the 7-branes in these backgrounds.  It is possible, however, to impose that the SL(2,Z) monodromy matrix fix one eigenvector in the space of 2-form potential fields $B_2^{NSNS},C_2^{RR}$; this is not compatible with fixing the axio-dilaton at a constant value, and in particular is away from the Sen limit of weakly coupled orientifolds.

It is perhaps simpler to work with the S-duality invariant potential field $C_4$.
Let us consider obtaining the Chern-Simons term in (\ref{eq:LifshitzEFT}) for example from the worldvolume coupling
\be\label{eq:CSseven}\int f_2\wedge C_4 \ee
arising on D5-branes.  We can consider a D5-brane wrapped on a 2-cycle in $Y$, at a point in the $S^1$, with its charge cancelled by flux, by an O5-plane, or by an anti-D5-brane at the diametrically opposite point on the circle (though the latter is likely to suffer from a disallowed tachyon).  Equivalently, we may dissolve 5-branes into 7-branes and consider 7-branes which wrap the $S^2\times S^1$ times a contractible circle in the dual $S^3$, with worldvolume flux through the $S^2$ and use the $\int f\wedge f\wedge C_4$ coupling.
In order to match to the parameters of the effective theory, we must normalize the fields accordingly.  To start, define
\be\label{eq:hatted} \hat F_2\equiv f_2, ~~~~ \hat B_2\equiv \int_{S^2}C_4  \ee

Rescaling the fields so as to normalize them as in (\ref{eq:LifshitzEFT}) reveals that
\be\label{eq:Lifshitzc} c\sim \frac{V_2^{1/2}}{V^{1/2}\sqrt{\alpha'}} \ee
where $V_2\sim R^2$ is the volume of the wrapped $S^2$ in string units.
Putting this together with the structure of the moduli potential, we obtain
\be\label{eq:crelation} \frac{4z}{z^2+z+4}\equiv \frac{c^2}{\Lambda}\sim \frac{V_2R^2}{V\epsilon}\sim
\frac{1}{Q_1(N_c\epsilon)^{1/2}} \ee
where in the last step we plugged this into the stabilized solution (\ref{eq:scalingsfour}).
%If the 4-cycle volume $V_4$ is of order $V/R^2$, then the ratio (\ref{eq:crelation}) is of order $1/\epsilon$ and %could potentially provide a working Lifshitz construction for $\epsilon$ of order 1 depending on the precise %coefficients in the model.  More generically, in a situation with richer topology, the volume $V_4$ will be less than %this and we can tune the ratio (\ref{eq:crelation}) using $\epsilon$.
Again it is interesting to note that the available values of the dynamical critical exponent $z$ are determined by the available values of the cosmological constant in the string landscape.

Alternatively, one can consider the Chern-Simons coupling $\int f\wedge C_6$ on the 7-branes already present in the model, as long as there is a monodromy-invariant combination of $C_6$ and its S-duality partner in the 7-brane background.  There are also bulk (closed string) Chern-Simons couplings that may be used depending on the compatibility of their form fields with the 7-branes in the model.

Finally, we must ensure that the added fluxes do not destabilize the moduli.  Note that the rescaling we did to normalize the fields as in (\ref{eq:LifshitzEFT}) puts all the moduli-dependence of the new degrees of freedom into the Chern-Simons term, which depends on one combination of moduli (\ref{eq:Lifshitzc}); let us denote this combination $e^{\sigma_c/M_P}$ in terms of a canonically normalized scalar field $\sigma$.  For $z$ of order 1, this Chern-Simons term is of the same order as the leading terms in the moduli potential, and for larger $z$ it becomes less important.  Canceling the variation of the action with respect to $\sigma_c$, in the presence of this term, therefore shifts $\sigma$ by at most an amount of order $M_P$.  Since the radii are large to begin with, this leaves us near the original solution.
Furthermore, the Chern-Simons term does not contribute to the variation of the action with respect to the four-dimensional metric, so the effective four dimensional cosmological term is also close to its original value, still negative.

%\subsubsection{The field theory side}

%Given the $AdS_4$ holographic dual pair on which our Lifshitz solutions are based, it is interesting to ask what is %the effect on the field theory side of the further ingredients we add to produce the Lifshitz scaling on the gravity %side. If we trade the flux for branes, or obtain the fluxes via a geometric transition, do we obtain something like %the 6-brane density in our other construction of Lifshitz? {\tt comment on the relation after S6 construction is %incorporated.}

%\section{Comments on fermions?}

\subsection{Landscape of holographic Lifshitz superconductors}

Having now constructed Lifshitz solutions from the top down, we can consider systems with superconducting instabilities and study how their parameters vary as we scan over a corner of the landscape (c.f. \cite{Denef:2009tp}).  In particular, it is interesting to ask whether the
non-Fermi liquid behaviors like (\ref{eq:resistivity}) that we find are correlated with higher temperature holographic superconductivity, analogously to what happens for some real world materials (although strange metallic behavior is also observed in non-high $T_c$ materials such as heavy fermions), or if instead these features are independently variable.  This type of question -- one which has been much discussed in the context of cosmology and particle physics -- is notoriously difficult to answer reliably.  Here we will restrict ourselves to a few comments in the context of the constructions given above.

As reviewed for example in \cite{Hartnoll:2009sz}, holographic superconductors arise when the normalizable mode of a bulk charged scalar field condenses.  Solutions of this type in Lifshitz geometries have been described recently in \cite{LifshitzSC}. We will not make a detailed study of such theories here but rather discuss general features of the superconducting instability.

A charged scalar field has a (radially dependent) contribution to its mass squared of order $g^{tt} \Phi(v)^2$ (where recall $\Phi=A_t$ in our notation).  Now in our probe brane solution (\ref{eq:sol}), $g^{tt} \Phi(v)^2$ grows toward the IR region of the geometry (toward larger $v$).  Roughly speaking, an instability toward condensation of the scalar field sets in at
the radial position for which the total mass squared of the charged scalar goes more negative than the Breitenlohner-Freedman (BF) bound \cite{Gubser:2008px, Hartnoll:2008kx, Denef:2009tp}.
The critical temperature $T_c$ is determined by this scale, since the temperature of the black hole needed to barely screen this instability depends on its radial position. This argument is incomplete as it is not the asymptotic BF bound that is relevant in general. For instance at low temperatures it is a near horizon BF bound that controls the instability.

One could ask the question of whether $T_c$ increases as we increase $z$, deviating further and further from Fermi liquid theory (\ref{eq:resistivity}).  The question is not precise until we decide what to hold fixed in making this comparison.  In the construction of \S\ref{subsec:LifshitzLandscape}, charged scalar fields with bare mass $m=0$ arise naturally from intersections of flavor branes. Therefore these constructions come with the potential to become superconducting. In \cite{Kachru:2008yh}\ the analogue of the BF bound for Lifshitz geometries was derived:
\be\label{eq:BFLifshitz} m^2L^2>-4 ~~\Rightarrow ~~ m^2 > -8 |\Lambda|/(z^2+z+4) \,.  \ee
As discussed above, we can vary $z$ by varying
the ratio (\ref{eq:cLambda}).
%At $z=1$, the central charge of the CFT is of order $M_P^2/|\Lambda|$.
If we hold $M_P^2/|\Lambda|$ fixed as we increase $z$ then it is clear from (\ref{eq:BFLifshitz}) that the scalar massless comes closer to an instability in the absence of an electric field background.
This may translate into a higher $T_c$, although the profile $\Phi(\r)$ itself depends on $z$ so one should perform a complete calculation to be sure.
%It may be motivated by the idea that changing the doping to go from a Fermi liquid to a non-Fermi-liquid %in the phase diagram of the cuprates, without changing the material otherwise, would not change the %underlying number of degrees of freedom $M_P^2/|\Lambda|$.  (It increases $J^t$, which in itself also %increases $T_c$.)
However, it is not clear that this is physically the correct ratio to fix. More abstractly we have enough parameters in the construction to increase $z$ while keeping fixed $|\Lambda|/(z^2+z+4)$ and hence $T_c$.  So even in the restricted corner of the landscape we have studied so far, varying over different theories (analogous to varying over materials in the real world) can independently change $z$ and $T_c$. It would be very interesting to delve deeper into this question to see if useful correlations arise among appropriately defined quantities in the landscape of holographic non-Fermi liquids.  We leave that for future work.

\subsection{Lifshitz from brane polarization}

In Eq.~(\ref{eq:brcondition}) we have seen that the backreaction of the brane fields on the metric becomes strong in the IR.  We have not yet solved the backreaction problem in general, but in studying it have found an unexpected and novel realization of a Lifshitz solution, which we describe here.

One mechanism by which singularities are resolved in string theory is brane polarization~\cite{Myers:1999ps,Polchinski:2000uf}, where a brane wrapped on the would be singularity expands to a finite radius due to the potential from form fields, and screens the diverging fields.  For hadronic systems at finite density, it has been noted that baryons can polarize in this way~\cite{Bergman:2007wp,Davis:2007ka,Rozali:2007rx}.  That work was in the probe approximation for 3+1 QCD; here we would like to take into account backreaction in a conformal background.

For a single static baryon the action is
\be
S = \int dt\, (-M \sqrt{-g_{tt}} + A_t)\ ,
\ee
where we have reduced on the dimensions in which the baryon is wrapped.
We couple a continuous distribution\footnote{The baryons are discrete, but we have found in brane models that their density per curvature volume is large, and so their backreaction will be smooth.}
 $\rho(v) \geq 0$ of baryons to Einstein-$U(1)$ theory,
\be\label{eq:bulkaction}
S = \int d^4x \left[ \sqrt{-g} \left( \frac{1}{2 \kappa^2} \left[{\cal R } + \frac{6}{L^2}\right] -  {\cal F}({|F|^2}) \right) + \rho \left(-M \sqrt{-g_{tt}} + A_t \right)
\right] \,,
\ee
with $F=dA$.  We have taken a general function of $|F|^2=F_{\mu\nu}F^{\mu\nu}/2$; for field strengths in the $vt$ plane this allows us to treat DBI and Maxwell together, as well as generalizations.  We have omitted many fields that could appear in a realistic string background, such as the dilaton and the compactification radius.  In a more top-down treatment we must include these; we will address this after first analyzing the simpler effective theory (\ref{eq:bulkaction}).\footnote{In fact, this truncated action has a nonsingular solution even without brane polarization, as we will discuss in a later subsection, though as we will also discuss this solution is subject to an instability in the presence of a bulk fermi sea.}

We are interested in solutions with metrics of the form
\be
ds^2 = L^2 \left(- e^{2\gamma_t (\r)} dt^2 +  e^{2\gamma_x(\r)} (dx^2 + dy^2) + \frac{d\r^2}{\r^2} \right) \,,
\ee
and gauge potential
$A_t(\r)$.

Whenever a solution to the equations of motion can be found that is compatible with the constraint $\rho \geq 0$, then we can eliminate $\rho$ from the action. The $\rho$ equation of motion implies that
\be
A_t = ML e^{\gamma_t(\r)} \ \ {\rm or}\ \ \rho(v)=0\ ,
\ee
representing either an extremum or an endpoint of the action.  In the former case, the resulting action for $\gamma_{t,x}$ is
\be
S = \frac{L^2}{ \k^2} \int \frac{d\r}{\r} e^{\gamma_t + 2\gamma_x} \left\{ 3 - (\k^2 L^2) { \cal F}(-M^2 v^2 \gamma_t'^2/L^2)  + 2 {\r^2 \gamma_t' \gamma_x'} + {\r^2 \gamma_x'^2} \right\} \, .
\ee
Also, the $A_t$ equation gives
\be
\rho = 2 M L \partial_v (e^{2\gamma_x} {\cal F}' v \gamma_t')\,.
\ee

In a scaling solution, $v \gamma_t' = f$ and $v\gamma_x' = k$ are constants.  In this case the equations of motion reduce to
\bea
0 & = & 3 - (\k^2 L^2){ \cal F}(-M^2 f^2/L^2) - f^2 -fk - k^2 \, ,\\
0 & = & -f^2 - fk + 2k^2 + (2\k^2 M^2) (f^2 + 2fk) { \cal F}'(-M^2 f^2/L^2) \,.
\ee
Let us specialize to the Maxwell action
\be\label{Maxconvention}
{\cal F}(x) = {x\over 2g^2} = {|F|^2\over 2g^2}
\ee
%${\cal F}(x) = x/4g^2$.
Then
\bea
0 & = & 3 + (\mu^2 - 1) f^2 -fk - k^2 \,, \label{eq:1} \\
0 & = & -f^2 - fk + 2k^2 + 2\mu^2 (f^2 + 2fk) \, ,\label{eq:2}
\eea
where $\mu^2 = M^2 \k^2/2 g^2$.  The dynamical critical exponent is given by the ratio
$z = f/k$.  The two possible values are readily obtained from Eq.~(\ref{eq:2}),
\be\label{eq:zmu}
z = -2\, ,\ \frac{1}{1-2\mu^2}
\ee
Note that
\be
\rho = \frac{2fk LM}{g^2 v} e^{2\gamma_x}  \label{eq:dens}
\ee
is positive only when $f$ and $k$ have the same sign, so we must take the second solution, and only for $1 > 2\mu^2$; also, $f$ and $k$ as given by Eq.~(\ref{eq:1}) are then real.  Similar results hold for the DBI and other actions.

The density~(\ref{eq:dens}) is constant per unit three-volume in the bulk, and so its integral diverges toward the boundary.  At finite baryon density, the charge density must be zero for $v$ less than some minimum value, and the full solution is $AdS_4$ near the boundary, with a transition region around the discontinuity in the density, and approaching to the Lifshitz solution in the IR.

This suggests an interesting element of the holographic dictionary.  Discrete remnants of translation symmetry is a familiar possibility in real space.  Holography maps the radial direction into scale.  Putting these together, a discrete remnant of scale symmetry may occur naturally in field theories with holographic duals.

To apply this mechanism to our construction, we have two cases: the $U(1)$ is electromagnetism, or it is a new gauge symmetry on another brane.  In the latter case, this can provide the Lifshitz bulk that we need in our construction.  The other case, where  the electromagnetic $U(1)$ is inducing the polarization, is interesting and brings up the question of the microscopic origin of our model.  Generally, there are color branes with strongly coupled gauge fields, and flavor branes, with charged fields living on the intersection.  The electron itself must be neutral under the emergent color group, and so is identified with the lightest color singlet electromagnetically charged state.  In many models, depending on the brane configuration, this will be the baryon.  Thus one can think of the electron in these models as separating into $N$ spinons  (in some others it will be a scalar-fermion bilinear).  In the phase we are discussing here, these would be confined into localized electrons, so this would likely be a normal rather than a strange state.

\subsubsection{Baryon-induced Lifshitz: top down considerations}

In the above discussion, there were two string-theoretic issues left unresolved: the moduli-dependence in the effective action (\ref{eq:bulkaction}) and the possible values of $z$ (equivalently $\mu^2$) in (\ref{eq:zmu}).  Let us analyze these next using more details of the internal structure of string and M theory compactifications.  We will start by determining $z$ in a class of string constructions assuming the moduli are still stabilized in the presence of the new sources generating Lifshitz.  Then we will explain how the stabilization of the moduli is affected by the new sources, finding that they remain stabilized for a range of $z$ which includes the value ($z=2$) that we obtain in our simplest models.

Let us take the Maxwell case for simplicity.  The baryon is a $p$-dimensional-brane wrapped on a $p$-cycle $\Sigma_p$
of volume $V_p$ in Planck or string units (for example one could consider the M theory case, with eleven-dimensional Planck length $\ell_P$).  The U(1) is $A\equiv A_{(1)}=\int_{\Sigma_p} A_{(p+1)}$ in terms of the $p+1$-form potential $A_{(p+1)}$ sourced by the $p$-brane.  Because we will be interested in computing $z$, let us normalize these elements explicitly:
\be\label{eq:formreduction}
A_{(p+1)}\equiv A_{(1)}\wedge \omega_{(p)}, ~~~~ \mu_p\int_{\Sigma_p}\omega_{(p)}\equiv 1
\ee
This way we reproduce the coupling $\int d^4x \rho A_t$ in the effective theory (\ref{eq:bulkaction}).  In M theory we have M2-branes ($p=2$) and M5-branes ($p=5$), with $\mu_p=(2\pi)^{-p}\ell_P^{-(p+1)}$.

As discussed above,
\be\label{eq:muandz}
z={1\over{1-2\mu^2}}, ~~~~~~ \mu^2={M^2\kappa^2\over{2 g^2}}.
\ee
Here
\be\label{eq:branemass}
M=V_p\ell_P^p\tau_p
\ee
in terms of the $p$-brane tension $\tau_p=\mu_p$.  Also
\be
\kappa^2={(2\pi)^8\ell_P^2\over 2 V},
\ee
and from the dimensional reduction of the kinetic term for $A_{(p+1)}$ we have
\be\label{eq:gsquared}
{1\over {2 g^2}} = {1\over{p! (2\pi)^8\ell_P^9}}\int d^7x\sqrt{G_7} ~ \omega_{(p)i_1\dots i_p} ~ \omega_{(p)}^{i_1\dots i_p}.  \label{eq:1over4g2}
\ee
where $\sqrt{G_7}$ is the square root of the determinant of the metric of the internal seven dimensions.
%{\tt Note that we took ``$x=F^2$" to be ${1\over 2}F_{\mu\nu}F^{\mu\nu}=g^{vv}g^{tt}\Phi'^2$ in evaluating it in terms %of $\Phi'$ in the brane polarization section.}
Note that for a simple product geometry, or anything close to it such as a fibration, this quantity will scale like $V/V_p^2$, leaving no dependence of $z$ (\ref{eq:muandz}) on $V$ and $V_p$.

In general, in order to evaluate $\mu^2$ (\ref{eq:muandz}), we need information about the geometry to compute (\ref{eq:gsquared}).  However, a very simple case to consider to begin with is that of D0-branes sitting at a point in the compactification on a six-manifold $X_6$.  Coming back to string theory then, we have for the D0-brane mass
\be\label{eq:Dzeromass}
M={1\over g_s\sqrt{\alpha'}}
\ee
The form $\omega_{(p)}$ \eqn{eq:formreduction}\ becomes simply $\omega_{(0)}=\sqrt{\alpha'}$.

The D0-brane sources the 1-form gauge field $A_{(1)}$ in type IIA string theory, with kinetic term $|F|^2 Vol(X_6)\alpha'/4\kappa_{10}^2$.  Plugging into \eqn{eq:muandz}, using that $\kappa^2=\kappa_{10}^2g_s^2/Vol(X_6)$, we obtain
\be\label{eq:ztwo}
\mu^2={1\over 4} ~~~ \Rightarrow ~~~ z=2
\ee
It is interesting that this calculation lands us on the simple value of $z$ which also leads to linear resistivity in our simplest setup.  (In order to connect to that discussion, we need to include flavor branes in the compactification on $X_6$).  One obtains the same result $z=2$ for a Dp-brane wrapped on one factor of a product manifold.  On more complicated manifolds there are effects that go in both directions.  If there are components of $\omega_{(p)}$ not tangent to the wrapped brane, as in nontrivial fibrations, they increase the kinetic term~(\ref{eq:1over4g2}) and therefore also $\mu^2$ and $z$; if $\mu^2 \geq 1/2$ then the branes do not polarize at all.  On the other hand, if the brane wraps in a region with warping, its tension, as well as $\mu^2$ and $z$, are reduced.

Specific examples of IIA compactifications with known field theory duals which we can use for this purpose include the near horizon limit of D2-branes and flavor D6-branes (equivalently a certain orbifold of M theory on $S^7$) and the IIA limit of the ABJM theory \cite{ABJM}\ (another orbifold of M theory on $S^7$).  Another class of models arises at large radius from type IIA on $S^6$ with 6-form flux and flavor D4-branes.  Altogether, this simple construction provides concrete, though ultimately discrete, Lifshitz geometries from string theory.

More generally, we could consider baryons in M theory on a spectrum of different manifolds such as $Q(1,1,1)$ and $Y_{p,k}$ \cite{Ypks}.  See \cite{Igor}\ for a recent discussion of baryons from wrapped 5-branes in $Q(1,1,1)$; these do not condense in their simplest setup as can be seen from similar reasoning to that given above.  It will be interesting to analyze the spectrum of available values of $z$ from brane-polarization induced Lifshitz in the Landscape (Freund-Rubin and beyond). Of course, these are at best toy models for several reasons; for example, the supersymmetry preserved by some of the underlying theories \cite{Ypks,Q111,ABJM}\ renders them distant from direct  condensed-matter applications.

Let us next address the question of whether the moduli are destabilized by the polarized branes and the field strength $F_{vt}$ turned on in our background.  Let us begin with a stabilized $AdS_4$ solution in the absence of these additional ingredients, and analyze their effects on the moduli, in a metric that takes a product form as just discussed.
After substituting the solution to the $\rho$ equation of motion, we have a term in the Lagrangian proportional to
\be\label{eq:extra}
  -\sqrt{-g}{1\over g^2}F^2\sim\sqrt{-g} {M^2 f^2\over {g^2 L^2}}\sim \sqrt{-g}{\tau_p^2 V f^2\over L^2}
\ee
In the last step we used (\ref{eq:branemass})
and that in the product-like form of the metric, ${1\over g^2}\sim {V\over V_p^2}$ (the inverse powers of $V_p$ coming from the fact that $A$ descends from the higher-rank $p$-form potential $A_{(p)}$).  This term scales like the 4-dimensional curvature term.  Let us focus on the type IIA examples mentioned above.  If we Weyl rescale, sending $g_{\mu\nu}\to g_{\mu\nu} g_s^2/Vol(X_6)$ to go to four-dimensional Einstein frame, we remove the moduli dependence from the Einstein term in the standard way.  In so doing, we also {\it almost} remove the moduli dependence from the extra term \eqn{eq:extra}, since it scales the same way in terms of powers of $g_{\mu\nu}$.  Remembering that $f=d\gamma_t/d \log v$ and that $g_{tt}\sim e^{2\gamma_t}$, and integrating by parts once, one finds a residual dependence on one combination of the moduli which is proportional to $\mu^2 f(f+2k)\log(Vol(X_6)/g_s^2)$.  We find in explicit examples, such as type IIA on $S^6$ with 6-form flux and flavor D4-branes, that this extra term shifts, but does not eliminate, the solution for the moduli for sufficiently small $\mu^2$.  In particular, the value of interest \eqn{eq:ztwo}\ for our explicit construction is in this range.

Another question is whether there are relevant perturbations of these Lifshitz solutions as in \cite{Kachru:2008yh}, so that one must tune to reach the fixed point.  We leave this question for future work.
There is yet another type of potential instability to consider in holographic models which we will describe in the final subsection below.  The present models are subject to this instability, but as we will see the effect is negligible -- the instability is very slow -- in certain limits (large $N$, large radius or finite temperature).

\subsection{Backreaction in a Fermi surface model}

Refs.~\cite{MIT}\ considered a possible holographic model of a Fermi liquid, in which a Reissner-Nordstrom black hole is surrounded by a bulk density of charged fermions.  Although somewhat different from the baryon gas considered in the previous model, we are led to examine the backreaction of the fermions on the solution.

The Einstein-Maxwell action
\begin{equation}
\label{eq:bulkaction2}
S = \frac{1}{2 \kappa^2} \int d^4x \sqrt{-g} \left(  {\cal R} + \frac{1}{L_2^2}  -  \frac{L_2^2}{2e_3^2} F_{\mu\nu} F^{\mu\nu} \right) \,,
\end{equation}
has an $AdS_2\times \R^2$ solution
\begin{eqnarray}
ds^2 &=&{L_2^2}\left(  \frac{-d\tau^2 + d\r^2}{\r^2} + dx^i dx^i \right)\ , \nonumber\\
A_{\tau} &=& e_3/\r\ .
\end{eqnarray}
In the constructions of \cite{MIT}, this near-horizon geometry goes over to $AdS_4$ toward the boundary, but our discussion will only involve the $AdS_2$ region. Coupling in a Dirac fermion of mass $m$ and charge $q$, there is a condensation of bulk fermions when $q e_3 > m L_2$.

Let us calculate the total charge carried by these fermions.  This is simple in the case $qe_3 \gg 1$, $qe_3 \gg mL_2$.  The WKB approximation gives coordinate momentum $k_\r \sim qe_3 /\r$, which is large on the scale $\r$ on which the geometry varies.  Thus we study the bulk Fermi sea in a locally flat geometry.  The Fermi energy seen by an inertial observer is
\begin{equation}
 q A_{\tilde\tau} = q A_\tau/\sqrt{-g_{\tau\tau}} = qe_3 /L_2\ ,
\end{equation}
and the number density for relativistic fermions, in inertial coordinates, is
\begin{equation}
\tilde n = \frac{q^3  A_{\tilde\tau}^3}{3\pi^2} = \frac{ q^3 e_3^3}{3 \pi^2 L_2^3}\ .
\end{equation}
To obtain the total density in the field theory we must integrate with the invariant volume element in the radial direction,
\begin{equation}\label{eq:rhoCFT}
\rho_{\rm CFT} = q \int \frac{L_2 d\r}{\r}\, \tilde n\ .
\end{equation}
The divergence at small $\r$ is cut off by the transition to the $AdS_4$ geometry, but the divergence at large $\r$ is real and implies that backreaction cannot be neglected.

This density should be compared with the charge density of the black hole itself,
\begin{equation}
\rho_{\rm RN} = \frac{1}{\kappa^2 e_3} \ ,
\end{equation}
so that
\begin{equation}
\label{ratio}
\frac{\rho_{\rm CFT} }{\rho_{\rm RN} } \approx q^4 e_3^4 \frac{\kappa^2}{L_2^2} \ln \infty\ .
\end{equation}
We have assumed $q e_3 $ to be large to permit the WKB approximation, but the conclusion is general as long as we are in the regime where there is a density of fermions.  The vector potential $A_\tau$ leaves a scale symmetry unbroken, so the density will always be constant in inertial coordinates, and thus diverge with the spatial volume of $AdS_2$.  For $qe_3 $ not large, the coefficient of the logarithm is small in the supergravity regime, and so the effect of the backreaction becomes significant only at the largest scales.

To figure out the true geometry we again adopt the metric Ansatz
\begin{equation}
ds^2 = {L_2^2}\left(  - e^{2\gamma_\tau(v)} d\tau^2 + \frac{ d\r^2}{\r^2} + e^{2\gamma_x(v)} dx^i dx^i \right)\ .
\end{equation}
The action is
\begin{equation}
S = S_\psi + \frac{L_2^2}{2\kappa^2} \int {d^3 x} \int dy \left(
e^{\gamma_\tau + 2\gamma_x}(1 + 4\dot \gamma_\tau \dot \gamma_x + 2 \dot\gamma_x^2)
+ \frac{1}{e_3^2} e^{-\gamma_\tau + 2\gamma_x} \dot A_\tau^2\right)\ ,
\end{equation}
where $y = \ln v$ and a dot denotes $\partial/\partial y$.
The field equations are
\begin{eqnarray}
1 - 4\ddot \gamma_x - 6 \dot \gamma_x^2 - \frac{1}{e_3^2} e^{-2\gamma_\tau} \dot A_\tau^2
&=& {2\kappa^2}{L_2^2} T^{\tilde\tau \tilde\tau}\ ,
\nonumber\\
1 - 2\ddot \gamma_x - 2\ddot \gamma_\tau - 2 \dot \gamma_x^2 -2 \dot \gamma_x \dot \gamma_\tau  - 2 \dot \gamma_\tau^2 + \frac{1}{e_3^2} e^{-2\gamma_\tau} \dot A_\tau^2
&=& -{2\kappa^2}{L_2^2} T^{\tilde x \tilde x}\ ,
\nonumber\\
1 - 4 \dot\gamma_\tau \dot \gamma_x - 2 \dot \gamma_x^2 - \frac{1}{e_3^2} e^{-2\gamma_\tau} \dot A_\tau^2
&=& -{2\kappa^2}{L_2^2} T^{\tilde \r \tilde \r}\ ,
\nonumber\\
e^{-\gamma_\tau} \ddot A_\tau - e^{-\gamma_\tau} (\dot \gamma_\tau - 2\dot \gamma_x) \dot A_\tau
&=&{\kappa^2 e_3^2}{L_2} j^{\tilde\tau}\ .
\end{eqnarray}
We will again assume $e_3 q$ to be large, so the inertial frame energy density, pressure, and charge density are given by the local equation of state of the Fermi gas,\footnote{Backreaction of a bulk Fermi gas in another situation was considered in Ref.~\cite{deBoer:2009wk}.}
\begin{equation}
T^{\tilde\tau \tilde\tau} = 3
T^{\tilde\r  \tilde\r} = 3 T^{\tilde x \tilde x} = \frac{q^4 A_{\tilde\tau}^4}{4\pi^2} \ ,
\quad
j^{\tilde\tau} = \frac{q^4 A_{\tilde\tau}^3}{3\pi^2}\ , \quad A_{\tilde\tau} \equiv
A_\tau/L_2 e^{\gamma_\tau}\ .
\end{equation}

It is natural to look for a solution with Lifshitz scaling, as the contraction of the transverse directions will regulate the divergence of the charge.  Inserting
\begin{equation}
\gamma_\tau = a y\ ,\quad \gamma_x = by\ ,\quad A_\tau = \mu e^{cy}\ ,
\end{equation}
one finds that the field equations require that $c = a$ and that
\begin{eqnarray}
1 - 6b^2 - {a^2 w} &=& 3\epsilon  w^2\ ,\nonumber\\
1 - 2 a^2 - 2ab - 2b^2 + {a^2 w} &=& - \epsilon w^2\ ,\nonumber\\
1 - 4ab - 2 b^2 -{a^2w} &=& - \epsilon w^2\ ,
\nonumber\\
{abw} &=& \epsilon w^2 \ .
\end{eqnarray}
Here we have defined $w = \mu^2/e_3^2$, while
\begin{equation}
\epsilon =
\frac{\kappa^2 }{L_2^2} \frac{q^4 e_3^4}{6\pi^2}
\end{equation}
is the same dimensionless expansion parameter appearing in the charge ratio~(\ref{ratio}).  These four equations for the three unknowns $a, b, w$ satisfy one linear relation, owing to the Bianchi identity.

Expanding for small $\epsilon$, we find
\begin{equation}
a \sim -1+ \epsilon\ ,\quad b \sim -\epsilon\ ,\quad w \sim 1 - \epsilon\ ,
\end{equation}
so we recover the unbackreacted solution as $\epsilon \to 0$.  The Lifshitz exponent $z = a/b \sim 1/\epsilon$ is large.  Even for small $\epsilon$, however, there is a large qualitative effect.  The total flux emerging from $\r = \infty$ now vanishes, and the charge previously attributed to the black hole is now carried entirely by the bulk fermions.  One can also see this from the charge~(\ref{eq:rhoCFT}): at small $\epsilon$ the charge density is essentially unchanged, but the total volume is now
\begin{equation}
\int \frac{d\r}{\r^{1+2\epsilon}} = \frac{1}{2\epsilon}\ ,
\end{equation}
with the result that the bulk contribution to the charge density is exactly equal to that which had been attributed to the black hole.  Although the local density of bulk fermions is suppressed by $\kappa^2/L_2^2$, which is an inverse power of $N$, the bulk volume gives the reciprocal power.  Further, the horizon area is now zero, eliminating the widely discussed puzzle regarding the zero temperature entropy.

Note that for $m L_2 > q e_3$ (which is outside the regime of Fermi surface behavior~\cite{MIT}) there is no bulk charge and the horizon is still present.   As $e_3$ is increased, it becomes energetically favorable for the charge instead to be carried by explicit bulk fermions.  In this phase, the black hole is unstable to radiating all its charge into the bulk.  There is no intermediate regime where the charge is shared between the black hole and the bulk.  It would be interesting to understand the behavior of the entropy during the transition.

The Fermi liquid pole identified in the second and fourth papers of~\cite{MIT} arises from a state in the domain wall region between the $AdS_4$ and $AdS_2$ geometries, and so is not strongly affected by this modification in the extreme IR.\footnote{Note that the existence of these states in the domain wall, and the presence of the Fermi sea down to the $AdS_2$ horizon, are not directly connected.  Over most of the parameter space of Ref.~\cite{MIT}, the former implies the latter.  However, T. Faulkner informs us that there are situations where this will not be the case, and so the $AdS_2$ geometry will still be present.}  Also, the modification of the throat involves energy scales exponentially small in $L^2/\kappa^2$, and so will have little effect at temperatures larger than this.  However, given there may be no small parameter $L^2/\kappa^2$ in the real systems, the backreaction effect is likely to be important.  The relation between the bulk and boundary Fermi surfaces should be better understood.

\subsection{Charged black holes versus probe branes}

For many applications of the holographic correspondence to condensed matter systems, it is essential to introduce a finite charge density. Some of the deepest questions in the field are concerned with strongly coupled physics at finite density and the rearrangement of the Fermi surface that appears to occur at quantum critical points in the cuprate and heavy fermion phase diagrams. The strange metallic physics of these materials, that has been the inspiration and focus of this paper, is likely to be tied to an exotic non-Fermi liquid description of matter at finite density.

To capture the correct new physics, it is expected to be important to be in a regime in which the strongly coupled charged degrees of freedom are not dilute, in the sense of including nonlinearities in the charge density. While holography automatically works in a strongly coupled regime, without quasiparticles, there have been two approaches to finite density in the applied holography literature, and it may be useful to compare and contrast.

The charge density is the time component of a conserved current. Via the basic holographic dictionary, a conserved current has a dual description as a Maxwell field. Nonlinearities in the charge density in the field theory will be captured by interactions of the Maxwell field in the bulk. In one approach (e.g. \cite{Hartnoll:2007ai, Hartnoll:2007ih, Hartnoll:2007ip, Hartnoll:2008kx, MIT}) the Maxwell action takes the usual simple quadratic form $F^2$. However, this Maxwell field is then coupled to a dynamical metric and possibly charged fields which induce interactions in the Maxwell sector. One fairly robust feature of this setup (in the large $N$ limit) is that at the quantum critical point (i.e. without relevant operators turned on or condensates) the finite density theory is dual to a charged black hole which at zero temperature becomes an extremal black hole with a near horizon $AdS_2 \times \R^2$ region.

A second approach is to consider a nonlinear action for the Maxwell field (e.g. \cite{myers, kob, KulaxiziParnachev, Karch:2008fa}) and ignore the interactions of the Maxwell field with the metric. This is the probe approximation we have used in this paper. There is some ambiguity in choice of an action here, a favorite is the DBI action \cite{Born:1934gh}, as this arises naturally on D-branes in string theory. It also has the appealing property of a maximal field strength.  Within string theory, the square-root (DBI) action we have been using to govern the gauge fields is dual to the action $m\int dt\sqrt{1-\dot x^2}$ for the motion of relativistic particles.
An interesting aspect of the D-brane actions is that it may be possible to construct explicitly the gravitational dual of a theory in which at weak coupling the charge density is carried entirely by fermions.

Ultimately to move into an experimentally interesting regime it is likely to be necessary to combine these approaches. For instance, a phenomenological 3+1 dimensional bulk model incorporating aspects from both approaches is a gravitating DBI action
\be\label{eq:bulkactionDBI}
S =  \frac{1}{2 \kappa^2} \int d^4x \sqrt{-g}\left( {\cal R} + \frac{8}{L^2} \right)
- \frac{1}{\kappa^2 L^2} \int d^4x \sqrt{- \det \left(g_{ab} + {\textstyle{\frac{\kappa L}{g}}} F_{ab} \right)}  \,,
\ee
with $F=dA$. The numerator of the cosmological constant term is $8=6+2$ in order to cancel the contribution of $-2$ from the DBI action at $F=0$. One solution of this theory that can be found explicitly is the metric
\be
ds^2 = \frac{4 L^2}{15 \rho^2} \left(- \left(1 - \frac{\rho^2}{\rho_+^2} \right)^2 dt^2 + d\rho^2 \right) + L^2 \left( dx^2 + dy^2 \right) \,,
\ee
and field strength
\be
F = \frac{g L}{\sqrt{15} \kappa} \left(1 - \frac{\rho^2}{\rho_+^2} \right) \frac{d \rho \wedge dt}{\rho^2} = \frac{g}{2 \kappa L} \text{vol}_2 \,.
\ee
These are candidate solutions for the near horizon geometry of low temperature black holes. The temperature is given by $T = \frac{1}{\pi \rho_+} \,.$
In the zero temperature limit, $\rho_+ \to \infty$, the metric becomes $AdS_2 \times \R^2$. Thus we see, as one should have anticipated, once gravitational backreaction is included the DBI theory has important features in common with the Einstein-Maxwell approach.  As derived in the previous subsection, bulk fermions will condense in this background, leading to a Lifshitz solution all told.

\subsection{Fermi seasickness}

For completeness, let us briefly note another source of instabilities which we have come across in analyzing string-theoretic models of finite-density field theory.\footnote{A similar class of instabilities is under investigation in other works \cite{otherinstabilities}.}  We will start by describing the effect, and then explain that it is a very long-time instability in many finite density systems, including those that we study.

String theory contains branes of various dimensionalities \cite{Polchinski:1998rr}.  Within the $d+2$ noncompact dimensions, some of these are domain walls lying at some radial position $v(t)$ in the gravity side warped throat geometry; the radial position is a motion collective coordinate of the brane.  We include time dependence here because generically a domain wall feels a nontrivial potential driving it to a larger or smaller value of $v$.  If the potential provides a restoring force driving all domain walls toward the IR (large $v$), then these branes do not represent an additional instability.  But if the potential drives any of the branes up toward the UV end of the throat, this represents a new instability.

Such a potential may get contributions from many sources.  In a quantum critical theory, its form is limited by the scaling symmetry; an example of this is the $\lambda\phi^4=L^4/\alpha'^4v^4$ potential for anti-D3-branes in $AdS_5$. Our main interest here is that at finite density, the bulk gauge field dual to the quantum field theory chemical potential in general contributes to this potential.

The effect of interest is rather simple.  Consider for example a set of color $p$-branes and flavor $q$-branes, with $q>p$.  For massive charge carriers, the two are displaced from each other.  Flavors are $p$-$q$ strings.  A chemical potential for these strings introduces a finite density of them.  They pull out on the color branes, moving them away from the origin in the space of adjoint scalar fields, toward the flavor branes.  So far we have described this in the weakly coupled D-brane picture, so let us translate this to strong coupling.
The flavor branes become space filling branes on the gravity side of the holographic correspondence, the strong 't Hooft coupling limit of the field theory.  A flavor $U(1)$ gauge field with an $F_{vt}$ field strength implements the chemical potential as we discussed at length in the above analysis.  This field costs energy, but it can reduce its energy by ending on a domain wall $p$-brane at a finite radial position $v$ in the bulk.  This introduces a potential driving the brane up the throat.  The potential suggested by this effect is linear, and hence tends to beat the underlying quantum critical potential near the origin even if the latter leads to a restoring force.

More generally, independently of the presence of flavor branes, this instability can arise if the gravitational pull toward the bottom of the throat is overcompensated by electric fields from fluxes which are presence in the construction.  In the models of brane-polarization-induced Lifshitz discussed above, we find that this instability arises for the D2-brane theories describe there, with the exception of D2-D4.  But the latter has an instability in a transverse direction:  a D2-brane brought out to a finite radial position is unstable to dissolving into the flavor D4-branes in that case.

As we have mentioned, this instability involves a scalar field collective coordinate of the domain wall branes.
We might expect this problem to be absent in realistic field theories for condensed matter physics, which may not include such scalar fields.  However, in holographic systems controlled by a general relativistic approximation and UV completed by string theory, the ubiquity of branes suggests that this is a fairly generic issue.  From the point of view of the dual field theory, most CFT's with geometric duals have nontrivial moduli spaces or pseudomoduli spaces, and at finite density it often turns out to the case that the moduli are driven away from the IR point of interest.  It may be possible to prevent this problem with model building maneuvers, such as projecting out scalar modes via orbifolding (e.g. by requiring that all the $p$ branes be fractional branes).

Once this problem is present, is there any way to obtain relief from Fermi seasickness without eliminating the Fermi sea itself and returning to the dry landscape?
Firstly, it is interesting to note that the motion of the brane collective coordinate toward or away from the origin is limited by the speed of light in the bulk \cite{DBIsky}; in the large-$N_c$ approximation in which we are working this process takes forever.
Warming the system up also helps: the black hole pulls the color branes back toward the origin.  This renders the system metastable.

%Also, as alluded to above, other contributions to the potential might apply away from the origin in scalar field %space, preventing the branes from coming all the way up the throat.

\section*{Acknowledgments}
We would like to thank A. O'Bannon, M. Beasley, G. Horowitz, T. Faulkner, E. Fradkin, S. Kachru, A. Karch, S. Kivelson, D. Mateos, C. Nayak, M. Roberts, S. Sachdev, T. Senthil and A. Sinha.
E.S. and D.T. thank KITP and S.A.H., J.P., and S. Sachdev for the stimulating workshop
``Quantum Criticality and the AdS/CFT Correspondence" where this collaboration began.  J.P. transfers his share of the foregoing thanks to S.A.H. for doing all the work, both conceptual and organizational, in making the workshop happen.  We also thank the organizers and participants of the KITP program ``The Physics of Higher Temperature Superconductivity".
The research of E.S. is supported by NSF grants
PHY-0244728 and PHY05-51164, by the DOE under contract DE-AC03-76SF00515, and by the BSF.
The research of S.A.H. is partially supported by DOE grant DE-FG02-91ER40654 and by the FQXi foundation. D.T. is supported by
the Royal Society.


\begin{thebibliography}{10}





\bibitem{stewart}
  G.~R.~Stewart,
  ``Non-Fermi-liquid behavior in d- and f- electron metals,''
  Rev. Mod. Phys. {\bf 73}, 797 (2001).

\bibitem{hussey1}
R~A.~Cooper, Y.~Wang, B.~Vignolle, O.~J.~Lipscombe, S.~M.~Hayden, Y.~Tanabe, T.~Adachi,
Y.~Koike, M.~Nohara, H.~Takagi, C.~Proust and N.~E.~Hussey,
``Anomalous criticality in the electrical resistivity of La$_{2-x}$Sr$_x$CuO$_4$'',
Science, {\bf 323}, 603 (2009).

\bibitem{hussey2}
N.~E.~Hussey,
``Phenomenology of the normal state in-plane transport properties of high-T$_c$ cuprates,''
J. Phys.:Condens. Matter {\bf 20}, 123201 (2008)
[arXiv:0804.2984 [cond-mat.supr-con]].

\bibitem{martin}
S.~Martin, A.~T.~Fiory, R.~M.~Fleming, L.~F.~Schneemeyer and J.~V.~Waszczak,
``Normal state transport properties of Bi${}_{2+x}$Sr${}_{2-y}$CuO${}_{6 \pm \delta}$ crystals,''
Phys. Rev. {\bf B 41}, 846 (1990).

\bibitem{optical}
D. van der Marel, H. J. A. Molegraaf, J. Zaanen, Z. Nussinov, F. Carbone, A. Damascelli, H. Eisaki, M. Greven, P. H. Kes and M. Li,
``Quantum critical behavior in a high-T$_c$ superconductor,''
Nature {\bf 425}, 271, (2003).

\bibitem{hall}
A.~W.~Tyler and A.~P.~Mackenzie,
``Hall effect of single layer, tetragonal Tl$_2$Ba$_2$CuO$_{6+\d}$ near optimal doping,''
Physica C {\bf 282-287}, 1185 (1997).

%\cite{Lee:2008xf}
\bibitem{MIT}
  S.~S.~Lee,
  ``A Non-Fermi Liquid from a Charged Black Hole: A Critical Fermi Ball,''
  arXiv:0809.3402 [hep-th].
  %%CITATION = ARXIV:0809.3402;%%

%\cite{Liu:2009dm}
%\bibitem{Liu:2009dm}
  H.~Liu, J.~McGreevy and D.~Vegh,
  ``Non-Fermi liquids from holography,''
  arXiv:0903.2477 [hep-th].
  %%CITATION = ARXIV:0903.2477;%%

%\cite{Cubrovic:2009ye}
%\bibitem{Cubrovic:2009ye}
  M.~Cubrovic, J.~Zaanen and K.~Schalm,
  ``Fermions and the AdS/CFT correspondence: quantum phase transitions and the
  emergent Fermi-liquid,''
  arXiv:0904.1993 [hep-th].
  %%CITATION = ARXIV:0904.1993;%%

%\cite{Faulkner:2009wj}
%\bibitem{MIT}
  T.~Faulkner, H.~Liu, J.~McGreevy and D.~Vegh,
  ``Emergent quantum criticality, Fermi surfaces, and AdS2,''
  arXiv:0907.2694 [hep-th].
  %%CITATION = ARXIV:0907.2694;%%

T. Faulkner, N. Iqbal, H. Liu, J. McGreevy, D. Vegh, ...,
Work in progress.

%\cite{Aharony:1999ti}
\bibitem{Aharony:1999ti}
  O.~Aharony, S.~S.~Gubser, J.~M.~Maldacena, H.~Ooguri and Y.~Oz,
  ``Large N field theories, string theory and gravity,''
  Phys.\ Rept.\  {\bf 323}, 183 (2000)
  [arXiv:hep-th/9905111].
  %%CITATION = PRPLC,323,183;%%

%\cite{Hartnoll:2009sz}
\bibitem{Hartnoll:2009sz}
  S.~A.~Hartnoll,
  ``Lectures on holographic methods for condensed matter physics,''
  arXiv:0903.3246 [hep-th].
  %%CITATION = ARXIV:0903.3246;%%

%\cite{Herzog:2009xv}
\bibitem{Herzog:2009xv}
  C.~P.~Herzog,
  ``Lectures on Holographic Superfluidity and Superconductivity,''
  J.\ Phys.\ A  {\bf 42}, 343001 (2009)
  [arXiv:0904.1975 [hep-th]].
  %%CITATION = JPAGB,A42,343001;%%

  %\cite{McGreevy:2009xe}
\bibitem{McGreevy:2009xe}
  J.~McGreevy,
  ``Holographic duality with a view toward many-body physics,''
  arXiv:0909.0518 [hep-th].
  %%CITATION = ARXIV:0909.0518;%%

%\cite{Hartnoll:2009qx}
\bibitem{Hartnoll:2009qx}
  S.~A.~Hartnoll,
  ``Quantum Critical Dynamics from Black Holes,''
  arXiv:0909.3553 [cond-mat.str-el].
  %%CITATION = ARXIV:0909.3553;%%

%\cite{Karch:2007pd}
\bibitem{kob}
  A.~Karch and A.~O'Bannon,
  ``Metallic AdS/CFT,''
  JHEP {\bf 0709}, 024 (2007)
  [arXiv:0705.3870 [hep-th]].
  %%CITATION = JHEPA,0709,024;%%

%\cite{Kulaxizi:2008jx}
\bibitem{KulaxiziParnachev}
  M.~Kulaxizi and A.~Parnachev,
  ``Holographic Responses of Fermion Matter,''
  Nucl.\ Phys.\  B {\bf 815}, 125 (2009)
  [arXiv:0811.2262 [hep-th]].
  %%CITATION = NUPHA,B815,125;%%

%\cite{Kulaxizi:2008kv}
%\bibitem{Kulaxizi:2008kv}
  M.~Kulaxizi and A.~Parnachev,
  ``Comments on Fermi Liquid from Holography,''
  Phys.\ Rev.\  D {\bf 78}, 086004 (2008)
  [arXiv:0808.3953 [hep-th]].
  %%CITATION = PHRVA,D78,086004;%%

%\cite{Karch:2009eb}
\bibitem{Karch:2009eb}
  A.~Karch, M.~Kulaxizi and A.~Parnachev,
  ``Notes on Properties of Holographic Matter,''
  arXiv:0908.3493 [hep-th].
  %%CITATION = ARXIV:0908.3493;%%

%\cite{Kachru:2008yh}
\bibitem{Kachru:2008yh}
  S.~Kachru, X.~Liu and M.~Mulligan,
  ``Gravity Duals of Lifshitz-like Fixed Points,''
  Phys.\ Rev.\  D {\bf 78}, 106005 (2008)
  [arXiv:0808.1725 [hep-th]].
  %%CITATION = PHRVA,D78,106005;%%

%\cite{Koroteev:2007yp}
\bibitem{Koroteev:2007yp}
  P.~Koroteev and M.~Libanov,
  ``On Existence of Self-Tuning Solutions in Static Braneworlds without
  Singularities,''
  JHEP {\bf 0802}, 104 (2008)
  [arXiv:0712.1136 [hep-th]].
  %%CITATION = JHEPA,0802,104;%%



  \bibitem{hertz}
J.~A.~Hertz,
``Quantum critical phenomena,"
Phys. Rev. {\bf B} 14 (1976) 1165.

\bibitem{Ando}
Y. Ando, A.N. Lavrov, S. Komiya, K. Segawa, and X.F. Sun,
``Mobility of the Doped Holes and the Antiferromagnetic Correlations in Underdoped High-$T_c$ Cuprates"
Phys.\ Rev.\ Lett. {\bf 87}, 017001-1 (2001).

\bibitem{Phillips}
P. Phillips,
``Breakdown of One-Paramater Scaling in Quantum Critical Scenarios for the High-Temperature Copper-oxide Superconductors"
Phys.\ Rev.\ Lett. {\bf 95}, 107002.

\bibitem{lifbh}
  U.~H.~Danielsson and L.~Thorlacius,
  ``Black holes in asymptotically Lifshitz spacetime,''
  JHEP {\bf 0903}, 070 (2009)
  [arXiv:0812.5088 [hep-th]].
  %%CITATION = JHEPA,0903,070;%%

\bibitem{ohmann}  R.~B.~Mann,
  ``Lifshitz Topological Black Holes,''
  JHEP {\bf 0906}, 075 (2009)
  [arXiv:0905.1136 [hep-th]].
  %%CITATION = JHEPA,0906,075;%%

\bibitem{bertoldi}
 G.~Bertoldi, B.~A.~Burrington and A.~Peet,
  ``Black Holes in asymptotically Lifshitz spacetimes with arbitrary critical
  exponent,''
  arXiv:0905.3183 [hep-th].
  %%CITATION = ARXIV:0905.3183;%%

\bibitem{bm}
  K.~Balasubramanian and J.~McGreevy,
  ``An analytic Lifshitz black hole,''
  arXiv:0909.0263 [hep-th].
  %%CITATION = ARXIV:0909.0263;%%

\bibitem{hr}
  S.~W.~Hawking and S.~F.~Ross,
  ``Duality between Electric and Magnetic Black Holes,''
  Phys.\ Rev.\  D {\bf 52}, 5865 (1995)
  [arXiv:hep-th/9504019].
  %%CITATION = PHRVA,D52,5865;%%

\bibitem{Burgess:1986dw}
  C.~P.~Burgess,
  ``Open string instability in background electric fields,''
  Nucl.\ Phys.\  B {\bf 294}, 427 (1987).
  %%CITATION = NUPHA,B294,427;%%
  V.~V.~Nesterenko,
  ``The dynamics of open strings in a background electric field,''
  Int.\ J.\ Mod.\ Phys.\  A {\bf 4}, 2627 (1989).
  %%CITATION = IMPAE,A4,2627;%%
  C.~Bachas and M.~Porrati,
  ``Pair creation of open strings in an electric field,''
  Phys.\ Lett.\  B {\bf 296}, 77 (1992)
  [arXiv:hep-th/9209032].
  %%CITATION = PHLTA,B296,77;%%

\bibitem{Gopakumar:2000na}
  R.~Gopakumar, J.~M.~Maldacena, S.~Minwalla and A.~Strominger,
  ``S-duality and noncommutative gauge theory,''
  JHEP {\bf 0006}, 036 (2000)
  [arXiv:hep-th/0005048].
  %%CITATION = JHEPA,0006,036;%%

  \bibitem{Seiberg:2000ms}
  N.~Seiberg, L.~Susskind and N.~Toumbas,
  ``Strings in background electric field, space/time noncommutativity  and a
  new noncritical string theory,''
  JHEP {\bf 0006}, 021 (2000)
  [arXiv:hep-th/0005040].
  %%CITATION = JHEPA,0006,021;%%

\bibitem{DBIsky}
%\cite{Alishahiha:2004eh}
  M.~Alishahiha, E.~Silverstein and D.~Tong,
  ``DBI in the sky,''
  Phys.\ Rev.\  D {\bf 70}, 123505 (2004)
  [arXiv:hep-th/0404084].
  %%CITATION = PHRVA,D70,123505;%%

%\cite{Horava:2009vy}
\bibitem{Horava:2009vy}
  P.~Horava and C.~M.~Melby-Thompson,
  ``Anisotropic Conformal Infinity,''
  arXiv:0909.3841 [hep-th].
  %%CITATION = ARXIV:0909.3841;%%

\bibitem{Klebanov:1999tb}
 I.~R.~Klebanov and E.~Witten,
 ``AdS/CFT correspondence and symmetry breaking,''
 Nucl.\ Phys.\  B {\bf 556}, 89 (1999)
 [arXiv:hep-th/9905104].
 %%CITATION = NUPHA,B556,89;%%

 \bibitem{Witten:2003ya}
  E.~Witten,
  ``SL(2,Z) action on three-dimensional conformal field theories with Abelian
  symmetry,''
  arXiv:hep-th/0307041.
  %%CITATION = HEP-TH/0307041;%%

%\cite{Witten:2001ua}
\bibitem{Witten:2001ua}
  E.~Witten,
  ``Multi-trace operators, boundary conditions, and AdS/CFT correspondence,''
  arXiv:hep-th/0112258.
  %%CITATION = HEP-TH/0112258;%%

\bibitem{obhall}
  A.~O'Bannon,
  ``Hall Conductivity of Flavor Fields from AdS/CFT,''
  Phys.\ Rev.\  D {\bf 76}, 086007 (2007)
  [arXiv:0708.1994 [hep-th]].
  %%CITATION = PHRVA,D76,086007;%%

  %\cite{Hartnoll:2007ih}
\bibitem{Hartnoll:2007ih}
  S.~A.~Hartnoll, P.~K.~Kovtun, M.~Muller and S.~Sachdev,
  ``Theory of the Nernst effect near quantum phase transitions in condensed
  matter, and in dyonic black holes,''
  Phys.\ Rev.\  B {\bf 76}, 144502 (2007)
  [arXiv:0706.3215 [cond-mat.str-el]].
  %%CITATION = PHRVA,B76,144502;%%

  %\cite{Hartnoll:2007ip}
\bibitem{Hartnoll:2007ip}
  S.~A.~Hartnoll and C.~P.~Herzog,
  ``Ohm's Law at strong coupling: S duality and the cyclotron resonance,''
  Phys.\ Rev.\  D {\bf 76}, 106012 (2007)
  [arXiv:0706.3228 [hep-th]].
  %%CITATION = PHRVA,D76,106012;%%

%\cite{Son:2002sd}
\bibitem{Son:2002sd}
  D.~T.~Son and A.~O.~Starinets,
  ``Minkowski-space correlators in AdS/CFT correspondence: Recipe and
  applications,''
  JHEP {\bf 0209}, 042 (2002)
  [arXiv:hep-th/0205051].
  %%CITATION = JHEPA,0209,042;%%

\bibitem{sachdev}
S.~Sachdev,
{\it Quantum phase transitions}, Cambridge University Press,
UK (2001).

%\cite{Herzog:2007ij}
\bibitem{Herzog:2007ij}
  C.~P.~Herzog, P.~Kovtun, S.~Sachdev and D.~T.~Son,
  ``Quantum critical transport, duality, and M-theory,''
  Phys.\ Rev.\  D {\bf 75}, 085020 (2007)
  [arXiv:hep-th/0701036].
  %%CITATION = PHRVA,D75,085020;%%

  \bibitem{ACreview}
D.N. Basov and T. Timusk,
``Electrodynamics of high-Tc superconductors"
Rev. Mod. Phys.
{\bf 77}, 721 (2005).

  %\cite{Horowitz:2009ij}
\bibitem{Horowitz:2009ij}
  G.~T.~Horowitz and M.~M.~Roberts,
  ``Zero Temperature Limit of Holographic Superconductors,''
  arXiv:0908.3677 [hep-th].
  %%CITATION = ARXIV:0908.3677;%%

 %\cite{Bertoldi:2009vn}
\bibitem{Bertoldi:2009vn}
  G.~Bertoldi, B.~A.~Burrington and A.~Peet,
  ``Black Holes in asymptotically Lifshitz spacetimes with arbitrary critical
  exponent,''
  arXiv:0905.3183 [hep-th].
  %%CITATION = ARXIV:0905.3183;%%

  %\cite{Ross:2009ar}
\bibitem{Ross:2009ar}
  S.~F.~Ross and O.~Saremi,
  ``Holographic stress tensor for non-relativistic theories,''
  JHEP {\bf 0909}, 009 (2009)
  [arXiv:0907.1846 [hep-th]].
  %%CITATION = JHEPA,0909,009;%%

 %\cite{Karch:2008fa}
\bibitem{Karch:2008fa}
  A.~Karch, D.~T.~Son and A.~O.~Starinets,
  ``Zero Sound from Holography,''
  Phys.\ Rev.\ Lett.\  {\bf 102}, 051602 (2009)
  [arXiv:0806.3796 [hep-th]].
  %%CITATION = ARXIV:0806.3796;%%

\bibitem{Rey}
S.~J.~Rey,
  ``String Theory On Thin Semiconductors: Holographic Realization Of Fermi
  Points And Surfaces,''
  Prog.\ Theor.\ Phys.\ Suppl.\  {\bf 177}, 128 (2009)
  [arXiv:0911.5295 [hep-th]].
  %%CITATION = PTPSA,177,128;%%


%\cite{Karch:2002sh}
\bibitem{KarchKatz}
  A.~Karch and E.~Katz,
  ``Adding flavor to AdS/CFT,''
  JHEP {\bf 0206}, 043 (2002)
  [arXiv:hep-th/0205236].
  %%CITATION = JHEPA,0206,043;%%

\bibitem{myers}   S.~Kobayashi, D.~Mateos, S.~Matsuura, R.~C.~Myers and R.~M.~Thomson,
  ``Holographic phase transitions at finite baryon density,''
  JHEP {\bf 0702}, 016 (2007)
  [arXiv:hep-th/0611099].
  %%CITATION = JHEPA,0702,016;%%

%\cite{Herzog:2006gh}
\bibitem{Herzog:2006gh}
  C.~P.~Herzog, A.~Karch, P.~Kovtun, C.~Kozcaz and L.~G.~Yaffe,
  ``Energy loss of a heavy quark moving through N = 4 supersymmetric
  Yang-Mills plasma,''
  JHEP {\bf 0607}, 013 (2006)
  [arXiv:hep-th/0605158].
  %%CITATION = JHEPA,0607,013;%%

%\cite{Gubser:2006bz}
\bibitem{Gubser:2006bz}
  S.~S.~Gubser,
  ``Drag force in AdS/CFT,''
  Phys.\ Rev.\  D {\bf 74}, 126005 (2006)
  [arXiv:hep-th/0605182].
  %%CITATION = PHRVA,D74,126005;%%

%\cite{Mateos:2007vc}
\bibitem{Mateos:2007vc}
  D.~Mateos, S.~Matsuura, R.~C.~Myers and R.~M.~Thomson,
  ``Holographic phase transitions at finite chemical potential,''
  JHEP {\bf 0711}, 085 (2007)
  [arXiv:0709.1225 [hep-th]].
  %%CITATION = JHEPA,0711,085;%%

  %\cite{Karch:2007br}
\bibitem{kob2}
  A.~Karch and A.~O'Bannon,
  ``Holographic Thermodynamics at Finite Baryon Density: Some Exact Results,''
  JHEP {\bf 0711}, 074 (2007)
  [arXiv:0709.0570 [hep-th]].
  %%CITATION = JHEPA,0711,074;%%

%\cite{Azeyanagi:2009pr}
\bibitem{Azeyanagi:2009pr}
M.~Taylor,
  ``Non-relativistic holography,''
  arXiv:0812.0530 [hep-th];
  %%CITATION = ARXIV:0812.0530;%%

T.~Azeyanagi, W.~Li and T.~Takayanagi,
  ``On String Theory Duals of Lifshitz-like Fixed Points,''
  JHEP {\bf 0906}, 084 (2009)
  [arXiv:0905.0688 [hep-th]].
  %%CITATION = JHEPA,0906,084;%%

\bibitem{KachruTrivedi}
S.~S.~Gubser and F.~D.~Rocha,
  ``Peculiar properties of a charged dilatonic black hole in $AdS_5$,''
  arXiv:0911.2898 [hep-th];
  %%CITATION = ARXIV:0911.2898;%%


K.~Goldstein, S.~Kachru, S.~Prakash and S.~P.~Trivedi,
  ``Holography of Charged Dilaton Black Holes,''
  arXiv:0911.3586 [hep-th].
  %%CITATION = ARXIV:0911.3586;%%

%\cite{Li:2009pf}
\bibitem{Li:2009pf}
  W.~Li, T.~Nishioka and T.~Takayanagi,
  ``Some No-go Theorems for String Duals of Non-relativistic Lifshitz-like
  Theories,''
  arXiv:0908.0363 [hep-th].
  %%CITATION = ARXIV:0908.0363;%%

\bibitem{Aharonyetal}
O. Aharony, S. Kachru, X. Liu, and M. Mulligan, in progress.


%\cite{Polchinski:2009ch}
\bibitem{Polchinski:2009ch}
  J.~Polchinski and E.~Silverstein,
  ``Dual Purpose Landscaping Tools: Small Extra Dimensions in AdS/CFT,''
  arXiv:0908.0756 [hep-th].
  %%CITATION = ARXIV:0908.0756;%%
%\cite{Denef:2009tp}

\bibitem{Denef:2009tp}
  F.~Denef and S.~A.~Hartnoll,
  ``Landscape of superconducting membranes,''
  Phys.\ Rev.\  D {\bf 79}, 126008 (2009)
  [arXiv:0901.1160 [hep-th]].
  %%CITATION = PHRVA,D79,126008;%%



\bibitem{LifshitzSC}
%\cite{Brynjolfsson:2009ct}
%\bibitem{Brynjolfsson:2009ct}
  E.~J.~Brynjolfsson, U.~H.~Danielsson, L.~Thorlacius and T.~Zingg,
  ``Holographic Superconductors with Lifshitz Scaling,''
  arXiv:0908.2611 [hep-th];
  %%CITATION = ARXIV:0908.2611;%%

S.~J.~Sin, S.~S.~Xu and Y.~Zhou,
  ``Holographic Superconductor for a Lifshitz fixed point,''
  arXiv:0909.4857 [hep-th].
  %%CITATION = ARXIV:0909.4857;%%

%\cite{Gubser:2008px}
\bibitem{Gubser:2008px}
  S.~S.~Gubser,
  ``Breaking an Abelian gauge symmetry near a black hole horizon,''
  Phys.\ Rev.\  D {\bf 78}, 065034 (2008)
  [arXiv:0801.2977 [hep-th]].
  %%CITATION = PHRVA,D78,065034;%%

  %\cite{Hartnoll:2008kx}
\bibitem{Hartnoll:2008kx}
  S.~A.~Hartnoll, C.~P.~Herzog and G.~T.~Horowitz,
  ``Holographic Superconductors,''
  JHEP {\bf 0812}, 015 (2008)
  [arXiv:0810.1563 [hep-th]].
  %%CITATION = JHEPA,0812,015;%%

\bibitem{Myers:1999ps}
  R.~C.~Myers,
  ``Dielectric-branes,''
  JHEP {\bf 9912}, 022 (1999)
  [arXiv:hep-th/9910053].
  %%CITATION = JHEPA,9912,022;%%

\bibitem{Polchinski:2000uf}
  J.~Polchinski and M.~J.~Strassler,
  ``The string dual of a confining four-dimensional gauge theory,''
  arXiv:hep-th/0003136.
  %%CITATION = HEP-TH/0003136;%%

\bibitem{Bergman:2007wp}
  O.~Bergman, G.~Lifschytz and M.~Lippert,
  ``Holographic Nuclear Physics,''
  JHEP {\bf 0711}, 056 (2007)
  [arXiv:0708.0326 [hep-th]].
  %%CITATION = JHEPA,0711,056;%%

\bibitem{Davis:2007ka}
  J.~L.~Davis, M.~Gutperle, P.~Kraus and I.~Sachs,
  ``Stringy NJL and Gross-Neveu models at finite density and temperature,''
  JHEP {\bf 0710}, 049 (2007)
  [arXiv:0708.0589 [hep-th]].
  %%CITATION = JHEPA,0710,049;%%

\bibitem{Rozali:2007rx}
  M.~Rozali, H.~H.~Shieh, M.~Van Raamsdonk and J.~Wu,
  ``Cold Nuclear Matter In Holographic QCD,''
  JHEP {\bf 0801}, 053 (2008)
  [arXiv:0708.1322 [hep-th]].
  %%CITATION = JHEPA,0801,053;%%

\bibitem{ABJM}
%\cite{Aharony:2008ug}
  O.~Aharony, O.~Bergman, D.~L.~Jafferis and J.~Maldacena,
  ``N=6 superconformal Chern-Simons-matter theories, M2-branes and their
  gravity duals,''
  JHEP {\bf 0810}, 091 (2008)
  [arXiv:0806.1218 [hep-th]].
  %%CITATION = JHEPA,0810,091;%%


\bibitem{Ypks}

J.~P.~Gauntlett, D.~Martelli, J.~F.~Sparks and D.~Waldram,
  ``A new infinite class of Sasaki-Einstein manifolds,''
  Adv.\ Theor.\ Math.\ Phys.\  {\bf 8}, 987 (2006)
  [arXiv:hep-th/0403038].
  %%CITATION = 00203,8,987;%%


%\cite{Herzog:2009gd}
\bibitem{Igor}
  C.~P.~Herzog, I.~R.~Klebanov, S.~S.~Pufu and T.~Tesileanu,
  ``Emergent Quantum Near-Criticality from Baryonic Black Branes,''
  arXiv:0911.0400 [hep-th].
  %%CITATION = ARXIV:0911.0400;%%


\bibitem{Q111}
D.~Fabbri, P.~Fre', L.~Gualtieri, C.~Reina, A.~Tomasiello, A.~Zaffaroni and A.~Zampa,
  ``3D superconformal theories from Sasakian seven-manifolds: New  nontrivial
  evidences for AdS(4)/CFT(3),''
  Nucl.\ Phys.\  B {\bf 577}, 547 (2000)
  [arXiv:hep-th/9907219].
  %%CITATION = NUPHA,B577,547;%%


S.~Franco, A.~Hanany, J.~Park and D.~Rodriguez-Gomez,
  ``Towards M2-brane Theories for Generic Toric Singularities,''
  JHEP {\bf 0812}, 110 (2008)
  [arXiv:0809.3237 [hep-th]].
  %%CITATION = JHEPA,0812,110;%%

S.~Franco, I.~R.~Klebanov and D.~Rodriguez-Gomez,
  ``M2-branes on Orbifolds of the Cone over $Q^{1,1,1}$,''
  JHEP {\bf 0908}, 033 (2009)
  [arXiv:0903.3231 [hep-th]].
  %%CITATION = JHEPA,0908,033;%%



\bibitem{deBoer:2009wk}
  J.~de Boer, K.~Papadodimas and E.~Verlinde,
  ``Holographic Neutron Stars,''
  arXiv:0907.2695 [hep-th].
  %%CITATION = ARXIV:0907.2695;%%


%\cite{Hartnoll:2007ai}
\bibitem{Hartnoll:2007ai}
  S.~A.~Hartnoll and P.~Kovtun,
  ``Hall conductivity from dyonic black holes,''
  Phys.\ Rev.\  D {\bf 76}, 066001 (2007)
  [arXiv:0704.1160 [hep-th]].
  %%CITATION = PHRVA,D76,066001;%%

%\cite{Born:1934gh}
\bibitem{Born:1934gh}
  M.~Born and L.~Infeld,
  ``Foundations Of The New Field Theory,''
  Proc.\ Roy.\ Soc.\ Lond.\  A {\bf 144}, 425 (1934).
  %%CITATION = PRSLA,A144,425;%%

\bibitem{otherinstabilities}

B.~McInnes,
  ``Instability of Near-Extremal Black Holes Dual to Strongly Coupled Field
  Theories on Flat Spacetime,''
  arXiv:0905.1180 [hep-th];
  %%CITATION = ARXIV:0905.1180;%%

G. Horowitz and M. Rangamani, work in progress.

%\cite{Polchinski:1998rr}
\bibitem{Polchinski:1998rr}
  J.~Polchinski,
  ``String theory. Vol. 2: Superstring theory and beyond,''
%\href{http://www.slac.stanford.edu/spires/find/hep/www?irn=4634802}{SPIRES entry}
{\it  Cambridge, UK: Univ. Pr. (1998) 531 p}

%
%\cite{DeWolfe:2001nz}
%\bibitem{Garyproduct}
%  O.~DeWolfe, D.~Z.~Freedman, S.~S.~Gubser, G.~T.~Horowitz and I.~Mitra,
%  ``Stability of AdS(p) x M(q) compactifications without supersymmetry,''
%  Phys.\ Rev.\  D {\bf 65}, 064033 (2002)
%  [arXiv:hep-th/0105047].
  %%CITATION = PHRVA,D65,064033;%%




\end{thebibliography}
\end{document}